\documentclass[reprint, amsmath, amssymb,aps,prd,floatfix, eqsecnum, longbibliography]{revtex4-1}

\usepackage{amsmath, graphicx,natbib, bm ,array,hyperref}
\hypersetup{colorlinks = true,allcolors = blue }

\newcolumntype{L}{>{$}l<{$}}
\newcolumntype{C}{>{$}c<{$}}
\newcolumntype{R}{>{$}r<{$}}

\DeclareMathOperator\sgn{sgn}
\DeclareMathOperator\sgnleft{\overleftarrow{\text{sgn}}}
\DeclareMathOperator\sgnright{\overrightarrow{\text{sgn}}}
\DeclareMathOperator{\sinc}{sinc}

\begin{document}

\newcommand{\pd}{\partial}
\newcommand{\beq}{\begin{equation}}
\newcommand{\eeq}{\end{equation}}
\newcommand{\bseq}{\begin{subequations}}
\newcommand{\eseq}{\end{subequations}}
\newcommand{\bal}{\begin{align}}
\newcommand{\eal}{\end{align}}

\newcommand{\coeffI}{Z_I}
\newcommand{\coeffP}{Z_P}
\newcommand{\fermifn}{f}

\renewcommand{\arraystretch}{1.2}

\title{Many-body wavefunctions for quantum impurities out of equilibrium. I. The nonequilibrium Kondo model}

\author{Adrian B. Culver}
 \email[Present address: Mani L. Bhaumik Institute for Theoretical Physics, Department of Physics and Astronomy, University of California, Los Angeles, California 90095, USA; ]{adrianculver@physics.ucla.edu}
\author{Natan Andrei}
 \email{natan@physics.rutgers.edu}
\affiliation{Center for Materials Theory, 
Department of Physics and Astronomy, Rutgers University, Piscataway, New Jersey 08854, USA
}

\begin{abstract}
We present here the details of a method [A. B. Culver and N. Andrei, \href{https://doi.org/10.1103/PhysRevB.103.L201103}{Phys. Rev. B {\bf 103}, L201103 (2021)}] for calculating the time-dependent many-body wavefunction that follows a local quench.  We apply the method to the voltage-driven nonequilibrium Kondo model to find the exact time-evolving wavefunction following a quench where the dot is suddenly attached to the leads at $t=0$.  The method, which does not use Bethe ansatz, also works in other quantum impurity models and may be of wider applicability.  We show that the long-time limit (with the system size taken to infinity first) of the time-evolving wavefunction of the Kondo model is a current-carrying nonequilibrium steady state that satisfies the Lippmann-Schwinger equation.  We show that the electric current in the time-evolving wavefunction is given by a series expression that can be expanded either in weak coupling or in strong coupling, converging to all orders in the steady-state limit in either case.  The series agrees to leading order with known results in the well-studied regime of weak antiferromagnetic coupling and also reveals a universal regime of \emph{strong ferromagnetic} coupling with Kondo temperature $T_K^{(F)} = D e^{-\frac{3\pi^2}{8} \rho |J|}$ ($J<0$, $\rho|J|\to\infty$).  In this regime, the differential conductance $dI/dV$ reaches the unitarity limit $2e^2/h$ asymptotically at \emph{large} voltage or temperature.
\end{abstract}

\maketitle

\section{Introduction}\label{sec: Introduction}

In a quantum quench, the ground state of an initial Hamiltonian $H_i$ is evolved in time by a final Hamiltonian $H_f$ following a sudden change of parameters.  As this time evolution is unitary, quench calculations are usually applied to closed systems; however, the quench formalism can also be used to make predictions for open, driven systems.  In the case of a sudden and spatially localized quench, the long-time limit (with the system size always large enough so that the effect of the quench does not reach the boundaries) may yield a nonequilibrium steady state (NESS) that carries current and generates entropy.  The study of quenches that result in a NESS is a promising direction for gaining insights into nonequilibrium phenomena.

A simple physical quantity to characterize  a quench is the expectation value of an observable: $\mathcal{O}(t) = \langle \Psi| e^{i H t} \widehat{\mathcal{O}} e^{-i H t} | \Psi\rangle$, where $|\Psi\rangle$ is the initial state and $H=H_f$ is the Hamiltonian that is switched on suddenly at $t=0$.  Some basic questions arise: Does $\mathcal{O}(t)$ reach a limit as $t\to\infty$?  If so, does this limit coincide with the expectation value in the NESS state, that is, do we have $\lim_{t \to \infty} \mathcal{O}(t) = \langle \Psi_{\text{NESS}}| \widehat{\mathcal{O}} |\Psi_{\text{NESS} }\rangle$  ?  In the case of the electric current in the Kondo model, we answer both questions with ``yes.''  The methods of calculation that we introduce to arrive at these answers could be of wider use. 

In the nonequilibrium Kondo model, a localized quantum impurity (the dot) is coupled via spin exchange to two reservoirs of electrons (the leads).  Experimentally, this system is realized in quantum dot systems, in which electrons are confined to a nanoscale region and a single unpaired electron acts as the impurity in the Coulomb blockade regime \cite{Goldhaber-Gordon,Cronenwett,vanderWiel,KretininEtAl}.

The universal antiferromagnetic regime of the nonequilibrium Kondo model has been studied theoretically by a variety of approaches, including Keldysh perturbation theory \cite{KaminskiNazarovGlazman, RoschEtAlPRL, DoyonAndrei}, flow equations \cite{Kehrein}, the real-time renormalization group \cite{Schoeller, PletyukhovSchoeller}, and the variational principle \cite{AshidaEtAl}; the Kondo regime has also been studied in the Anderson model using perturbation theory \cite{HershfieldDaviesWilkins}, Fermi liquid theory \cite{Oguri}, integrability \cite{KonikSaleurLudwig}, time-dependent density matrix renormalization group \cite{Heidrich-MeisnerEtAl, EckelEtAl}, scattering Bethe ansatz \cite{ChaoPalacios}, dynamical mean field theory \cite{DordaEtAl}, quantum Monte Carlo \cite{ProfumoEtAl}, and numerical renormalization group combined with time-dependent density matrix renormalization group \cite{SchwarzEtAl}.  A much more complete list of theoretical works on this subject is found in the references in \cite{AshidaEtAl}.  The strong ferromagnetic regime that we explore with our method (and show the universality of) has received little attention \footnote{In Ref. \cite{LloydEtAl}, a mapping between the weak antiferromagnetic regime of the spin $S>1/2$ Kondo model and the strong ferromagnetic regime of the spin $S’=S-1/2$ model was constructed and used to study the antiferromagnetic model. This mapping has some similarity to the correspondence that we note in the $S=1/2$ model in this paper. Here, however, we propose the strong ferromagnetic regime itself as universal.}.

\begin{figure}[htp]
    \includegraphics[width=.9\linewidth]{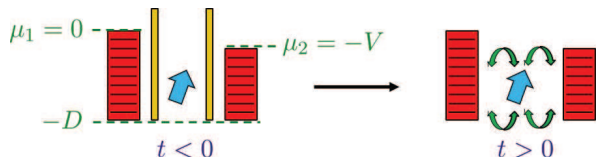}
    \caption{Schematic of the quench process.  Prior to $t=0$, the leads are filled with free electrons, with no tunneling to the dot allowed.  From $t=0$ onward, the system evolves with the many-body Hamiltonian $H_{\text{Kondo}}$, with tunneling to and from the leads resulting in an electric current.}
    \label{fig: quench}
\end{figure}

We consider a quench setup in which the uncoupled system consists of Fermi seas in each lead;  the difference in chemical potentials represents an externally imposed bias voltage.  The quench at $t=0$ consists of switching on the coupling to the dot, after which the system evolves by the full Kondo Hamiltonian and an electric current develops (see Fig. \ref{fig: quench}).

In this paper, we present a method \cite{CulverAndrei_PRBLetter} for \nocite{LloydEtAl}calculating the wavefunction following a local quench, which also applies to other quantum dot problems and may have wider applicability.  We calculate the exact many-body wavefunction following the quench described above, then use it to find a series expression for the electric current as a function of time.  We concentrate on the steady-state current measured at energy scales much smaller than the bandwidth, where it is a universal function governed by an emergent scale: the Kondo temperature $T_K$.  We compare with prior work in the much-studied weak coupling antiferromagnetic regime, then proceed to identify another universal regime: \emph{strong ferromagnetic} coupling, with its own scale $T^{(\text{F})}_K$.

With universality in mind, we study the two-lead Kondo model in the wide-band limit \cite{DoyonAndrei}:
\begin{multline}
    H_{\text{Kondo}} = -i \int_{-L/2}^{L/2 } dx\ \sum_{\gamma=1,2} \psi_{\gamma a}^\dagger(x) \frac{d}{dx}\psi_{\gamma a}(x) \\
    + \sum_{\gamma,\gamma' =1,2}\frac{1}{2} J \psi_{\gamma a}^\dagger(0) \bm{\sigma}_{a a'} \psi_{\gamma' a'}(0) \cdot \mathbf{S} - B  S^z. \label{eq: H for two-lead Kondo}
\end{multline}
This one dimensional Hamiltonian \footnote{Sum over repeated spins is implied throughout, and we use natural units: $\hbar= e = k_B = v_F =\mu_S = 1$.} captures the universal low-energy physics of more realistic models, and can be obtained by following the standard steps of linearizing the energy spectrum about the Fermi level and unfolding to obtain right-moving electrons.  We have taken the coupling of the dot to the leads to be symmetric, and put a magnetic field $B\hat{z}$ on the dot.  The Kondo coupling $J$ is dimensionless in our convention; we can make contact with the usual convention by expressing our final results in terms of the usual quantity $g\equiv \rho J$ (where $\rho=\frac{1}{2\pi}$ is the density of states per unit length in our convention.)

Prior to the quench, we assume that the bias voltage is applied but the tunneling to the dot is blocked.  That is, the initial density matrix is a product $\rho = \exp \left[ -\frac{1}{T_1}\sum_{|k|<D}(k-\mu_1) c_{1ka}^\dagger c_{1ka} \right] \otimes \exp \left[ -\frac{1}{T_2} \sum_{|k|<D}(k-\mu_2) c_{2ka}^\dagger c_{2ka} \right]$ of filled Fermi seas in each lead (cut off by the bandwidth $D$), with the bias voltage appearing as $V= \mu_1 - \mu_2$.  At $t=0$, we turn on the Kondo coupling $J$ and the system evolves via the many-body Hamiltonian $H$, with the time-evolving density matrix $\rho(t) = e^{-i H t} \rho e^{i H t}$.
Since the total number of electrons in the system is conserved, the (average) electric current at time $t$ is the time derivative of the number of electrons in one of the leads:
\beq
    I(t) \equiv - \frac{d}{dt} \text{Tr}\left[ \rho(t) \hat{N}_1 \right]/ \text{Tr} \rho, \label{eq: I(t) basic def}
\eeq
where $\hat{N}_1=  \int_{-L/2}^{L/2}dx\ \psi_{1a}^\dagger(x) \psi_{1a}(x)$.  (We note here that although we focus on the current, our formalism can also be used to calculate other quantities.)  Since we have linearized the spectrum, the answers we obtain for small numbers of electrons have no physical meaning.  Rather than evaluate our results for a large but finite number of electrons, we find it more convenient to take the thermodynamic limit: the system size $L\to \infty$ with fixed density.  In this limit, the time $t$ is held fixed.  This guarantees that the effects of the quench, which travel at the Fermi velocity, never reach the (artificial) periodic boundaries of the system.  As shown in more detail in Ref. \cite{DoyonAndrei}, this order of limits permits us to describe what is physically an open, driven system using a formalism of unitary time evolution.

One of the main results of this paper is the exact solution of the many-body density matrix $\rho(t)$.  The method we introduce allows us to find the exact time-evolving wavefunction starting from any number of electrons with arbitrary lead indices, momenta, and spins; this suffices to construct the density matrix.  We show that in the long-time limit, with the system size always larger, the time-evolving wavefunction becomes a Lippmann-Schwinger ``in'' state, that is, an eigenstate of the Hamiltonian that satisfies the incoming boundary condition of $N$ plane waves with the specified quantum numbers.  This provides an exact and explicit example of a nonequilibrium steady state (NESS) in a many-body problem.  We can also solve for this NESS directly using a time-independent version of our formalism.

With the many-body wavefunction in hand, we turn to the calculation of the current at time $t$ following the quench.  A lengthy calculation based on Wick's theorem brings the current to a form in which is suitable for taking the thermodynamic limit; this limit yields a series expression for the current.  This series has the interesting property that it really yields two series: one in powers of $J$ for small $J$, and one in powers of $1/J$ for large $|J|$.

We use the series to answer the two basic questions raised earlier in this Introduction.  We show that our series expression for the current reaches a long-time limit to all orders (in either $J$ or $1/J$), and that this limit agrees with the expectation value of the current operator in the NESS. 
We then evaluate the first several terms of the series, focusing on both the usual universal regime of weak antiferromagnetic coupling and a universal regime of \emph{strong ferromagnetic} coupling.  In each regime, we allow the external parameters $T_1$, $T_2$, and $V$ to be arbitrary in order to investigate the scaling properties of the steady-state current using the Callan-Symanzik equation.  We find the standard scaling at leading order for weak antiferromagnetic coupling, and a Kondo temperature $T_K^{(F)}$ for strong ferromagnetic coupling.

This paper is organized as follows.  In Sec. \ref{sec: Time-dependent many-body wavefunction}, we present a formalism for quench dynamics and apply it to the two-lead Kondo model.  We take the long-time limit to find the NESS.  In Sec. \ref{sec: The electric current}, we use the time-evolving wavefunction to find a series expression for the current following the quench.  We discuss the power counting for $J\to0$ and $|J|\to\infty$, then consider the steady-state limit.

A considerable amount of technical material is deferred to the Appendices, in which we develop a number of techniques for manipulating the many-body wavefunction and calculating its matrix elements.  The efficient notation we introduce in Appendix \ref{sec: Notation for calculations} is essential for comprehending the remaining appendices.  Further details on the work presented in this paper are available in Ref. \cite{Culver_thesis}.

\section{Time-dependent many-body wavefunction}\label{sec: Time-dependent many-body wavefunction}
We present the exact wavefunction $e^{-i H t}|\Psi\rangle$ of the two-lead Kondo model given an initial state $|\Psi\rangle$ built from an arbitrary number of momentum creation operators.  We show that the wavefunction goes to a NESS at large time (with the system size taken to infinity first) and present the NESS explicitly.  The general formalism we develop for finding the wavefunction may be of wider use, though we have so far only applied it to quantum impurity models with linearized leads.

We begin in Sec. \ref{sec: Time evolution: General formalism} with the general formalism, using the Kondo model as an example.  The formalism replaces the many-body Schrodinger by an equivalent set of differential equations.  With minor adjustments, this formalism can also be applied to finding NESS wavefunctions directly, without following the full time evolution.  In Sec. \ref{sec: Two-lead Kondo model: Reduction to the one-lead case}, we show that the differential equations we need to solve in the two-lead Kondo model reduce to those of the one-lead Kondo model; we then solve these equations in Sec. \ref{sec: Crossing states of the Kondo model}, completing the solution.  In Sec. \ref{sec: Solution in an alternate basis}, we present the same solution in an alternate basis that makes the physics of large coupling more transparent.  In Sec \ref{sec: The nonequilibrium steady state},  we find the NESS explicitly as the long-time limit of the time-evolving wavefunction.

\subsection{Time evolution: General formalism}\label{sec: Time evolution: General formalism}
The general formalism we now set up is a way of reducing the original many-body Schrodinger equation to an infinite family of differential equations that we call ``inverse problems.''  For a generic Hamiltonian, this family of inverse problems may be just as intractable as the Schrodinger equation; however, they can be solved in closed form in the Kondo model.  The extension of the formalism to models with charge fluctuations is deferred to our next paper.

We first illustrate the idea by considering the simple case of a quench of two electrons ($N=2$) in the Kondo model \eqref{eq: H for two-lead Kondo}.  Suppressing spin and lead indices and ignoring antisymmetrization for the moment, we write the two-electron wavefunction in position space as a function $\phi(t,x_1,x_2)$.  Since the quench occurs precisely at $x=0$, and since the electrons of the model travel rightward at the Fermi velocity (which has been set to unity), the effect of the quench is contained in the ``light cone'' from $x=0$ to $t$.  Thus, if both $x_1$ and $x_2$ are inside the light cone, then the function $\phi(t,x_1,x_2)$ is complicated; if both are outside, then the function is simple; and if one is inside and the other outside, then the function is a product of a simple function and a complicated function (each of $t$ and of one position variable).  This discussion generalizes to the $N$-particle case.  Our method is an exact reformulation of the many-body Schrodinger equation which takes care of all the simple parts of the problem (outside of the light cone) and isolates the hard part of the problem, namely, the differential equations for the complicated functions inside the light cone.  We note here that the discussion above applies to the linearized models we consider, since there is a light cone and the non-interacting problem is simple; however, the reformulation we present below is more general, and could potentially be of use in a wider class of problems.

We begin with the Kondo Hamiltonian, separated into non-interacting and interacting parts:
\bseq
\begin{align}
    H^{(0)} &= -i \int_{-L/2}^{L/2 } dx\ \sum_{\gamma=1,2} \psi_{\gamma a}^\dagger(x) \frac{d}{dx}\psi_{\gamma a}(x)  - B  S^z,\label{eq: H0 two-lead Kondo} \\
    H^{(1)} &= \sum_{\gamma,\gamma' =1,2}\frac{1}{2} J \psi_{\gamma a}^\dagger(0) \bm{\sigma}_{a a'} \psi_{\gamma' a'}(0) \cdot \mathbf{S}, \label{eq: H1 two-lead Kondo}\\
    H &= H^{(0)} + H^{(1)}.\label{eq: H=H0+H1 two-lead Kondo}
\end{align}
\eseq
The problem is to calculate the time evolution of an initial state with arbitrary quantum numbers $\gamma_1 k_1 a_1,\dots, \gamma_N k_N a_N$ for the leads and $a_0$ for the impurity:
\beq
    |\Psi(t) \rangle \equiv e^{-i H t} c_{\gamma_N k_N a_N}^\dagger \dots c_{\gamma_1 k_1 a_1}^\dagger |a_0 \rangle,
\eeq
where $c_{\gamma k a}^\dagger \equiv \frac{1}{\sqrt{L}} \int_{-L/2}^{L/2}dx\ e^{i k x} \psi_{\gamma a}^\dagger(x)$ and where the impurity state, $|a_0\rangle = |\pm\frac{1}{2}\rangle$, implicitly includes a tensor product with the vacua of the leads (i.e we have $c_{\gamma k a}|a_0\rangle =0$).  Equivalently, we need to solve the differential equation
\beq
    \left( H - i \frac{d}{dt} \right)|\Psi(t) \rangle = 0,
\eeq
with the initial condition
\beq
    |\Psi(t=0) \rangle = \left( \prod_{j=1}^N c_{\gamma_j k_j a_j}^\dagger \right) |a_0 \rangle.
\eeq
To begin our construction of the solution, we define time-evolving impurity states that evolve by $H^{(0)}$ only:
\beq
    |a_0(t) \rangle = e^{-i H^{(0)} t} |a_0\rangle = e^{i  a_0 B t} |a_0\rangle ,
\eeq
and we also define a set of time-dependent operators $c_{\gamma k a}^\dagger(t)$ that describe the free evolution of the electron quantum numbers:
\beq
    c_{\gamma k a}^\dagger(t) = e^{-i H^{(0)} t} c_{\gamma k a}^\dagger e^{i H^{(0)} t} = e^{-i k t} c_{\gamma k a}^\dagger.\label{eq: cdagger(t) simplest}
\eeq
Note that the signs in the exponents are the opposite from the interaction picture.  The motivation for these definitions is that in the simplest case, $J=0$ (no interaction), the full solution for the time evolution is
\beq
    |\Psi^0(t)\rangle \equiv \left(\prod_{j=1}^N  c_{\gamma_j k_j a_j }^\dagger(t) \right) |a_0(t)\rangle,
\eeq
as can be seen by canceling each factor of $1 = e^{i H^{(0)}t } e^{-i H^{(0)} t}$.  So far, this is essentially the approach used by Gurvitz to study transport in non-interacting Floquet models \cite{Gurvitz}.  To allow interactions, we will systematically add a \emph{finite} number of correction terms to $|\Psi^0(t)\rangle$ to arrive at the full, exact solution $|\Psi(t)\rangle$.

We define an operator $A_{\gamma k a}(t)$ that plays a large role in the following calculations.  The idea is that it measures the amount by which the $c_{\gamma k a}^\dagger(t)$ operators fail to describe the full time evolution:
\bseq
\begin{align}
    &A_{\gamma k a}(t) \equiv [ H, c_{\gamma k a}^\dagger(t)] - i \frac{\partial }{\partial t} c_{\gamma k a}^\dagger(t)\\
    &= \frac{1}{2\sqrt{ L}}  J e^{-i kt}  \left( \psi_{1b}^\dagger(0) + \psi_{2b}^\dagger(0) \right) \bm{\sigma}_{b a} \cdot \mathbf{S}.\label{eq: Agamma(t) first formula}
\end{align}
\eseq
Using this operator, we proceed to show the equivalence of the many-body Schrodinger equation to a set of ``inverse problems.''  We present the exact solution of these inverse problems in the Kondo model in the next section; the equivalence, though, can be better seen by working in a more general setting, without yet using some of the more specific details of the Kondo model.  We thus consider a model in which the Hilbert space consists of ``impurity states'' $|\beta\rangle$ (for simplicity $\beta$ ranges over a finite set) and any states produced by ``field operators'' $c_\alpha^\dagger$ acting on impurity states, where $\alpha$ may stand for any quantum numbers.  We assume that the impurity states are vacua of the field operators (i.e., $c_\alpha |\beta\rangle = 0$).  In the Kondo model \eqref{eq: H=H0+H1 two-lead Kondo}, the impurity states are the two possible configurations of the impurity spin along the $z$-axis ($\beta\equiv a_0=\pm 1/2$), and the field operators are the electron creation operators for the leads ($\alpha\equiv \gamma k a$, $c_{\alpha}^\dagger \equiv c_{\gamma k a}^\dagger$).

Given initial quantum numbers $\alpha_1,\dots,\alpha_N$ and $\beta$, we wish to solve the Schrodinger equation:
\beq
    \left( H- i \frac{d}{dt} \right)|\Psi(t)\rangle =0,
\eeq
with the initial condition
\beq
    |\Psi(t=0)\rangle = c_{\alpha_N}^\dagger \dots c_{\alpha_1}^\dagger |\beta\rangle. \label{eq: initial condition}
\eeq
We write the Hamiltonian as $H = H^{(0)} + H^{(1)}$ and define time-dependent impurity states $|\beta(t)\rangle \equiv e^{-i H^{(0)} t} |\beta\rangle$, time-dependent field operators $c_\alpha^\dagger(t) \equiv e^{-i H^{(0) } t} c_\alpha^\dagger e^{i H^{(0)} t}$, and operators $A_\alpha(t) \equiv  [ H, c_{\alpha}^\dagger(t)] - i \frac{\partial }{\partial t} c_{\alpha}^\dagger(t)$.  The key conditions for our formalism are the following (all of which are easily verified in the Kondo model):
\begin{itemize}
    \item $H^{(0)}$ maps any impurity state into some linear combination of impurity states:
    \beq
        H^{(0)} |\beta \rangle = \sum_{\beta' } u_{\beta \beta'} |\beta'\rangle. \label{eq: condition 1}
    \eeq
    \item  $H^{(1)}$ annihilates any impurity state:
    \beq
        H^{(1)} |\beta\rangle = 0. \label{eq: condition 2}
    \eeq
    \item Any $A(t)$ anticommutes with any $c^\dagger(t)$: 
    \beq
        \{ A_{\alpha_2}(t) , c_{\alpha_1}^\dagger(t) \} = 0.\label{eq: condition 3}
    \eeq
\end{itemize}
The third condition effectively restricts this paper to models in which the interaction term $H^{(1)}$ is quadratic in field operators.  (Note that the model can still be interacting if there are non-commuting operators acting on impurity states, as occurs with the Pauli matrices in the Kondo model.)  In quantum impurity models, this means that only spin fluctuations are allowed.  In our next paper, we present a version of our formalism that applies to models with a number-conserving quartic interaction term, thus allowing us to explore charge fluctuations, as well.   

In the special case of no interaction ($H^{(1)} =0$), the solution is a product of time-evolving field operators acting on the time-evolving impurity state:
\beq
    |\Psi^0(t)\rangle = c_{\alpha_N}^\dagger(t) \dots c_{\alpha_1}^\dagger(t) |\beta(t)\rangle.
\eeq
Our main result of this section, the reformulation of the many-body Schrodinger equation in the interacting case, is stated below in Eqs. \eqref{eq: wavefn construction}, \eqref{eq: chi condition 1}, and \eqref{eq: chi condition 2}.  We proceed to build up to this result by presenting several special cases.  

\begin{figure}[htp]
    \includegraphics[width=.7\linewidth]{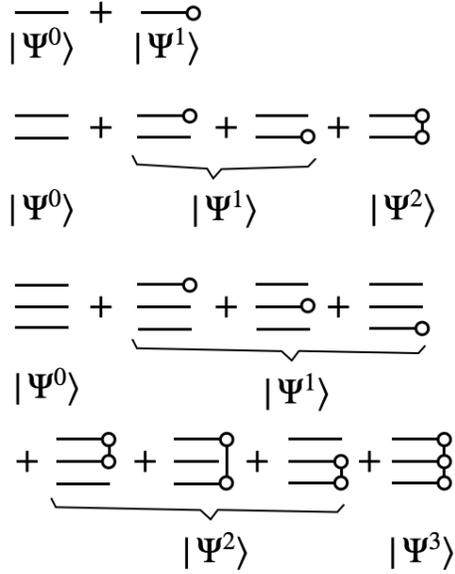}
    \caption{The wavefunction $|\Psi(t)\rangle = |\Psi^0(t)\rangle  + \dots + |\Psi^N(t)\rangle $ for $N=1,2,$ and $3$.  Each line represents a quantum number $\alpha_j$ of the initial state ($j=1,\dots,N)$.  Ordinary lines represent $c_\alpha^\dagger(t)$ operators, while each line that ends on a circle represents a quantum number assigned to a ``crossing state'' (see main text).  Sign factors, antisymmetrizations, and dependence on the time $t$ (and on the impurity quantum number $\beta$) are all implicit.}
    \label{fig: Wavefunction examples}
\end{figure}

The next step is to see ``by how much'' the freely evolving state $|\Psi^0(t)\rangle$ fails to satisfy the Schrodinger equation; that is, to compute $(H- i \frac{d}{dt})|\Psi^0(t)\rangle$.  Note that due to conditions \eqref{eq: condition 1} and \eqref{eq: condition 2}, the state $|\beta(t)\rangle$ is annihilated by $H- i\frac{d}{dt}$.  Our approach will be to bring $H$ past all of the $c_\alpha^\dagger(t)$ operators to its right at the cost of commutators [$A_\alpha(t)$ operators].  We then find differential equations that characterize a finite number of terms $|\Psi^1(t)\rangle,\dots,|\Psi^N(t)\rangle$ that are to be added to $|\Psi^0(t)\rangle$ to obtain the full wavefunction.  The state $|\Psi^0(t)\rangle$ already satisfies the correct initial condition \eqref{eq: initial condition}, so each of the added terms will be required to vanish at $t=0$.  We present the cases of $N=1$, $2$, and $3$ as a warm-up (see Fig. \ref{fig: Wavefunction examples} for illustration), then proceed to general $N$.

\subsubsection{N=1}
In this case, the freely evolving state is $|\Psi^0(t)\rangle = c_{\alpha_1}^\dagger(t)|\beta(t)\rangle$.  Bringing $(H- i\frac{d}{dt})$ past the $c_{\alpha_1}^\dagger(t)$ operator to annihilate $|\beta(t)\rangle$ yields
\beq
    \left(H - i \frac{d}{d t} \right) |\Psi^0(t)\rangle = A_{\alpha_1}(t) | \beta(t) \rangle.
\eeq
Let us suppose we can construct a state $|\chi_{\alpha_1,\beta}(t)\rangle$ which is the ``inverse of $A_{\alpha_1}(t)|\beta(t)\rangle$'' in the following precise sense:
\bseq
    \begin{align}
        \left(H - i \frac{d}{dt}\right) |\chi_{\alpha_1,\beta}(t)\rangle &= - A_{\alpha_1}(t) |\beta(t)\rangle, \label{eq: chi1 Schrod}\\
        |\chi_{\alpha_1,\beta}(t=0)\rangle &= 0. \label{eq: chi1 initial condition}
    \end{align}
\eseq
Given such a state (which we explicitly construct in the Kondo model in Sec. \ref{sec: Crossing states of the Kondo model}), the full solution is immediate:
\beq
    |\Psi(t) \rangle = |\Psi^0(t)\rangle + |\Psi^1(t)\rangle,
\eeq
where $|\Psi^1(t)\rangle = |\chi_{\alpha_1,\beta}(t)\rangle$.  The point of these manipulations is that the state $|\chi_{\alpha_1,\beta}(t)\rangle$ appears again in the solution for larger $N$.  For reasons that become clear once we find explicit expressions in the Kondo model, we refer to $|\chi_{\alpha_1,\beta}(t)\rangle$ as a ``crossing state.''

\subsubsection{N=2}
The freely evolving state is $|\Psi^0(t)\rangle = c_{\alpha_2}^\dagger(t)c_{\alpha_1}^\dagger(t)|\beta(t)\rangle$, and we find
\bseq
\begin{align}
    &\left(H - i \frac{d}{dt} \right) |\Psi^0(t)\rangle = A_{\alpha_2}(t) c_{\alpha_1}^\dagger(t) | \beta(t) \rangle\notag\\
    &\qquad \qquad \qquad\qquad \qquad  +  c_{\alpha_2}^\dagger(t)  A_{\alpha_1}(t)| \beta(t) \rangle\\
    &= c_{\alpha_2} ^\dagger(t)A_{\alpha_1}(t) |\beta(t)\rangle - c_{\alpha_1}^\dagger(t)A_{\alpha_2}(t)|\beta(t) \rangle\label{eq: leftover terms N=2}
\end{align}
\eseq
where we used the third condition above [Eq. \eqref{eq: condition 3}].  To cancel these leftover terms, we reuse the same crossing state $|\chi_{\alpha_1,\beta}(t)\rangle$ that appeared in the $N=1$ case, defining
\beq
    |\Psi^1(t)\rangle = c_{\alpha_2}^\dagger(t) |\chi_{\alpha_1, \beta} (t) \rangle-  c_{\alpha_1}^\dagger(t)  |\chi_{\alpha_2, \beta}(t) \rangle.
\eeq
The point is that, if we bring $(H-i\frac{d}{dt})$ to the right of the $c_\alpha^\dagger(t)$ operators in $|\Psi^1(t)\rangle$, then by the condition \eqref{eq: chi1 Schrod} that the crossing state satisfies, we obtain exactly what we need to cancel the leftover terms on the right-hand side of Eq. \eqref{eq: leftover terms N=2}.  Bringing $(H-i\frac{d}{dt})$ to the right generates new commutators:
\begin{multline}
    \left( H-  i\frac{d}{dt} \right) \left( |\Psi^0(t)\rangle + |\Psi^1(t)\rangle \right) = A_{\alpha_2}(t) |\chi_{\alpha_1, \beta} (t) \rangle\\
    -  A_{\alpha_1}(t)  |\chi_{\alpha_2, \beta}(t) \rangle.
\end{multline}
We are presented with a new ``inverse problem,'' namely to find a state $|\chi_{\alpha_1 \alpha_2,\beta}(t)\rangle$ that satisfies
\bseq
    \begin{align}
        \left(H - i \frac{d}{dt} \right) |\chi_{\alpha_1 \alpha_2,\beta}(t) \rangle &= - A_{\alpha_2}(t) |\chi_{\alpha_1 , \beta}(t)\rangle,\\
        |\chi_{\alpha_1 \alpha_2,\beta}(t=0) \rangle &= 0. 
    \end{align}
\eseq
Given such a state, the full solution is $|\Psi(t)\rangle = |\Psi^0(t)\rangle+|\Psi^1(t)\rangle+|\Psi^2(t)\rangle$, where
\beq
    |\Psi^2(t)\rangle = |\chi_{\alpha_1 \alpha_2,\beta}(t) \rangle - |\chi_{\alpha_2 \alpha_1,\beta}(t) \rangle.
\eeq
This exhibits the pattern that continues to all $N$: the states $|\Psi^1(t)\rangle,\dots, |\Psi^{N-1}(t)\rangle$ are built from crossing states that have been encountered already (up to $N-1$), while $|\Psi^N(t)\rangle$ requires a new crossing state.

\subsubsection{N=3}
Following the same steps for $|\Psi^0(t)\rangle = c_{\alpha_3}^\dagger(t)c_{\alpha_2}^\dagger(t)c_{\alpha_1}^\dagger(t)|\beta(t)\rangle$, we obtain
\beq
    |\Psi(t) \rangle = |\Psi^1(t) \rangle+|\Psi^2(t) \rangle+|\Psi^3(t) \rangle
\eeq
where:
\begin{widetext}
    \bseq
        \begin{align}
            |\Psi^1(t) \rangle &= c_{\alpha_3}^\dagger(t) c_{\alpha_2}^\dagger(t) |\chi_{\alpha_1,\beta}(t) \rangle -  c_{\alpha_3}^\dagger(t) c_{\alpha_1}^\dagger(t) |\chi_{\alpha_2,\beta}(t) \rangle  +c_{\alpha_2}^\dagger(t) c_{\alpha_1}^\dagger(t) |\chi_{\alpha_3,\beta}(t) \rangle,   \\
            |\Psi^2(t)\rangle &= c_{\alpha_3}^\dagger(t) \left( |\chi_{\alpha_1 \alpha_2,\beta} (t)\rangle - |\chi_{\alpha_2 \alpha_1,\beta} (t)\rangle \right) - c_{\alpha_2}^\dagger(t)\left(  |\chi_{\alpha_1 \alpha_3,\beta} (t) \rangle -|\chi_{\alpha_3 \alpha_1,\beta} (t) \rangle \right)\notag\\
            &\qquad + c_{\alpha_1}^\dagger(t) \left( |\chi_{\alpha_2 \alpha_3,\beta} (t) \rangle -|\chi_{\alpha_3 \alpha_2,\beta} (t) \rangle\right),\\
            |\Psi^3(t)\rangle &= |\chi_{\alpha_1 \alpha_2 \alpha_3,\beta} (t) \rangle \pm \left( 5 \text{ permutations} \right),
        \end{align}
    \eseq
\end{widetext}
where $|\chi_{\alpha_1 \alpha_2 \alpha_3,\beta}(t)\rangle$ is a new crossing state we must construct, satisfying
\bseq
    \begin{align}
        \left( H - i \frac{d}{d t} \right) &|\chi_{\alpha_1 \alpha_2 \alpha_3,\beta} (t) \rangle = \notag\\
        &\qquad - A_{\alpha_3}(t) |\chi_{\alpha_1 \alpha_2,\beta}(t) \rangle, \\
        &|\chi_{\alpha_1 \alpha_2 \alpha_3,\beta}(t=0) \rangle =0.
    \end{align}
\eseq

\subsubsection{General N}
Evidently, there are many sums and permutations to keep track of in the case of general $N$.  For this purpose, we have developed a compact notation (see Appendix \ref{sec: Notation for calculations}) which allows us to do the calculation for general $N$ in a few lines (see Appendix \ref{sec: Proof of general formalism}).  Here, we give an overview of the general $N$ case in conventional notation.

We commute $H$ past each $c_\alpha^\dagger(t)$ operator to find
\bseq
    \begin{align}
        &\left(  H - i \frac{d }{dt} \right) |\Psi^0(t)\rangle = \sum_{m=1}^N c_{\alpha_N}^\dagger(t) \dots\notag\\
        &\qquad \times \left( [H, c_{\alpha_m}^\dagger(t)] - i \frac{\partial}{\partial t} c_{\alpha_m}^\dagger(t)\right) \dots c_{\alpha_1}^\dagger(t) |\beta(t)\rangle \\
        &= \sum_{m=1}^N (-1)^{m-1} \left( \prod_{j =1,j \ne m}^N c_{\alpha_j}^\dagger(t)\right)A_{\alpha_m}(t) |\beta(t)\rangle,\label{eq: leftover from Psi0}
    \end{align}
\eseq
where the second equation follows from the condition \eqref{eq: condition 3}, which permits us to bring $A_{\alpha_m}(t)$ past all of the field operators to its right at the cost of a sign factor.  We then define a state $|\Psi^1 (t)\rangle$ as
\beq
    |\Psi^1(t)\rangle = \sum_{m=1}^N (-1)^{m-1}  \left( \prod_{j =1,j\ne m}^N c_{\alpha_j}^\dagger(t)\right) |\chi_{\alpha_m ,\beta}(t) \rangle,
\eeq
where the crossing state $|\chi_{\alpha,\beta}(t)\rangle$ is as in the $N=1$ case.  The point is that if $H-i\frac{d}{dt}$ were to act only on the crossing state, then $(H- i\frac{d}{dt})|\Psi^{(1)}(t)\rangle$ would exactly cancel the right-hand side of Eq. \eqref{eq: leftover from Psi0}.  To reach the crossing state, though, $H - i\frac{d}{dt}$ must commute past each $c_\alpha^\dagger(t)$ operator; we therefore obtain
\begin{multline}
    \left( H - i \frac{d}{d t} \right) \left(|\Psi^0(t)\rangle + |\Psi^1(t)\rangle \right)\\
    = \sum_{1\le m_1 < m_2 \le N} (-1)^{m_1 + m_2 -1}
    \left( \prod_{\substack{j=1 \\ j\ne m_1, m_2 }}^N c_{\alpha_j }^\dagger(t) \right) \\
    \times \Biggr( A_{\alpha_{m_2}}(t) |\chi_{\alpha_{m_1} ,\beta}(t) \rangle 
    - (m_1 \leftrightarrow m_2) \Biggr). \label{eq: leftover from Psi0 + Psi1}
\end{multline}
Note that this equation has a similar structure to Eq. \eqref{eq: leftover from Psi0}, but with $N-2$ of the $c_\alpha^\dagger(t)$ operators appearing instead of $N-1$.  To cancel the new leftover terms, we use the crossing state $|\chi_{\alpha_1 \alpha_2,\beta}(t)\rangle$ that appeared in the $N=2$ case, defining
\begin{multline}
    |\Psi^2(t) \rangle = \sum_{1\le m_1 < m_2 \le N} (-1)^{m_1 + m_2 -1}
    \left( \prod_{\substack{j=1 \\ j\ne m_1, m_2 }}^N c_{\alpha_j }^\dagger(t) \right) \\
    \times \Biggr( |\chi_{\alpha_{m_1} \alpha_{m_2} ,\beta}(t) \rangle - (m_1 \leftrightarrow m_2) \Biggr).
\end{multline}
The action of $H- i\frac{d}{dt}$ on $|\Psi^2(t)\rangle$ then cancels the right-hand side of Eq. \eqref{eq: leftover from Psi0 + Psi1}, leaving an expression of a similar form but with $N-3$ field operators instead of $N-2$.  The new leftover terms are canceled by $|\Psi^3(t)\rangle$ which is built from permutations of the crossing state $|\chi_{\alpha_1 \alpha_2 \alpha_3,\beta}(t)\rangle$, and so on.  This process terminates when all $N$ field operators are eliminated.

In Appendix \ref{sec: Proof of general formalism}, we prove that the full time-evolving wavefunction can be written as
\begin{multline}
        |\Psi(t) \rangle = |\Psi^0(t)\rangle + \sum_{n=1}^N \sum_{1\le m_1 < \dots < m_n \le N} \\
        \times (-1)^{m_1+\dots+m_n + 1}
        \left( \prod_{\substack{ j=1 \\ j \ne m_{\ell}\ \forall \ell }}^N c_{\alpha_j}^\dagger(t) \right) \\
        \times\sum_{\sigma \in \text{Sym}(n) }(\sgn \sigma)
        |\chi_{\alpha_{m_{\sigma_1}} \dots \alpha_{m_{\sigma_n}} ,\beta} (t) \rangle,\label{eq: wavefn construction}
\end{multline}
where the terms in the summation over $n$ are exactly the $|\Psi^1(t)\rangle$, $|\Psi^2(t)\rangle$, etc. states discussed above, and where each crossing state satisfies the appropriate inverse problem:
\bseq
    \begin{align}
        \left(H- i\frac{d}{dt} \right)& |\chi_{\alpha_1 \dots \alpha_n ,\beta}(t) \rangle =\notag\\
        &-A_{\alpha_n}(t) |\chi_{\alpha_1 \dots \alpha_{n-1}, \beta}(t), \label{eq: chi condition 1}\\
        |\chi_{\alpha_1 \dots \alpha_n,\beta}(t=0)\rangle &= 0,\label{eq: chi condition 2}
    \end{align}
\eseq
with $|\chi_{, \beta}(t) \equiv |\beta(t)\rangle$ [so that setting $n=1$ in Eq. \eqref{eq: chi condition 1} reproduces Eq. \eqref{eq: chi1 Schrod}].  We emphasize that this representation of the many-body wavefunction is exact given only the three conditions \eqref{eq: condition 1}, \eqref{eq: condition 2}, and \eqref{eq: condition 3}.

We have thus transformed the original many-body Schrodinger equation to the problem of finding crossing states satisfying Eqs. \eqref{eq: chi condition 1} and \eqref{eq: chi condition 2}.  We turn next to the explicit solution for these crossing states in the Kondo model.

\subsection{Two-lead Kondo model: Reduction to the one-lead case}\label{sec: Two-lead Kondo model: Reduction to the one-lead case}
We return to the particular case of the two-lead Kondo Hamiltonian given in Eqs. \eqref{eq: H0 two-lead Kondo}-\eqref{eq: H=H0+H1 two-lead Kondo}.  Our task in this section is to show that the crossing states of the two-lead model are related in a simple way to those of the one-lead model.  We find the crossing states of the one-lead model in the next section, completing the solution for the wavefunction.

We make the usual transition from the lead $1$/lead $2$ basis to the odd/even ($o/e$) basis:
\beq
    \begin{pmatrix}
        \psi_{oa} \\ \psi_{ea}
    \end{pmatrix}
    =\frac{1}{\sqrt{2}}
    \begin{pmatrix}
        1 & -1 \\
        1 & 1
    \end{pmatrix}
    \begin{pmatrix}
        \psi_{1a} \\ \psi_{2a}
    \end{pmatrix}.
\eeq
Then the odd sector is non-interacting, and the even sector is a copy of the one-lead model:
\bseq
    \begin{align}
        H_o &= -i \int_{-L/2}^{L/2} \psi_{oa}^\dagger(x) \frac{d}{dx} \psi_{oa}(x),\\
        H_e^{(0)} &= -i \int_{-L/2}^{L/2} \psi_{ea}^\dagger(x) \frac{d}{dx} \psi_{ea}(x) - B S^z,\\
        H_e^{(1)} &=  J \psi_{e a}^\dagger(0) \bm{\sigma}_{aa'}\psi_{e a'}(0)\cdot \mathbf{S},\\
        H_e &= H_e^{(0)} + H_e^{(1)},\\
        H^{(0)} &= H_o + H_e^{(0)},\ H^{(1)} = H_e^{(1)}.
    \end{align}
\eseq
In either basis, the time-dependent field operators evolve by phases: \beq
    c_{\gamma k a}^\dagger(t) = e^{-i k t} c_{\gamma k a}^\dagger \qquad( \gamma=1,2, o, \text{ or } e).
\eeq
It is then straightforward to calculate the $A(t)$ operators in either basis:
\beq
    A_{\gamma ka }(t) =
    \begin{cases}
        \frac{1}{\sqrt{L}}  J e^{-i kt}  \psi_{eb}^\dagger(0) \bm{\sigma}_{b a} \cdot \mathbf{S} & \gamma = e, \label{eq: Ae(t)} \\
        \frac{1}{\sqrt{2}} A_{e k a }(t) &  \gamma=1,2,\\
        0 & \gamma = o.
    \end{cases}
\eeq
From the previous section, we know that the solution for the many-body wavefunction follows immediately from the construction of crossing states that satisfy Eqs. \eqref{eq: chi condition 1} and \eqref{eq: chi condition 2}. 
Our primary interest is in the time evolution of two Fermi seas, in particular, a state with quantum numbers in the original lead $1$/lead $2$ basis.  As the interaction is entirely in the even sector, one way to proceed would be to write the original state as a linear combination of states in the odd/even basis, solve the time-evolution problem for states with \emph{even} quantum numbers, and then add the non-interacting odd parts that evolve by phases only.  We instead take a more efficient route: we solve the time evolution problem for a state with even quantum numbers, then we reuse the same crossing states to construct the lead $1$/lead $2$ solution \emph{directly}.  The essential point is that the crossing states for the lead $1$/lead $2$ problem are related to the crossing states for the even problem in a simple way.

If the quantum numbers of the initial state are all in the even sector, then the family of inverse problems [Eqs. \eqref{eq: chi condition 1} and \eqref{eq: chi condition 2}] is
\begin{multline}
    \left(H- i\frac{d}{dt} \right) |\chi_{e k_1 a_1 \dots e k_n a_n,a_0}(t) \rangle =\\
    - A_{e k_n a_n}(t) |\chi_{e k_1 a_1 \dots e k_{n-1} a_{n-1} ,a_0}(t), \label{eq: chi condition even}
\end{multline}
where each $|\chi_e(t)\rangle$ state must vanish at $t=0$.  If instead the quantum numbers of the initial state are in the lead $1$/lead $2$ basis, then the family of inverse problems is
\begin{multline}
    \left(H- i\frac{d}{dt} \right) |\chi_{\gamma_1 k_1 a_1 \dots \gamma_n k_n a_n ,a_0}(t) \rangle =\\
    - \frac{1}{\sqrt{2}} A_{e k_n a_n}(t) |\chi_{\gamma_1 k_1 a_1 \dots \gamma_{n-1} k_{n-1} a_{n-1},a_0}(t), \label{eq: chi condition 1 lead1/lead2}
\end{multline}
where we have used the relation \eqref{eq: Ae(t)} between the $A(t)$ operators in the two bases.  It follows that the crossing states in this case are related to those in the even case by simple numerical prefactors:
\beq
    |\chi_{\gamma_1 k_1 a_1 \dots \gamma_n k_n a_n,a_0 }(t)\rangle = 2^{-n/2} |\chi_{e k_1 a_1 \dots e k_n a_n,a_0 }(t)\rangle.
\eeq
We have therefore reduced the time evolution problem of the two-lead model to the construction of the $|\chi_e(t)\rangle$ states that solve Eq. \eqref{eq: chi condition even}.  For completeness, we write the full wavefunction starting from quantum numbers in the lead $1$/lead $2$ basis (see Fig. \ref{fig: Wavefn lead1lead2}):
\begin{figure}[htp]
    \includegraphics[width=\linewidth]{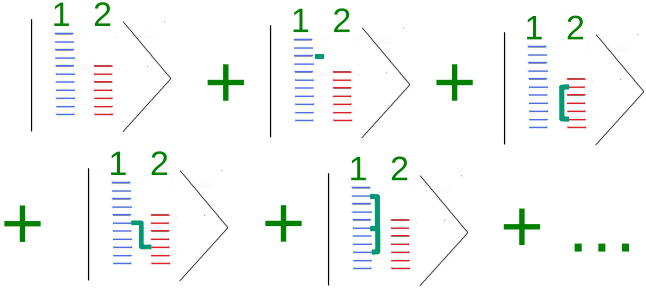}
    \caption{The $N$-body wavefunction of the two-lead Kondo model, either at arbitrary time [Eq. \eqref{eq: Psi(t) starting with lead1/lead2}] or the NESS [Eq. \eqref{eq: PsiNESS final}] that is reached at long time with the system size taken to infinity first.  Lines represent the momenta and spin quantum numbers of electrons in each lead.  Any number of electrons, from lead $1$ or lead $2$, can be put into a crossing state (indicated by connecting lines), which is built from even sector operators only.  For a fixed number $N$ of electrons, the wavefunction is a \emph{finite} sum.}
    \label{fig: Wavefn lead1lead2}
\end{figure}
\begin{widetext}
    \begin{multline}
        e^{-i H t} \left(\prod_{j=1}^N c_{\gamma_j k_j a_j}^\dagger \right) |a_0\rangle = |\Psi^0(t)\rangle + \sum_{n=1}^N 2^{-n/2} \sum_{1\le m_1 < \dots < m_n \le N} (-1)^{m_1+\dots+m_n+1} \left( \prod_{\substack{ j=1 \\ j \ne m_{\ell}\ \forall \ell}}^N c_{\gamma_j k_j a_j}^\dagger(t) \right)\\
        \times \sum_{\sigma \in \text{Sym}(n) }(\sgn \sigma) |\chi_{e k_{m_{\sigma_1}}a_{m_{\sigma_1}} \dots e k_{m_{\sigma_n}}a_{m_{\sigma_n}},a_0}(t)\rangle. \label{eq: Psi(t) starting with lead1/lead2}
    \end{multline}
\end{widetext}
To complete the solution of the wavefunction, we have to construct the crossing states $|\chi_{e k_1 a_1 \dots e k_n a_n, a_0}(t)\rangle$ of the even sector.  This is the core difficulty of the problem, and is presented in the following section.

\subsection{Crossing states of the Kondo model}\label{sec: Crossing states of the Kondo model}
We present the crossing states of the Kondo model.  We find that they are built from products of the single-particle $\mathcal{T}$ matrix for an electron crossing the impurity (hence the name ``crossing'').  We show the calculation in detail for the simplest case, $n=1$.  We then state the result for arbitrary $n\ge1$ and refer the reader to Appendix \ref{sec: Kondo crossing states in the general case} for the proof (which is similar to the $n=1$ calculation).

Taking $n=1$ in Eq. \eqref{eq: chi condition even}, we see that the first ``inverse problem'' is to find a state $|\chi_{e k_1 a_1,a_0}(t)\rangle$ satisfying
\beq
    \left( H- i \frac{d}{dt} \right) |\chi_{e k_1 a_1,a_0}(t) \rangle = - A_{e k_1 a_1}(t) |a_0\rangle, \label{eq: chi condition 1 for Kondo n=1}
\eeq
with the initial condition
\beq
    |\chi_{e k_1 a_1, a_0}(t=0) \rangle = 0.\label{eq: chi condition 2 for Kondo n=1}
\eeq
We make the following ansatz:
\begin{multline}
    |\chi_{e k_1 a_1, a_0}(t)\rangle = \frac{1}{\sqrt{L} } \int_{-L/2}^{L/2} dx_1\ F_{k_1 a_1,a_0}^{b_1,b_0}(t-x_1)\\
    \times \Theta(0<x_1<t) \psi_{e b_1}^\dagger(x_1) e^{i b_0 B t}|b_0\rangle, \label{eq: n=1 crossing state}
\end{multline}
where $F$ is a smooth function that we soon determine, $\Theta(0< x_1 < t) = \Theta(x_1)\Theta(t-x_1)$, and $0 \le t < L/2$.  Evolution to later times is unnecessary, seeing as the regime of interest is $t\ll L$ (so that the effect of the quench does not explore the boundaries of the system); we may as well restrict to $t < L/2$ to avoid the ``coordinate singularity'' at $x=\pm L/2$. 

The state \eqref{eq: n=1 crossing state} vanishes at $t=0$ by construction \footnote{The position space wavefunction of the crossing state at $t=0$ is $\langle b_0| \psi_{b_1}(x_1) |\chi_{e k_1 a_1, a_0}(t=0)\rangle= \frac{1}{\sqrt{L}} F_{k_1 a_1,a_0}^{b_1 ,b_0}(x_1)\Theta(x_1)\Theta(-x_1)$, which could be non-zero at $x_1=0$ if one assigns a non-zero value to the Heaviside function evaluated at the origin.  Even so, the position space wavefunction is certainly zero everywhere except on a set of measure zero (the point $x_1=0$), and there it is non-singular; thus, $|\chi_{e k_1 a_1, a_0}(t=0)\rangle$ has zero overlap with any physically reasonable state.  This in turn means that $|\chi_{e k_1 a_1, a_0}(t=0)\rangle$ is exactly the same as the zero state vector.  Similar comments apply to the general $n$ case discussed in Appendix \ref{sec: Kondo crossing states in the general case}.}, so the initial condition \eqref{eq: chi condition 2 for Kondo n=1} is satisfied.  A short computation yields
\begin{multline}
    \left( H - i \frac{d}{dt} \right) |\chi_{e k_1 a_1 ,a_0}(t)\rangle = \\
    \frac{1}{\sqrt{L}} \left( - i I_{d_1 d_0}^{b_1 b_0} + \frac{1}{4} J \bm{\sigma}_{b_1 d_1}\cdot \bm{\sigma}_{b_0 d_0} \right)\\
    \times F_{k_1 a_1,a_0}^{d_1,d_0}(t)e^{i d_0 B t} \Theta(t) \psi_{e b_1}^\dagger(0) |b_0\rangle,
\end{multline}
where we have made the following replacement:
\beq
    \delta(x_1) \Theta(0<x_1<t) =  \frac{1}{2}\delta(x_1)\Theta(t). \label{eq: delta Theta simplest}
\eeq
Equation \eqref{eq: delta Theta simplest} is equivalent to the regularization $\delta(x)\Theta(x) = \frac{1}{2}\delta(x)$ that has been used in Bethe ansatz calculations in the equilibrium case \cite{AndreiFuruyaLowenstein}; it corresponds to averaging the limits as $x\to0^\pm$ of a function (discontinuous at $x=0$) that is multiplied by $\delta(x)$.

From Eq. \eqref{eq: Ae(t)}, we see
\begin{multline}
    A_{e k_1 a_1}(t)|a_0(t)\rangle = \frac{1}{\sqrt{L}} \frac{1}{2} J e^{-i k_1 t}e^{i a_0 B t} \\
    \times \bm{\sigma}_{b_1 a_1} \cdot \bm{\sigma}_{b_0 a_0} \psi_{e b_1}^\dagger(0)|b_0 \rangle.\label{eq: A(t) on a_0(t)}
\end{multline}
Thus, the differential equation \eqref{eq: chi condition 1 for Kondo n=1} is satisfied for $0< t <L/2$ provided that
\begin{multline}
    \left( - i I_{d_1 d_0}^{b_1 b_0} + \frac{1}{4} J \bm{\sigma}_{b_1 d_1}\cdot \bm{\sigma}_{b_0 d_0} \right) F_{k_1 a_1,a_0}^{d_1,d_0}(t) e^{i d_0 B t}=\\
    -\frac{1}{2} J e^{-i k_1 t} e^{i a_0 B t}\bm{\sigma}_{b_1 a_1}\cdot \bm{\sigma}_{b_0 a_0}.\label{eq: F(t) condition}
\end{multline}
To remove any concern about the differential equation \eqref{eq: chi condition 1 for Kondo n=1} strictly at $t=0$, we consider evolution to arbitrary time $t$ (with $|t|< L /2$) in Appendix \ref{sec: Kondo crossing states in the general case}, and we find that the condition \eqref{eq: F(t) condition} is correct and sufficient.

Our subsequent calculations refer to the identity and spin-flip tensors, defined as
\beq
    I_{a_1 a_0}^{b_1 b_0} = \delta_{a_1}^{b_1}\delta_{a_0}^{b_0},\ P_{a_1 a_0}^{b_1 b_0} = \delta_{a_1}^{b_0}\delta_{a_0}^{b_1}.
\eeq
Using the identity $\bm{\sigma}_{b_0 a_0}\cdot \bm{\sigma}_{b_1 a_1} = 2 P_{b_1 b_0}^{a_1 a_0} - I_{b_1 b_0}^{a_1 a_0}$ and some matrix inversion, we find the following answer:
\beq
    F_{k_1 a_1,a_0}^{b_1, b_0}(t) = e^{ -i [k_1 + (b_0 - a_0) B] t} \left( -i \mathcal{T}_{a_1 a_0}^{b_1 b_0} \right),\label{eq: F}
\eeq
where we have introduced the bare single-particle $\mathcal{T}$ matrix:
\beq
    \mathcal{T} = \frac{\frac{1}{2} J}{1 - i \frac{1}{2} J + \frac{3}{16} J^2} \left[ - \left(1 + i \frac{3}{4}J\right) I + 2 P  \right]. \label{eq: T-matrix J=0 basis}
\eeq
As a check, we note that the corresponding bare $\mathcal{S}$ matrix,
\beq
    \mathcal{S} = I - i \mathcal{T},
\eeq
agrees precisely with the bare $\mathcal{S}$ matrix that appears in the Bethe ansatz solution for the stationary states of the one-lead model (see \cite{Andrei_lecture_notes}, for example).

The generalization of the $n=1$ crossing state \eqref{eq: n=1 crossing state} to general $n\ge1$ is
\begin{widetext}
        \beq
        |\chi_{e k_1 a_1 \dots e k_n a_n,a_0}(t) \rangle = L^{-n/2}\delta_{a_0}^{c_0}\delta_{c_n}^{b_0}\int_0^t dx_1 \dots dx_n    \left( \prod_{j=1}^n F_{k_j a_j, c_{j-1}}^{b_j, c_j}(t-x_j) \psi_{e b_j}^\dagger(x_j)  \right) \Theta(x_n < \dots < x_1) e^{i b_0 B t}|b_0\rangle. \label{eq: Kondo crossing state}
    \eeq
\end{widetext}
In Appendix \ref{sec: Kondo crossing states in the general case}, we show that the construction \eqref{eq: Kondo crossing state} satisfies the appropriate inverse problem, Eq. \eqref{eq: chi condition even}; the calculation reduces to the same condition \eqref{eq: F(t) condition}.  This completes the solution.

We can use the same $|\chi_e(t)\rangle$ crossing states given in Eq. \eqref{eq: Kondo crossing state} to write the exact wavefunction for initial quantum numbers in the even sector; this is the exact time-evolving wavefunction for the one-lead model.  For the case of zero magnetic field, this wavefunction was first found by Tourani \cite{Tourani} using the Yudson contour method \cite{Yudson}; our result here agrees exactly.

It is interesting to note that the integrability of the Kondo model (i.e., the factorization of scattering amplitudes via the Yang-Baxter equation) does not make any obvious appearance in our calculation.  

\subsection{Solution in an alternate basis}\label{sec: Solution in an alternate basis}
Above, we have written the exact wavefunction $|\Psi(t)\rangle$ for the Kondo model starting from field operators that evolve by the free Hamiltonion $H^{(0)}$; we refer to this as the solution in the $J=0$ basis.  It is interesting to note (though not essential for obtaining the results we present later in the paper) that $|\Psi(t)\rangle$ can be written in a $|J|\to\infty$ basis that is more suited to the strong coupling limit.

If the Kondo coupling is sent to infinity (with either sign), then the spin-flip term in the $\mathcal{T}$ matrix \eqref{eq: T-matrix J=0 basis} vanishes: 
\beq
    \lim_{|J|\to \infty} \mathcal{T}_{a_1 a_0}^{b_1 b_0} = - 2i  I_{a_1 a_0}^{b_1 b_0}.
\eeq
In this limit, we have an essentially single-particle problem.  The free particles are not the original electrons with zero phase shift as they cross the impurity, but quasiparticles with a $\pi/2$ phase shift.  The same phase shift is obtained if the Kondo interaction term is replaced by a potential scattering term of infinite strength.

With this motivation, we make an alternate definition of the $c_{\gamma k a}^\dagger(t)$ operators; instead of evolving them by the free ($J=0$) Hamiltonian, we evolve them by the free Hamiltonian plus a potential scattering term of infinite strength:
\beq
    c_{\gamma k a}^\dagger(t) = \lim_{|J'|\to \infty} e^{-i H_{J'}^{(0)} t} c_{\gamma k a }^\dagger e^{i H_{J'}^{(0)} t},
\eeq
where
\beq
    H_{J'}^{(0)} = H^{(0)}+ J' \psi_{eb}^\dagger(0) \psi_{eb }(0). 
\eeq
We can think of this as an alternate choice of what we call $H^{(0)}$ and $H^{(1)}$, or we can note that the calculations  we have done so far also work for any time-evolving $c_\alpha^\dagger(t)$ operators that agree with $c_\alpha^\dagger$ at $t=0$, as long as they anticommute with the resulting $A_\alpha(t)$ operators [the condition \eqref{eq: condition 3}].

We then find that the odd sector operators evolve by phases, as before [$c_{o ka}^\dagger(t) = e^{-i k t} c_{ok a}^\dagger$], while the even operators include a phase shift of $\pi/2$ for crossing the impurity:
\begin{multline}
    c_{e k a}^\dagger(t) = \frac{1}{\sqrt{L}}\int_{-L/2}^{L/2}dx\ e^{-i k(t-x)}\\
    \times \left[  1 - 2 \Theta(0< x< t) \right]\psi_{ea}^\dagger(x)\label{eq: c_e(t) quasiparticle basis},
\end{multline}
where we have taken $0\le t<L/2$.

Proceeding with the method, we find:
\beq
    A_{e k a}(t) = \frac{1}{\sqrt{L}} 2i e^{-i kt} \psi_{ea}^\dagger(0).
\eeq
This in turn leads to a different requirement on the function $F$; Eq. \eqref{eq: F(t) condition} is replaced by
\begin{multline}
    \left( - i I_{d_1 d_0}^{b_1 b_0} + \frac{1}{4} J \bm{\sigma}_{b_1 d_1}\cdot \bm{\sigma}_{b_0 d_0} \right) F_{k_1 a_1,a_0}^{d_1,d_0}(t) e^{i d_0 B t}=\\
    -2i  e^{-i k_1 t} e^{i a_0 B t}I_{a_1 a_0}^{b_1 b_0},\label{eq: F(t) condition qp basis}
\end{multline}
which has the solution
\bseq
    \begin{align}
        &F_{k_1 a_1,a_0}^{b_1, b_0}(t) = e^{ -i [k_1 + (b_0 - a_0) B] t}  i \mathcal{T}_{a_1 a_0}^{b_1 b_0} ,\\
        \mathcal{T} &=   \frac{\frac{1}{2}\widetilde{J} }{1 + i \frac{1}{2}\widetilde{J} + \frac{3}{16}\widetilde{J}^2 }\left[\left( 1 - i \frac{3}{4} \widetilde{J} \right)I + 2P  \right],
    \end{align}
\eseq
where $\widetilde{J} \equiv - \frac{16}{3J}$.  The difference in sign compared to Eq. \eqref{eq: F} is due to the $\pi/2$ phase shift; it can be verified that $\mathcal{T}$ as defined here leads to a unitary $\mathcal{S}$ matrix (while $-\mathcal{T}$ does not).

We emphasize that $|\Psi(t)\rangle$ is the same state vector as before; we are just writing it differently.  The $\mathcal{T}$ matrix in this basis describes the scattering of a single quasiparticle off the impurity.  The electron $\mathcal{T}$ matrix \eqref{eq: T-matrix J=0 basis} found earlier is linear in $J$ for small $J$, while the quasiparticle $\mathcal{T}$ matrix is linear in $1/J$ for large $|J|$; this explains why we find (below) a series for the electric current either in powers of $J$ or of $1/J$.  Either basis can be used for the calculation: the $J=0$ basis makes the $J$ series more manifest and the $1/J$ series less so, while the $|J|=\infty$ basis does the opposite.  We use the $J=0$ basis in the rest of the main text.

\subsection{The nonequilibrium steady state}\label{sec: The nonequilibrium steady state}
In the long-time limit, with the system size taken to infinity first, the time-evolving wavefunction of the Kondo model reaches a nonequilibrium steady state (NESS), as we show in this section.  The NESS can also be solved for directly using a time-independent version of our formalism: one replaces $H- i \frac{d}{dt}$ by $H - E$ and uses time-independent scattering operators that are closely related to the time-dependent field operators.

We begin by writing the exact wavefunction \eqref{eq: Psi(t) starting with lead1/lead2} in a form that makes the time dependence more clear.  Substituting in the explicit construction \eqref{eq: Kondo crossing state} of the crossing states and collecting all phase factors that depend on time, we obtain
\begin{widetext}
    \begin{multline}
        |\Psi(t) \rangle = e^{-i E t} \Biggr[ |\Psi(t=0)\rangle + \sum_{n=1}^N (2L)^{-n/2} \sum_{1\le m_1 < \dots < m_n \le N} (-1)^{m_1+\dots+m_n + 1} \left( \prod_{j=1, j \ne m_{\ell}\ \forall \ell}^N c_{\gamma_j k_j a_j}^\dagger \right)\\
        \times \sum_{\sigma \in \text{Sym}(n) }(\sgn \sigma) \delta_{a_0}^{c_0} \delta_{c_n}^{b_0} \int_0^t  \left( \prod_{j=1}^n F_{k_{m_{\sigma_j}}a_{m_{\sigma_j}}  ,c_{j-1}}^{b_j, c_j}(-x_j) \psi_{e b_j}^\dagger(x_j) dx_j \right)
        \Theta(x_n < \dots < x_1)  |b_0\rangle \Biggr], \label{eq: Psi(t) final}
    \end{multline}
\end{widetext}
where $E = -a_0 B + \sum_{j=1}^N k_j$ is the energy of the initial state.  The time dependence of the wavefunction appears only in the phase factor $e^{-iE t}$ and in the upper limit of $x$ integration.

\begin{figure}[htp]
    \includegraphics[width=.9\linewidth]{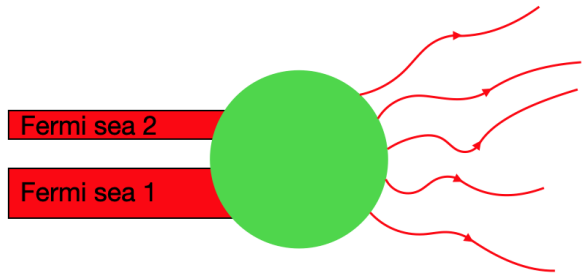}
    \caption{Schematic of the NESS obtained by taking the steady-state limit of $e^{-i H t} e^{i H^{(0)} t} |\Psi\rangle$.  The initial condition at $t=0$ becomes a boundary condition of two incoming Fermi seas, with a complicated result following the scattering off the dot.}
    \label{fig: LS_scattering}
\end{figure}

In the language of wavefunctions, the open system limit \cite{MehtaAndrei_NQI} corresponds to the pointwise limit: that is, we take the long-time limit of the wavefunction at each point $x$ (or more generally, $x_1,\dots,x_N$) without requiring that the limit is reached uniformly for all $x$.  Schematically, letting $|x\rangle$ stand for an $N$-body position state, we have
\beq
    \langle x |\Psi_{\text{NESS}} \rangle =  \lim_{\substack{t\to\infty, L\to \infty \\ t \ll L}} L^{N/2} e^{i E t} \langle x | \Psi(t) \rangle.
\eeq
The phase factor removes the effect of free time evolution (formally, an ``in'' state in scattering theory is the long-time limit of $e^{-i Ht} e^{i H^{(0)} t}|\Psi\rangle$), while the factor of $L^{N/2}$ is a conversion from Kronecker delta normalization to Dirac delta normalization.  Applying this to the time-evolving wavefunction \eqref{eq: Psi(t) final}, we obtain
\begin{widetext}
    \begin{multline}
        |\Psi_{\text{NESS}} \rangle = \left( \prod_{j=1}^N c_{\gamma_j k_j a_j}^\dagger \right)|a_0\rangle + \sum_{n=1}^N 2^{-n/2}\sum_{1\le m_1 < \dots < m_n \le N} (-1)^{m_1+\dots+m_n + 1} \left( \prod_{j=1, j \ne m_{\ell}\ \forall \ell}^N c_{\gamma_j k_j a_j}^\dagger \right)\\
        \times \sum_{\sigma \in \text{Sym}(n) }(\sgn \sigma) \delta_{a_0}^{c_0} \delta_{c_n}^{b_0} \int_0^\infty  \left( \prod_{j=1}^n F_{k_{m_{\sigma_j}}a_{m_{\sigma_j}}  ,c_{j-1}}^{b_j, c_j}(-x_j) \psi_{e b_j}^\dagger(x_j) dx_j \right)
        \Theta(x_n < \dots < x_1)  |b_0\rangle, \label{eq: PsiNESS final}
    \end{multline}
\end{widetext}
where the $c_{\gamma k a}^\dagger$ operators are, in this equation only, Dirac delta normalized [i.e., $c_{\gamma ka}^\dagger = \int dx\ e^{i k x} \psi_{\gamma a}^\dagger(x)$].  This is precisely the form of the Lippmann-Schwinger equation, with $\left( \prod_{j=1}^N c_{\gamma_j k_j a_j}^\dagger \right)|a_0\rangle$ being the free scattering state that encodes the boundary condition of $N$ incoming plane waves.  The initial condition of $|\Psi(t=0)\rangle=|\Psi\rangle$ in the time-dependent view has become a boundary condition (see Fig. \ref{fig: LS_scattering}).
The NESS given by \eqref{eq: PsiNESS final} is a many-body scattering state.  Its structure is very similar to the full solution $|\Psi(t)\rangle$, and it has the same interpretation in terms of free electrons and crossing states.  We can solve for the NESS directly, without following the full time evolution, by using a time-independent version of the formalism of Sec. \ref{sec: Time evolution: General formalism}.

While it is not necessary for understanding our results, we would like to mention the origin of our formalism.  We applied Yudson's contour method \cite{Yudson} to calculate the time-evolving wavefunction and NESS for two electrons ($N=2$) in the infinite-$U$ Anderson impurity model (we later became aware of Ref. \cite{ImamuraEtAl}, which finds the $N=2$ NESS for arbitrary $U$); the form of the NESS was an invaluable clue for us to develop a more general approach.

\section{The electric current in the Kondo model}\label{sec: The electric current}
When the full Kondo Hamiltonian $H$ is turned on at $t=0$, electrons begin to tunnel back and forth from the leads to the dot, and an electric current $I(t)$ develops over time.  Our task in this section is to calculate a series expression for $I(t)$, then to focus in particular on the steady-state limit.  This calculation provides a road map for the evaluation of other observables.

Since the wavefunction is a sum over subsets of the initial $N$ quantum numbers, one would expect an expectation value such as the current to be a double sum over subsets; we show that the double sum diagonalizes to a single sum (over subsets).  The terms in the sum are normal-ordered overlaps (normal ordering is defined below) that can be computed using only the even sector of the model.  We find that $n$-fold sums over momenta have precisely the right $1/L^n$ prefactor so that it is clear how to take the thermodynamic limit, turning sums into integrals.  We arrive at a series answer for the time-evolving current, and we show that it encompasses both a series in $J$ as $J\to0$ and a series in $1/J$ as $|J|\to \infty$.  We show that all orders of either series converge in the steady-state limit.  We then examine the steady-state current in two universal regimes (weak antiferromagnetic and strong ferromagnetic coupling) and two non-universal regimes (weak ferromagnetic and strong antiferromagnetic), with our main focus being on the universal regimes.

Although we have solved for the wavefunction in the presence of an arbitrary magnetic field on the dot, we set the magnetic field to zero in the following calculations.  A non-zero magnetic field introduces infrared difficulties in this model, as noted in Refs. \cite{RoschEtAlPRL} and \cite{ParcolletHooley}.  We return to this topic in the concluding section.

In Sec. \ref{sec: N electrons, arbitrary quantum numbers}, we set up the calculation of the electric current for $N$ electrons and present the reduction to a sum of normal-ordered overlaps.  The essential tool is Wick's theorem.  In Sec. \ref{sec: The current in the thermodynamic limit}, we take the thermodynamic limit to arrive at our series answer.  In Sec. \ref{sec: Steady-state limit of the current}, we consider the steady-state limit of the series.  In Sec. \ref{sec: Antiferromagnetic regime: Universality}, we calculate the current for small $J$, focusing on the antiferromagnetic case.  In Sec. \ref{sec: Ferromagnetic regime: Universality}, we calculate the current for large $|J|$, focusing on the ferromagnetic case.  In Sec. \ref{sec: RG discussion}, we discuss the renormalization group (RG) flow of the model.

\subsection{The current for N electrons}\label{sec: N electrons, arbitrary quantum numbers}
We set up the calculation at zero temperature, then later generalize to allow arbitrary temperatures in the leads.  We have verified that starting with arbitrary temperatures from the beginning results in the same answer for the current \cite{Culver_thesis}.

Since the total number of electrons is constant, the average electric current from lead $1$ to lead $2$ is the time derivative of the number of electrons in lead $1$:
\beq
    I(t) = -\frac{d}{dt} \langle \Psi(t) | \hat{N}_1 |\Psi(t) \rangle,\label{eq: I(t) def}
\eeq
where $\widehat{N}_1 = \int_{-L/2}^{L/2}dx\ \psi_{1a}^\dagger(x) \psi_{1a}(x)$.  Let us first show that $I(t)$ reduces to the evaluation of the expectation value of the bilinear $\psi_{o a}^\dagger(x)\psi_{e a}(x)$.  We write the number operator in the odd/even basis,
\beq
    \hat{N}_1 = \frac{1}{2}\hat{N} + \frac{1}{2}\left( \int_{-L/2}^{L/2}dx\ \psi_{oa}^\dagger(x)\psi_{ea}(x) + \text{H.c.} \right),
\eeq
then use the fact that $\hat{N}\equiv \hat{N_1}+\hat{N_2}$ is conserved to obtain
\beq
    I(t) = - \text{Re} \left[ \frac{d}{dt} \int_{-L/2}^{L/2}dx\ \langle \Psi(t) | \psi_{oa}^\dagger(x) \psi_{ea}(x) | \Psi(t) \rangle \right]. \label{eq: current as integral over x}
\eeq
Although we have the many-body wavefunction for arbitrary initial quantum numbers, we are ultimately interested in taking these quantum numbers to describe two filled Fermi seas.  One might think that it would be simplest to specialize to this case immediately.  However, we find it more convenient to work with arbitrary quantum numbers because the expectation value turns out to be a sum of matrix elements having \emph{every possible subset of the quantum numbers of the originally given state}.

The expectation value of $\psi_{oa}^\dagger(x) \psi_{ea}(x)$ is a sum of terms of the form (schematically)
\beq
    \langle \chi(t) | \left( \prod c(t) \right) \psi_{oa}^\dagger(x) \psi_{ea}(x) \left( \prod c^\dagger(t) \right) |\chi (t)\rangle,
\eeq
where the time-evolving operators and crossing states have various quantum numbers (not necessarily the same assignment on both sides).  It is convenient to anticommute the annihilation operators past the creation operators.  To do this with Wick's theorem, we introduce the \emph{normal ordering} symbol $:X:$ that moves every $c(t)$ operator (in any expression $X$) to the right of every $c^\dagger(t)$ operator, with the appropriate fermionic sign factors.  By definition, the crossing states are unaffected; in other words, this is normal ordering relative to the impurity state $|a_0\rangle$ (\emph{not} relative to a filled Fermi sea), and it only affects the time-dependent single-particle operators (\emph{not} the $\psi_{e}^\dagger$ and $\psi_e$ operators found inside the crossing states).  When we compute the expectation value of $\psi_{oa}^\dagger(x) \psi_{ea}(x)$,  we declare that these two ``external'' operators behave the same way as $c^\dagger(t)$ and $c(t)$ do under the normal ordering symbol.

By Wick's theorem, the product $\prod c(t)  \prod c^\dagger(t)$ is equal to the normal-ordered sum of all contractions, where the contraction of two operators is defined as the product in the original order minus the normal-ordered product (and hence is either the anticommutator, or zero).  It is these contractions that diagonalize the double sum over subsets to a single sum.

As a warm-up to the calculation for general $N$, we consider the quench problem starting with one or two electrons:
\begin{align}
    e^{-i H t} c_{\gamma_1 k_1 a_1}^\dagger |a_0\rangle &\equiv|\Psi_1\rangle,\\
    e^{-i H t} c_{\gamma_2 k_2 a_2}^\dagger c_{\gamma_1 k_1 a_1}^\dagger |a_0\rangle &\equiv |\Psi_{12}\rangle,
\end{align}
where dependence on $t$ is suppressed, and where the numbers $1$ and $2$ on the right-hand side are not lead indices, but instead stand for the quantum numbers $\gamma_1 k_1 a_1$ and $\gamma_2 k_2 a_2$.  (After these warm-up examples, we do not use this shorthand again.)  In terms of time-evolving operators and crossing states, these wavefunctions are given by
\begin{align}
    |\Psi_1\rangle &= c_1^\dagger | a_0 \rangle + |\chi_1\rangle, \label{eq: one electron wavefn shorthand}\\
    |\Psi_{12}\rangle &= c_2^\dagger c_1^\dagger |a_0\rangle + \left( c_2^\dagger |\chi_1\rangle +|\chi_{12}\rangle -(1\leftrightarrow 2 ) \right).
\end{align}
The overlap of single electron states [we include the operator insertion $\psi_{oa}^\dagger(x) \psi_{ea}(x)$ later] can be written as
\beq
    \langle \Psi_{1'} | \Psi_1\rangle = \langle \Psi_{1'}^0 | \Psi_1^0 \rangle + : \langle \Psi_{1'} | \Psi_1\rangle :, \label{eq: one electron overlap shorthand}
\eeq
where $1'$ stands for another distinct set of quantum numbers $\gamma_1' k_1' a_1'$, and where $|\Psi_1^0 \rangle = c_1^\dagger |a_0\rangle$.  In $: \langle \Psi_{1'} | \Psi_1\rangle :$, we must expand the product $\langle \Psi_{1'} | \Psi_1\rangle$ to four terms using Eq. \eqref{eq: one electron wavefn shorthand}, then move every $c$ operator to the right of every $c^\dagger$ operator (with appropriate minus signs).  In this simple case, the normal ordering symbol guarantees that  $: \langle a_0' | c_{1'}c_1^\dagger |a_0\rangle :\ =0$, and this is exactly compensated by the first term on the right-hand side of Eq. \eqref{eq: one electron overlap shorthand}.

A less trivial example is the overlap of states with two electrons.  A straightforward calculation shows
\begin{multline}
    \langle \Psi_{1' 2'} | \Psi_{1 2} \rangle = \langle \Psi_{1' 2'}^0 |\Psi_{12}^0 \rangle + \biggr[ \{ c_{2'},c_2\} : \langle \Psi_{1'} | \Psi_{1 } \rangle :\\ - (1 \leftrightarrow 2 )
    - (1' \leftrightarrow 2' ) + (1 \leftrightarrow 2 ,1' \leftrightarrow 2'  )\biggr]\\
    + :\langle \Psi_{1' 2'}|\Psi_{12}\rangle:,
\end{multline}
where $|\Psi_{12}^0\rangle = c_2^\dagger c_1^\dagger|a_0\rangle$.  This is now a large enough number of electrons to illustrate all features of a general result which is stated and proven in the Appendix [Eq. \eqref{eq: overlap as sum of NO overlaps}].  The result is that the overlap of two states evolving from any quantum numbers can be written as a sum of normal-ordered terms multiplied by contractions of the $c$ and $c^\dagger$ operators.  The normal-ordered terms are overlaps between time-evolving states with any possible subset of the original quantum numbers.

A similar result is true if one inserts operators in between the two states; we have calculated it explicitly in the case of a bilinear insertion, which suffices for the evaluation of the current.  To state the precise result, we first introduce the following notation for the time evolution of an initial state with arbitrary quantum numbers:
\begin{multline}
    |\Psi_{\gamma_1 k_1 a_1 \dots \gamma_n k_n a_n,a_0}(t)\rangle \equiv
    e^{-i H t} c_{\gamma_n k_n a_n}^\dagger \dots\\
    \times c_{\gamma_1 k_1 a_1}^\dagger |a_0\rangle.
\end{multline}
Inside the normal ordering symbol, it is understood that any $|\Psi(t)\rangle$ as just defined is to be written in terms of time-evolving field operators and crossing states before the normal ordering is applied.  Then, with the quantum numbers written as $\alpha\equiv \gamma k a$, we have (see Appendix \ref{sec: Evaluation of bilinears} for proof)
\begin{multline}
    \langle \Psi_{\alpha_1 \dots \alpha_N,a_0}(t) | \psi_{oa}^\dagger(x)\psi_{e a}(x) | \Psi_{\alpha_1 \dots \alpha_N,a_0}(t) \rangle = \\
    \sum_{n=1}^N \frac{1}{(n-1)!} \sum_{m_1,\dots,m_n =1}^N \{ c_{\alpha_{m_n} }(t),\psi_{oa}^\dagger(x) \}\\
    \times :\langle \Psi_{\alpha_{m_1} \dots \alpha_{m_{n-1}} ,a_0}(t) | \psi_{ea}(x) | \Psi_{\alpha_{m_1} \dots \alpha_{m_n},a_0}(t) \rangle:\\
    +\sum_{j=1 }^N \{ c_{\alpha_j}(t), \psi_{oa}^\dagger(x) \} \{ \psi_{ea}(x), c_{\alpha_j}^\dagger(t) \}.\label{eq: bilinear main text}
\end{multline}
The second term is independent of $t$ and so does not contribute to the current.  Notice that in the first term, there is only a single sum over subsets (i.e., the $m_j$ variables); the contractions in Wick's theorem became Kronecker deltas that diagonalized the double sum over subsets to a single sum.

An advantage of using normal-ordered overlaps is that they can be written in terms of the even sector only.  To see this, write the free electron operators in the odd/even basis:
\bseq
    \begin{align}
        c_{\gamma k a}^\dagger(t) &= e^{-ik t}\frac{1}{\sqrt{2}} \left[(-1)^{\gamma-1} c_{o ka}^\dagger + c_{e k a}^\dagger \right]\\
        &= \frac{1}{\sqrt{2}}\left[ (-1)^{\gamma-1} c_{o k a}^\dagger(t)+c_{e k a}^\dagger(t)\right].
    \end{align}
\eseq
Inside the normal ordering symbol, every $c_{\gamma k a}^\dagger(t)$ must eventually contract with some $\psi_{e b}(x)$ operator inside some crossing state; hence, every $c_{\gamma k a}^\dagger(t)$ can be replaced by $\frac{1}{\sqrt{2}} c_{e k a}^\dagger(t)$.  The same argument holds for the annihilation operators, and so we obtain (after a short calculation):
\begin{multline}
    : \langle \Psi_{\alpha_1 \dots \alpha_{n-1},a_0}(t) | \psi_{e a}(0) | \Psi_{\alpha_1 \dots \alpha_n,a_0}(t) \rangle = \\
    2^{-n+1/2} : \langle \Psi_{e k_1 a_1 \dots ek_{n-1} a_{n-1} ,a_0}(t) | \psi_{e a}(0) \\
    \times | \Psi_{e k_1 a_1 \dots e k_n a_n , a_0}(t) \rangle:.
\end{multline}
Substituting this into Eq. \eqref{eq: bilinear main text}, and noting that the $x$ integral in Eq. \eqref{eq: current as integral over x} commutes with the normal ordering symbol, we obtain
\begin{widetext}
    \beq
        I(t) =  \text{Re}\left[ - \frac{d}{dt} \sum_{n=1}^N 2^{-n} \frac{1}{(n-1)!} \frac{1}{L^n}\sum_{m_1, \dots, m_n=1 }^N
        (-1)^{\gamma_{m_n}-1 } \Omega_{n,a_0}(t; k_{m_1} a_{m_1},\dots,k_{m_n} a_{m_n})\right],\label{eq: I(t) in terms of even overlaps}
    \eeq
    where
    \beq
        \Omega_{n,a_0}(t; k_1 a_1,\dots, k_n a_n)= L^n :\langle \Psi_{e k_1 a_1 \dots e k_{n-1}a_{n-1},a_0}(t) | c_{ e k_n a_n }(t) | \Psi_{e k_1 a_1 \dots e k_n a_n,a_0}(t) \rangle:.
    \eeq
\end{widetext}
(The powers of $L$ are chosen this way so that $\Omega_{n,a_0}$ is $L$ independent, as shown below.  In the first equation, the momenta and spins being summed are chosen from the full list of $N$ initial quantum numbers; the second equation defines the function $\Omega_{n,a_0}$ on \emph{arbitrary} momenta and spins.)  This is the expectation value of the current in the time-evolving state $|\Psi_{\gamma_1 k_1 a_1 \dots \gamma_N k_N a_N,a_0}(t)\rangle$, with any initial quantum numbers in the lead $1$/lead $2$ basis.  The normal-ordered overlap on the right-hand side involves the even sector only; the dependence on the lead indices appears in the sign factor $(-1)^{\gamma_{m_n}-1}$.  This reflects the fact that the interaction term of the model is in the even sector only.

\subsection{The current in the thermodynamic limit}\label{sec: The current in the thermodynamic limit}
While Eq. \eqref{eq: I(t) in terms of even overlaps} is valid for arbitrary quantum numbers of the initial state, we are particularly interested in quantum numbers describing two Fermi seas.  A Fermi sea containing a small number of electrons is not meaningful since we linearized the spectrum about the Fermi level.  We therefore take the thermodynamic limit, which turns sums into integrals.

The $n$th term in the sum on the right-hand side of Eq. \eqref{eq: I(t) in terms of even overlaps} is a sum over all choices of $n$ quantum numbers; this includes a sum over all choices of $n$ momenta, which becomes an $n$-dimensional integral in the thermodynamic limit.  We can then allow the leads to have arbitrary temperatures $T_1$ and $T_2$ by generalizing these integrals to include Fermi functions:
\beq
    \fermifn_\gamma(k) \equiv \fermifn(T_\gamma, \mu_\gamma, k)\equiv \frac{1}{e^{ (k-\mu_\gamma)/T_\gamma} +1 },
\eeq
where $\gamma=1,2$.  We have verified that starting the calculation with a density matrix with arbitrary temperatures and chemical potentials leads to the same results as making the natural generalization (which we describe below) from the zero-temperature case \cite{Culver_thesis}.

The generalization from the zero-temperature case proceeds as follows.  Write $\mathcal{K}_\gamma$ for the set of allowed momenta in lead $\gamma=1,2$ (i.e., ranging from $-D$ to $\mu_\gamma$ and spaced by $2\pi/L$).  Then the following example illustrates the idea:
\begin{multline}
    \frac{1}{L^2}\sum_{k_1, k_2 \in \mathcal{K}_1 }\frac{1}{L}\sum_{k_3 \in \mathcal{K}_2 }\overset{\text{therm. limit}}{\to} \int_{-D}^{\mu_1} \frac{dk_1}{2\pi}\frac{dk_2}{2\pi}\int_{-D}^{\mu_2} \frac{dk_3}{2\pi} \\
    \overset{T_1,T_2}{\to} \int_{-D}^D \frac{dk_1}{2\pi}\frac{dk_2}{2\pi}\frac{dk_3}{2\pi}\ \fermifn_1(k_1)\fermifn_1(k_2)\fermifn_2(k_3),\label{eq: example of therm limit}
\end{multline}
where the first arrow represents the thermodynamic limit at zero temperature, and the second arrow represents the generalization to allow the two leads to have arbitrary temperatures.  It is essential that whatever function of $k_1,k_2$, and $k_3$ that is being summed here does not grow with $L$.

The generalization of the above example is
\bseq
\begin{align}
    &\frac{1}{L^n} \sum_{m_1, \dots, m_n =1}^N = \frac{1}{L^n}\sum_{\gamma_1,\dots,\gamma_n=1,2} \sum_{\substack{k_j \in \mathcal{K}_{\gamma_j}   \\ 1\le j \le n}  } \sum_{a_1 \dots a_n} \\
    &\to\sum_{\gamma_1,\dots,\gamma_n=1,2} \int_{-D}^D \left[ \prod_{j=1}^n \frac{dk_j}{2\pi}\ \fermifn_{\gamma_j}(k_j) \right] \sum_{a_1 \dots a_n}, 
\end{align}
\eseq
where we have first written the sum over abstract quantum numbers as a sum over lead indices, momenta, and spins, and then taken the thermodynamic limit, going directly to the generalization to arbitrary temperatures in the leads.

The function $\Omega_{n,a_0}$ being summed in Eq. \eqref{eq: I(t) in terms of even overlaps} involves the even sector only, so it is independent of the lead indices being summed.  We can therefore do the sum over lead indices explicitly, finding the following in the thermodynamic limit:
\begin{multline}
    I(t) \to  \text{Re}\biggr\{ - \frac{d}{dt} \sum_{n=1}^\infty 2^{-n} \frac{1}{(n-1)!}  \\
    \times \int_{-D}^D \left[ \prod_{j=1}^{n-1} \frac{dk_j}{2\pi} \left[ \fermifn_1(k_j) + \fermifn_2(k_j) \right] \right] \frac{dk_n}{2\pi}\left[ \fermifn_1(k_n) - \fermifn_2(k_n) \right] \\
    \times \sum_{a_1 \dots a_n} \Omega_{n,a_0}(t; k_1 a_1,\dots,k_n a_n)\biggr\}. \label{eq: I(t) in terms of Omega}
\end{multline}
Explicit evaluation of the function $\Omega_{n,a_0}$ (see Appendix \ref{sec: Evaluation of the normal-ordered overlap}) shows that it is an antisymmetrization of another function $\Omega_{n,a_0}^{(\text{off-diag})}$:
\begin{multline}
    \Omega_{n,a_0}(t; k_1 a_1,\dots, k_n a_n) = \sum_{\substack{\sigma, \sigma' \in \text{Sym}(n) \\ \sigma'(n) = n} } (\sgn \sigma)(\sgn \sigma')\\
    \times \Omega^{(\text{off-diag})}_{n,a_0}(t; k_{\sigma_1'} a_{\sigma_1'},\dots,k_{\sigma_n'} a_{\sigma_n'};k_{\sigma_1} a_{\sigma_1},\dots,k_{\sigma_n} a_{\sigma_n}), \label{eq: Omega in terms of Omega off-diag}
\end{multline}
where the function $\Omega^{(\text{off-diag})}_{n,a_0}$ is given by
\begin{multline}
    \Omega^{(\text{off-diag})}_{n,a_0}(t; k_1' a_1', \dots, k_n' a_n'; k_1 a_1,\dots, k_n a_n ) =\\
    \Xi_n[a_1' \dots a_{n-1}' ; a_1 \dots a_{n-1} ]_{a_0 a_0}^{b_0 c_{n-1}} (- i \mathcal{T} )_{a_n c_{n-1}}^{a_n' b_0}\\
    \times \int_0^t \left[ \prod_{\ell=1}^n dx_{\ell}\ e^{-i (k_\ell- k_{\ell}') (t-x_\ell) }\right] \Theta(x_n < \dots < x_1),
\end{multline}
with the tensor $\Xi_n$ defined as
\begin{multline}
    \Xi_n[a_1' \dots a_n' ; a_1 \dots a_n ]_{c' c}^{c_n' c_n} =\\
    \delta_{c'}^{c_0'}\delta_{c}^{c_0}
    \prod_{j=1}^{n} \left( \mathcal{S}_{a_j' c_{j-1}'}^{* b_j c_j'} \mathcal{S}_{a_j c_{j-1}}^{b_j c_j} -I_{a_j' c_{j-1}'}^{b_j c_j'} I_{a_j c_{j-1}}^{b_j c_j}  \right) ,\label{eq: Xi}
\end{multline}
where $\mathcal{S} = I- i \mathcal{T}$ is the bare single-particle $\mathcal{S}$ matrix for an electron crossing the impurity.  Note in particular that $\Omega^{(\text{off-diag})}_{n,a_0}$ grows with $t$ (at most as $t^n)$ and not with $L$; the same is then true of $\Omega_{n,a_0}$, justifying our calculation of the thermodynamic limit.  Substituting Eq. \eqref{eq: Omega in terms of Omega off-diag} into Eq. \eqref{eq: I(t) in terms of Omega} and using the symmetry of the integrand to eliminate the sum over permutations $\sigma'$, we find one of our main results, a series expression for the current in the thermodynamic limit:
\begin{widetext}
    \begin{multline}
        I(T_1, \mu_1; T_2,\mu_2; t) = \text{Re} \Biggr\{ \frac{\partial}{\partial t} \sum_{n=1}^\infty \sum_{\sigma\in \text{Sym}(n) } W_n^{(\sigma)}(J)  \int_{-D}^D \frac{dk_1\dots dk_n}{(2\pi)^n} \left[ \prod_{j=1}^{n-1} \left[ \fermifn_1(k_j) + \fermifn_2(k_j) \right]  \right] \left[ \fermifn_1(k_n) - \fermifn_2(k_n) \right] \\
        \times \int_0^t dx_1 \dots dx_n\  \left( \prod_{\ell=1}^n e^{-i ( k_{\ell}-k_{\sigma_{\ell}}  )x_
        \ell} \right)\Theta(x_n < \dots < x_1)
        \Biggr\},\label{eq: final answer for current} 
    \end{multline}
    where we have defined $J$-dependent spin sums via
    \beq
        W_n^{(\sigma)}(J) = \sum_{\substack{a_0,a_1,\dots,a_n \\ b_0, c_0}}\left( \sgn \sigma\right) \frac{1}{2^{n+1}} \Xi_{n-1}[ a_1,\dots, a_{n-1}; a_{\sigma_1} ,\dots ,a_{\sigma_{n-1}} ]_{a_0 a_0}^{b_0 c_0} i \mathcal{T}_{a_{\sigma_n} c_0}^{a_n b_0}. \label{eq: W}
    \eeq
\end{widetext}
We have included a sum over the initial impurity spin $a_0$ (compensated by an additional prefactor of $1/2$) purely for notational simplicity, and it is easily verified that using a fixed $a_0$ produces the same answer.

This series answer Eq. \eqref{eq: final answer for current} has the interesting property that it yields not only a series in powers of $J$ for small $J$ (which follows straightforwardly from the power counting of the crossing states), but also a series in the inverse parameter $1/J$ for large $|J|$.  The fundamental reason for the $1/J$ series is the existence of the $|J|=\infty$ basis discussed in Sec. \ref{sec: Solution in an alternate basis}; however, we give below a self-contained argument using only the $J=0$ basis.

We write the coefficients of the identity and spin-flip terms of the bare $\mathcal{S}$ matrix as $\coeffI$ and $\coeffP$:
\beq
    \mathcal{S}_{a_1 a_0}^{b_1 b_0} \equiv \left(I - i \mathcal{T}\right)_{a_1 a_0}^{b_1 b_0} \equiv \coeffI \delta_{a_1}^{b_1}\delta_{a_0 }^{b_0} + \coeffP \delta_{a_1}^{b_0}\delta_{a_0}^{b_1}.
\eeq
Explicitly, these coefficients are
\bseq
    \begin{align}
        \coeffI &= \frac{ 1- \frac{3}{16} J^2}{  1- i \frac{1}{2} J + \frac{3}{16}J^2 }, \label{eq: coeffI}\\
        \coeffP &= \frac{-i J}{  1- i \frac{1}{2} J + \frac{3}{16}J^2 } \label{eq: coeffP}.
    \end{align}
\eseq
Note in particular that $\coeffP$ is $O(J)$ for small $J$ and $O(1/J)$ for large $J$.  In Appendix \ref{sec: Properties of spin sums}, we prove that for $n\ge 2$, the spin sum $W_n^{(\sigma)}(J)$ has at least $n+1$ powers of $\coeffP$ (where we consider $\coeffP^*$ and $\coeffP$ as equivalent for power counting purposes).  This confirms that the current series can be expanded in either parameter.

In Table \ref{tab: Spin sums}, we list all non-vanishing spin sums up to $n=4$, leaving out the seven permutations at $n=4$ that start at order $O(J^6)$ or $O(1/J^6)$.  The product structure of the tensor \eqref{eq: Xi} permits fairly quick evaluation of these sums; an ordinary computer can produce Table \ref{tab: Spin sums} from the definition \eqref{eq: W} in a matter of seconds.
\begin{table}[htp]
    \caption{\label{tab: Spin sums}
    First several non-vanishing spin sums.
    }
    \begin{ruledtabular}
    \begin{tabular}{ L C }
    \sigma \equiv (\sigma_1,\dots,\sigma_n)&
    W_n^{(\sigma)}(J)\\
    \colrule
    (1) & 1- \coeffI -\frac{1}{2} \coeffP\\
    (2,1) & \frac{3}{4} |\coeffP|^2 \coeffP\\
    (3,1,2) & \frac{3}{4} |\coeffP|^4\left(-\coeffI + \frac{1}{2}\coeffP \right)\\
    (2,3,1) & \frac{3}{4} |\coeffP|^4\left( \coeffI + \frac{1}{2}\coeffP \right)\\
    (3,2,1) & -\frac{3}{4} |\coeffP|^4 \coeffP \\
    (2,3,4,1) & \frac{3}{4} |\coeffP|^4\left[ -\coeffP + |\coeffP|^2 \left(\coeffI + \frac{5}{4}\coeffP \right)   \right]\\
    (2,4,1,3) \text{ and } (3,1,4,2) & \frac{3}{4}|\coeffP|^4 \coeffP \left( 1- \frac{3}{4} |\coeffP|^2 \right)\\
    (3,4,1,2) & \frac{3}{4}|\coeffP|^4 \coeffP \left( -1 + |\coeffP|^2 \right)\\
    (4,1,2,3) & \frac{3}{4} |\coeffP|^4 \left[ -\coeffP + |\coeffP|^2 \left( -\coeffI + \frac{5}{4}\coeffP \right) \right]\\
    (4,3,2,1) & \frac{3}{4} |\coeffP|^4 \coeffP \left( 1 - \frac{3}{2} |\coeffP|^2 \right)
    \end{tabular}
    \end{ruledtabular}
\end{table}

\subsection{Steady-state limit of the current}\label{sec: Steady-state limit of the current}
A basic question in quench problems is the existence of the steady limit of observable quantities, such as the current:
\beq
    I_{\text{steady-state}}(T_1,T_2, V) = \lim_{t \to\infty} I(T_1, \mu_1; T_2,\mu_2; t),\label{eq: Isteadystate as limit}
\eeq
where we set $\mu_1=0$ and $\mu_2=-V$ on the right-hand side.

We argue that the existence of the long-time limit of our series expression \eqref{eq: final answer for current} reduces to a certain spin sum identity, which we then prove in Appendix \ref{sec: Properties of spin sums}.  This confirms the existence of the steady-state current to all orders either in $J$ or in $1/J$.  Note that Doyon and Andrei have already shown that the Schwinger-Keldysh perturbation series for the current converges in time to all orders in $J$ \cite{DoyonAndrei}.  As discussed in more detail in \cite{DoyonAndrei}, the leads serve as thermal baths in the limit of infinite system size, even though there is no explicit relaxation mechanism (i.e., coupling to an external bath whose degrees of freedom appear in the Hamiltonian).

A natural question to ask at this point is: Why are we concerned with showing that the time-evolving current converges in the long-time limit if we have already shown that the wavefunction reaches a NESS?  The original definition \eqref{eq: I(t) def} of the current can be shown to be equivalent to the expectation value of a local operator: $ I(t) = \langle\Psi(t) | \widehat{I} |\Psi(t)\rangle$ with $\widehat{I} = \text{Re}\left[ iJ \psi_{1a}^\dagger(0) \bm{\sigma}_{aa'}\cdot \psi_{2a'}(0) \mathbf{S}   \right]$.  The long-time limit of $I(t)$ should be the same as the expectation value of this local operator in the NESS:
\beq
    \lim_{t\to\infty} I(t) = \langle \Psi_{\text{NESS}}| \widehat{I} | \Psi_{\text{NESS} }\rangle.\label{eq: current in NESS agreement}
\eeq
Since we have $|\Psi_{\text{NESS}}\rangle$ explicitly, one might think that this proves that the long-time limit exists.  However, this is not so.  Evaluating the right-hand side of Eq. \eqref{eq: current in NESS agreement} with the time-independent version of our formalism, we find that it contains many infrared divergences; introducing an infrared regulator, we find that the problem of showing that these divergences cancel is equivalent to the problem of showing that $I(t)$ converges for large time.  Indeed, having a finite $t$ is itself an example of an infrared regulator.  If the limit on the left-hand side of Eq.  \eqref{eq: current in NESS agreement} does exist, then the equality holds.

There are two ways to proceed with the analysis of the time-evolving current \eqref{eq: final answer for current}: we can do the $n-1$ integrations over position variables analytically, leaving $n$ integrations over momenta still to be done; or we can do the $n$ integrations over momenta analytically, leaving $n-1$ integrations over position variables still to be done.  The first option leaves us with momentum integrals of the same type that arise in loops in a Keldysh calculation.  We pursue the second option, both because it allows for better understanding of the steady-state limit and because it results in integrals that are easier to evaluate in the large-bandwidth regime.

Our approach is to use the following formula for the Fourier transform of a Fermi function $f(T,\mu,k)$ (with temperature $T$, chemical potential $\mu$, and cutoff $D$):
\beq
    \int_{-D}^D dk\ e^{-i k y }\fermifn(T,\mu, k) = \frac{1}{i} \left( \frac{e^{i D y}}{y}  - \frac{\pi T e^{-i \mu y} }{\sinh(\pi T y) } \right),\label{eq: FT of n sharp cutoff}
\eeq
where error terms of order $O(e^{- \frac{1}{T}(D\pm \mu)})$ have been dropped on the right-hand side.  This truncation is very accurate in the universal regime, in which the cutoff is much larger than all other energy scales.  To use this formula, we relabel some integration coordinates to obtain  
\begin{multline}
    \frac{\partial}{\partial t} \int_0^t dx_1 \dots dx_n\ \left( \prod_{\ell=1}^n e^{-i (k_{\ell} - k_{\sigma_{\ell}} )x_n} \right)\\
    \times \Theta(x_n < \dots < x_1) \\
    =\int_0^\infty dx_1\dots dx_{n-1}\ \left( \prod_{\ell=1}^n e^{- i k_{\ell} y_{\ell}^{(\sigma)}}  \right)\\
    \times\Theta(t- x_1 - \dots - x_{n-1}),
\end{multline}
where we have defined the following linear combinations of the $x_j$ variables:
\beq
    y_{\ell}^{(\sigma)} = \sum_{m = \ell}^{n-1} x_m - \sum_{m = \sigma^{-1}(\ell) }^{n-1} x_m. \label{eq: y_j variables}
\eeq
Using the Fourier transform \eqref{eq: FT of n sharp cutoff} and the identity $\sum_{j = 1}^n y_j^{(\sigma)} =0 $, we then obtain
\begin{widetext}
    \beq
        I(T_1, \mu_1; T_2,\mu_2; t) = \frac{1}{2\pi} \text{Re} \Biggr\{ \sum_{n=1}^\infty \frac{1}{(i \pi)^{n-1}} \sum_{\sigma\in \text{Sym}(n) } W_n^{(\sigma)}(J)  \varphi_n^{(\sigma)}(T_1,\mu_1; T_2,\mu_2; t)  \Biggr\},
    \eeq
    where [defining $\widetilde{D} = D+\frac{1}{2}(\mu_1 + \mu_2)$ and $V= \mu_1 - \mu_2$]
    \begin{multline}
        \varphi_n^{(\sigma)}(T_1,\mu_1; T_2,\mu_2; t) =\frac{1}{i} \int_0^\infty dx_1\dots dx_{n-1}\ \Theta(t- x_1 - \dots - x_{n-1}) \\
        \times \left[ \prod_{j=1}^{n-1} \left( \frac{e^{i \widetilde{D} y_j^{(\sigma)}} }{ y_j^{(\sigma)} } - \frac{\pi T_1 e^{-i \frac{1}{2}V y_j^{(\sigma)}}}{2 \sinh ( \pi T_1 y_j^{(\sigma)} ) } - \frac{\pi T_2 e^{i \frac{1}{2}V y_j^{(\sigma)}}}{2 \sinh ( \pi T_2 y_j^{(\sigma)} ) }  \right)   \right] \left[    \frac{\pi T_2 e^{i\frac{1}{2}V y_n^{(\sigma)}}}{\sinh ( \pi T_2 y_n^{(\sigma)} ) }  -\frac{\pi T_1 e^{-i\frac{1}{2} V y_n^{(\sigma)}}}{ \sinh ( \pi T_1 y_n^{(\sigma)} ) }  \right]. \label{eq: varphi sharp cutoff Kondo}
    \end{multline}
\end{widetext}
We can now address the convergence of the series in time.  The key point is to show that for any permutation $\sigma$ such that the corresponding spin sum $W_n^{(\sigma)}(J)$ is non-vanishing, there is a finite limit $\lim_{t\to\infty} \varphi_n^{(\sigma)}(T_1,\mu_1; T_2,\mu_2,t)$.  The qualification that the spin sum be non-vanishing is an important one, since there are many cases in which the integral $\varphi_n^{(\sigma)}$ does \emph{not} converge in time.  The simplest example is $\varphi_2^{(1,2)}(T_1,\mu_1; T_2,\mu_2; t) = \widetilde{D} t V$.  This linear divergence is of no consequence for the current because it is multiplied by a vanishing spin sum: $W_2^{(1,2)}(J) = 0$.

More generally, divergences for large time are to be expected if one or more of the integration variables $x_1,\dots,x_{n-1}$ appears only in the Heaviside function and nowhere else in the integrand.  [For example, for $\sigma=(1,2)$, we have $y_1^{(\sigma)} = y_2^{(\sigma)} = 0$, so $x_1$ only appears in the Heaviside function, and $\varphi_2^{(1,2)} \sim t$.]  If instead all $x_j$ variables appear explicitly (not including the Heaviside function), then the only possible sources of divergences in time are the oscillating phase terms (since the $1/\sinh$ terms are very small at large $x$).  The oscillating phase terms take the form of multi-dimensional generalizations of the one-dimensional integral $\int_1^b du\ \frac{e^{iu}}{u}$, which is finite as $b\to\infty$; thus, we can expect that there are no time divergences even from the oscillating phases.  (Asymptotic evaluation of several of these integrals confirms this expectation; see Appendix \ref{sec: Asymptotic evaluation of integrals}.)

Our task, then, is to show that for any permutation $\sigma \in \text{Sym}(n)$ such that one or more of the $x_j$ variables is absent from $y_1^{(\sigma)},\dots,y_n^{(\sigma)}$, the corresponding spin sum $W_n^{(\sigma)}(J)$ vanishes.  These permutations are exactly the \emph{reducible} ones: those for which the permutation rearranges the first $m$ entries independently of the last $n-m$ (for some $m<n$).  From Eq. \eqref{eq: W} and from the product structure \eqref{eq: Xi} of the tensor $\Xi_n$, we see that the spin sums for all reducible permutations vanish provided that the following identity holds for any $n\ge 1$, $\sigma \in \text{Sym}(n)$:
\begin{multline}
    \sum_{a_0, a_1, \dots, a_n} \Xi_n[a_1, \dots, a_n; a_{\sigma_1}, \dots, a_{\sigma_n}]_{a_0 a_0}^{c' c} = 0.\label{eq: spin sum identity for convergence Kondo}
\end{multline}
We prove this identity in Appendix \ref{sec: Properties of spin sums}.  (The proof does not rely on the detailed form of the coefficients $\coeffI$ and $\coeffP$, but only on the fact that they lead to a unitary $\mathcal{S}$ matrix.)  Thus, we have shown convergence in time to all orders in $J$ and $1/J$.
  
Having established the existence of the steady-state limit, we can write
\bseq
\begin{align}
    I(T_1,T_2,V) \equiv I_{\text{steady-state}}(T_1,T_2,V) \equiv\\
    \lim_{t\to\infty } I(T_1,\mu_1 =0; T_2,\mu_2= -V; t),
\end{align}
\eseq
where both sides depend implicitly on the cutoff $D$ through $\widetilde{D} = D - V/2$ [see Eq. \eqref{eq: varphi sharp cutoff Kondo} and below].  Setting $\mu_1=0$ is no real loss of generality, since working with arbitrary $\mu_1$ (given fixed voltage difference $V$) only means that $\widetilde{D} = D +(\mu_1 + \mu_2)/2$, instead, and we will see that $\widetilde{D}$ can be replaced by $D$ in the large-bandwidth limit.

The steady-state current, then, depends on the three external parameters $T_1$, $T_2$, and $V$.  It is convenient to work in spherical coordinates $(M,\theta,\phi)$ with $V$ as the ``$Z$-axis'':
\bseq    
    \begin{align}
        V&= M \cos \theta,\\
        T_1 &= \sqrt{2} M \sin \theta \cos \phi,\ T_2 = \sqrt{2} M \sin \theta \sin \phi,\\
        M &= \sqrt{V^2 + \frac{1}{2} (T_1^2 + T_2^2 ) }.
    \end{align}
\eseq
It is to be expected that for large bandwidth, the steady-state integrals $\varphi_n^{(\sigma)}(T_1,T_2,V)$ include logarithmic divergences in the limit $D/M\to\infty$.  These logarithmic divergences, together with the coupling constant dependence contained in the spin sums $W_n^{(\sigma)}(J)$, encode the scaling properties and the emergence of the Kondo temperature $T_K$ through the Callan-Symanzik equation, as we discuss in more detail in the next two sections.  Here, we present a technical discussion of the steady-state integrals and their logarithmic divergences.

The basic integral we need to consider is the steady-state limit of \eqref{eq: varphi sharp cutoff Kondo}, which is obtained simply by deleting the Heaviside function:
\beq
    \varphi_n^{(\sigma)}(T_1,T_2,V) \equiv \lim_{t\to\infty } \varphi_n^{(\sigma)}(T_1,\mu_1=0; T_2,\mu_2=-V; t).
\eeq
As discussed above, any permutations $\sigma$ for which this limit fails to exist are of no importance, since the corresponding spin sum $W_n^{(\sigma)}(J)$ vanishes.  From Eq. \eqref{eq: final answer for current} we obtain
\begin{multline}
    I_{\text{steady-state}}(T_1,T_2,V) = \frac{1}{2\pi} \text{Re} \Biggr\{ \sum_{n=1}^\infty \frac{1}{(i \pi)^{n-1}}  \\
    \times \sum_{\sigma\in \text{Sym}(n) } W_n^{(\sigma)}(J)  \varphi_n^{(\sigma)}(T_1, T_2,V)  \Biggr\}.\label{eq: I steady-state in terms of varphi Kondo}
\end{multline}
We express the steady-state integral $\varphi_n^{(\sigma)}$ in spherical coordinates, denoting it by the same symbol.  Rescaling to dimensionless variables $u_j \equiv \frac{1}{2}M x_j$ and $v_j^{(\sigma)} \equiv \frac{1}{2} M y_j^{(\sigma)}$, we obtain
\begin{multline}
    \varphi_n^{(\sigma)}(M, \theta, \phi ) = M \cos \theta \int_0^\infty du_1\dots du_{n-1}\\
    \times \left[ \prod_{j=1}^{n-1} \left( e^{i (2D/M -\cos \theta) v_j^{(\sigma)}} - f(\theta,\phi; v_j^{(\sigma)})  \right) / v_j^{(\sigma)}   \right]\\
    \times h(\theta,\phi; v_n^{(\sigma)} ),\label{eq: varphi spherical coo Kondo}
\end{multline}
where
\begin{multline}
    f(\theta, \phi; v) = \frac{\sqrt{2} \pi \sin \theta \cos \phi\ v e^{-i (\cos \theta) v} }{\sinh(2^{3/2} \pi \sin\theta \cos \phi\ v  )}\\
    + \frac{\sqrt{2} \pi \sin \theta \sin \phi\ v e^{i (\cos \theta) v} }{\sinh(2^{3/2} \pi \sin\theta \sin \phi\ v)  },\label{eq: fn f Kondo}
\end{multline}
and
\begin{multline}
    h(\theta,\phi; v) =  \frac{1}{i} \biggr( \frac{\sqrt{2} \pi \tan \theta \sin \phi\ e^{i (\cos \theta) v} }{\sinh(2^{3/2} \pi \sin\theta \sin \phi\ v  )} \\
    - \frac{\sqrt{2} \pi \tan \theta \cos \phi\ e^{-i (\cos \theta) v} }{\sinh(2^{3/2} \pi \sin\theta \cos \phi\ v)  } \biggr).\label{eq: fn h Kondo}
\end{multline}
Note that $f(\theta, \phi;v=0) = h(\theta, \phi; v=0) =1$.

We have explicitly calculated the asymptotic forms of the integral \eqref{eq: varphi spherical coo Kondo} in the large-bandwidth regime for all permutations $\sigma$ that we need in order to find the current up to and including the $J^5$ or $1/J^5$ terms.
We find that the rapidly oscillating phases generate logarithmic divergences: powers of $\ln (2D/M)$ with coefficients that depend on the ratios $T_1/V$ and $T_2/V$ through the angles $\theta$ and $\phi$.  In some cases, there are also linear divergences, but they cancel in the final answer for the current at this order.

To arrive at Eq. \eqref{eq: varphi spherical coo Kondo}, we assumed $V >0$; however, the special case of $V=0$ reduces to an integral of the same form with different functions $f$ and $h$.  For example, the linear response conductance $G(T)=\pd I / \pd V|_{T_1=T_2=T,V=0}$ involves the following integral:
\begin{multline}
    \frac{\partial}{\partial V}\biggr\rvert_{V=0} \varphi_n^{(\sigma)}(T_1=T,T_2=T, V) = \int_0^\infty du_1 \dots du_{n-1}\\
    \times h\left( v_n^{(\sigma)} \right)\prod_{j=1}^{n-1} \frac{e^{i \frac{D}{\pi T} v_j^{(\sigma)} } -f\left( v_j^{(\sigma)} \right) }{ v_j^{(\sigma)} },\label{eq: varphi steady-state G(T) case}
\end{multline}
where $f$ and $h$ are given in this case by $f(v)=h(v) = v / \sinh v$.  The case of the thermoelectric current ($V=0$ with arbitrary $T_1$ and $T_2$) is similar.  All cases thus reduce to the study of the large-$\lambda$ behavior of the following general form:
\beq
    \int_0^\infty du_1 \dots du_{n-1}\ h\left( v_n^{(\sigma)} \right)\prod_{j=1}^{n-1} \frac{e^{i \lambda v_j^{(\sigma)} } -f\left( v_j^{(\sigma)} \right) }{ v_j^{(\sigma)} }.\label{eq: general form}
\eeq
Appendix \ref{sec: Asymptotic evaluation of integrals} presents our asymptotic results for the general form given in \eqref{eq: general form} only using general properties of $f$ and $h$.  The simplest non-trivial example is the permutation $\sigma=(2,1)$, for which we have $v_1^{(\sigma)}= -v_2^{(\sigma)} = u_1$ and the following asymptotic result:
\begin{multline}
    \int_0^\infty du_1\ \frac{e^{i\lambda u_1} - f(u_1) }{u_1} h(-u_1) \overset{\lambda\to\infty}{\longrightarrow} - h(0) \ln \lambda \\
    - h(0)\left( \gamma - i\frac{\pi}{2}\right) + \int_0^\infty du\ \ln u\ \frac{d}{du} \left[ f(u) h(-u) \right],\label{eq: R21 asymptotic}
\end{multline}
where $\gamma$ is the Euler constant [not to be confused with the anomalous dimension $\gamma(g)$ that we discuss later].  In the steady-state current in the regime of small $J$, the $\ln \lambda$ divergence here will be multiplied by $J^3$: it is the equivalent of the one-loop divergence that appears in a Keldysh calculation.

Notice that the constant ($\lambda$-independent) term in \eqref{eq: R21 asymptotic} is a more complicated functional of $f$ and $h$ than the log term.  This is the beginning of a pattern that seems to persist to higher orders.  For example, in the case of $\sigma=(2,3,1)$ that is explicitly written out in Appendix \ref{sec: Asymptotic evaluation of integrals}, there is a $\ln^2\lambda$ term that depends only on $h(0)$, a $\ln \lambda$ term involving both $h(0)$ and the same single variable integral over $f$ and $h$ that appears in \eqref{eq: R21 asymptotic}, and then a $\lambda$-independent constant that depends on the same quantities already encountered in $\ln^2\lambda$ and $\ln \lambda$ and also on a double integral involving $f$ and $h$.  These terms then appear in the small $J$ current multiplied by $J^4$ (two loops).  This pattern of asymptotic expansion is the mechanism underlying the scaling that we find in the following two sections.

\subsection{Antiferromagnetic regime: Universality}\label{sec: Antiferromagnetic regime: Universality}
We evaluate our current series in the regime of weak antiferromagnetic coupling.  We first review what scaling properties are expected on general grounds, then present the results of our calculations.  For easier comparison with the literature, we refer to $g\equiv \rho J = \frac{1}{2\pi}J$ from now on.

It is expected that, when all other energy scales in the problem are much smaller than the bandwidth, the current becomes a universal function $f_{\text{universal}}(T_1/T_K, T_2/T_K, V/T_K)$, where the Kondo temperature $T_K = D e^{ - \frac{1}{2g} + \frac{1}{2} \ln g}$ is a dynamically generated scale.  The ``scaling limit'' consists of taking $D\to\infty$ and $g\to 0^+$ with $T_K$ fixed; the resulting $f_{\text{universal}}$ is then the same as that which would be obtained from taking the low-energy limit of a calculation done with a more realistic Hamiltonian, e.g., with a more complicated band structure than the wide-band limit we have considered.

Universal scaling should manifest itself in a pattern of logarithmic divergences as $D/M$ is sent to infinity.  In the regime of small $|g|$ and large $D/M$, the perturbative renormalizability of the Kondo model constrains the steady-state current to the form  $I(T_1,T_2,V) \to V \sum_{n=2}^\infty\sum_{m=0}^{n-2}\ a_{nm} g^n \ln^m \frac{2D}{M}$, where the coefficients $a_{nm}$ depend only on the ratios $T_1/V$ and $T_2/V$.  This is shown in a very general setting by Delamotte in Ref. \cite{Delamotte}.  Our choice of $V$ for the dimensionful prefactor and $2D/M$ for the argument of the log is one of convenience.  We have assumed that the current starts at order $g^2$, as is confirmed by calculation.

The current (assuming large bandwidth from now on) should satisfy the Callan-Symanzik equation $\left[ D \frac{\partial}{\partial D} + \beta(g) \frac{\partial}{\partial g} + \gamma(g) \right]I(T_1,T_2,V) = 0$, which is a differential form of the statement that all UV divergences can be absorbed by using a running coupling constant and rescaling the current operator.  The solution to the Callan-Symanzik equation takes the form $I(T_1,T_2,V) =f_{\text{universal}}(T_1/T_K, T_2/T_K, V/T_K) e^{ -\int_0^g dg'\ \frac{\gamma(g')}{\beta(g')} }$, and the anomalous dimension $\gamma(g)$ should start at the same order or higher in $g$ as $\beta(g)$ so that the $g$-dependent scale factor goes to unity in the scaling limit.  (Such a scale factor has been seen before in the Kondo problem; see Ref. \cite{BarzykinAffleck}.)

Most of these general expectations are met by our series.  Up to and including the equivalent of three loops (which is $g^5$ in this case), the current at large bandwidth is a scaling form that satisfies the Callan-Symanzik equation with $\beta(g)$ and $\gamma(g)$ that are independent of the ratios $T_1/V$ and $T_2/V$.  The leading order of the beta function [$\beta(g) = - 2g^2$], and the corresponding leading-order expression $T_K = De^{-\frac{1}{2g } }$, agree with the standard answer \cite{Hewson}.  The only surprise is that the first correction to the beta function, and hence to $T_K$, differs by a constant from the expected answer; that is, we obtain $\beta(g) = -2 g^2 + \beta_3 g^3$ with $\beta_3=16$ instead of the expected \cite{Hewson} $\beta_3=2$.

Let us present these results in more detail.  We begin by writing the scaling form that we find for the current.  We will write the series in a triangular structure \cite{Delamotte} in which the $n$th column contains the $g^{n+1}$ terms, while the $n$th row contains terms of the form $g^{n+j} \ln^{j-1}\frac{2D}{M}$ ($j\ge 1$).  The entries in the first row are called the ``leading logarithms,'' the second row the ``sub-leading logarithms,'' and so on.  For large bandwidth, we find 
\begin{widetext}
\begin{alignat}{3}
    I_{\text{steady-state}}(T_1,T_2,V)=\notag\\
    \frac{3\pi}{4} V \Biggr\{ g^2 &+4 g^3 \ln\frac{2D}{M}  &&+12 g^4 \ln^2 \frac{2D}{M}  &&+32 g^5 \ln^3 \frac{2D}{M} \notag\\
    &+ C_1\left( \theta,\phi \right) g^3   &&+6 C_1\left( \theta,\phi \right) g^4\ln\frac{2D}{M}  &&+\left[24C_1\left( \theta,\phi \right) -32 \right] g^5 \ln^2\frac{2D}{M} \notag \\
    & &&+ C_2\left( \theta,\phi \right) g^4  &&-
    \begin{pmatrix}
    16 C_1\left( \theta,\phi \right) - 8C_2\left( \theta,\phi \right) \\ +64 + 3\pi^2
    \end{pmatrix}
    g^5\ln\frac{2D}{M}\notag\\
    & &&  &&+C_3\left(  \theta,\phi \right)\ g^5 \qquad \qquad  \qquad +O(g^6)   \Biggr\} ,\label{eq: I(T1,T2,V)}
\end{alignat}
where $C_1$ and $C_2$ are (in the spherical coordinates introduced earlier)
\bseq
\begin{align}
    C_1\left(\theta,\phi\right) &= 4\ \text{Re}\left\{ \gamma - \int_0^\infty du\ \ln u \frac{\partial}{\partial u} \left[ f\left(\theta,\phi;u\right) h\left(\theta,\phi;-u\right) \right] \right\}\label{eq: C1}\\
    C_2\left(\theta,\phi\right) &= \text{Re}\biggr\{ 6\gamma C_1\left(\theta,\phi\right) - 12\gamma^2 + \frac{7}{12}\pi^2- 4\int_0^\infty du\ \ln^2 u \frac{\partial}{\partial u}\left[ f\left(\theta,\phi;u\right) h\left(\theta,\phi;-u \right) \right] \notag\\
    &+ 8 \int_0^\infty du_1 du_2\ \ln u_1 \ln u_2 \frac{\partial }{\partial u_1}\frac{\partial }{\partial u_2} \left[ f\left(\theta,\phi,u_1 \right) f\left(\theta,\phi,u_2\right) h\left(\theta,\phi;-u_1 -u_2 \right) \right]\notag \\
    & +8 \int_0^\infty du_1 du_2\ \frac{1}{u_2} \ln \frac{u_1 +  u_2}{u_1} \frac{\partial}{\partial u_1} \left[ f\left(\theta,\phi,u_1 + u_2\right) f\left(\theta,\phi,-u_1\right) h\left(\theta,\phi;-u_2\right) \right] \biggr\}.\label{eq: C2}
\end{align}
\eseq
\end{widetext}
We omit a very lengthy explicit form of $C_3$ (a sum of integrals over $f$ and $h$, including triple integrals).

As discussed in more detail by Delamotte \cite{Delamotte}, this triangular structure makes clear the operation of perturbative renormalizability.  [Delamotte does not consider anomalous scaling $\gamma(g)$, but this is a simple modification.]  One can see that the leading logs are built from pure numbers, the sub-leading logs include pure numbers and the constant $C_1$, and so on.  We emphasize that we do not \emph{require} the answer to take this form; we \emph{find} it as the result of a detailed calculation.

Equation \eqref{eq: I(T1,T2,V)} satisfies the Callan-Symanzik equation
\beq
    \left( D \frac{\pd }{\pd D} + \beta(g) \frac{\pd }{\pd g} + \gamma(g) \right) I(T_1,T_2,V) = 0,
\eeq
with
\beq
    \beta(g) = -2 g^2 + \beta_3 g^3 + \beta_4 g^4 + O(g^5) \qquad (\beta_3 = 16),\label{eq: beta small J}
\eeq
and
\beq
    \gamma(g) = \gamma_2 g^2 +(-32 + 3\pi^2 - 2 \beta_4)g^3 + O(g^4) \quad(\gamma_2 = -32),
\eeq
where the constant $\beta_4$ would be determined by the next order of the current ($g^6$, or the equivalent of four loops).  As expected on general grounds, $\beta(g)$ and $\gamma(g)$ are found to depend only on the coupling constant $g$; the terms $C_1$ and $C_2$ (which contain all dependence on the angles $\theta$ and $\phi$) drop out of the scaling equation entirely.  In the following calculations, we leave $\beta_3$ and $\gamma_2$ unspecified in order to see how they appear in the final answers.

The calculation now follows some standard steps, and we omit many details.  We write the current in a universal form in the scaling limit ($g\to0^+$ with $T_K$ fixed).  The Kondo temperature $T_K$ is determined by $[D \frac{\pd}{\pd D} + \beta(g) \frac{\pd}{\pd g} ]T_K =0$, and is given by
\beq
    T_K = \alpha^{-1}D \exp[ -\frac{1}{2g} + \frac{\beta_3}{4} \ln |g| + O(g)],\label{eq: def of TK}
\eeq
where $\alpha >0$ is an arbitrary normalization constant.  The running coupling at scale $M$, denoted $g_M$, is such that $(D,g)$ and $(M,g_M)$ correspond to the same $T_K$.  In the high energy regime ($M \gg T_K$, with $M\ll D$ as always), the running coupling is
\begin{multline}
    g_M = \frac{1}{2 \ln \frac{M}{T_K}} \Biggr[ 1 + \frac{\beta_3}{4}\frac{ \ln \ln \frac{M}{T_K} }{\ln \frac{M}{T_K} } + \left( \frac{\beta_3}{4} \ln 2 + \ln \alpha \right) \frac{1}{\ln\frac{M}{T_K}} \\
    +  \frac{\beta_3^2}{16} \frac{\ln^2 \ln \frac{M}{T_K} }{\ln^2 \frac{M}{T_K} }
    + \frac{\beta_3}{2} \left( \frac{\beta_3}{4} \ln 2 + \ln \alpha - \frac{\beta_3}{8} \right) \frac{\ln \ln \frac{M}{T_K} }{\ln^2 \frac{M}{T_K} }\Biggr] \\
    + O\left( \frac{1}{\ln^3 \frac{M}{T_K} }\right).\label{eq: gM Kondo}
\end{multline}
(See Ref. \cite{DoyonAndrei} for the case $\beta_3=2$.)  We set the normalization constant $\alpha=1$ for now.  Solving the Callan-Symanzik equation and taking the scaling limit yields
\begin{multline}
    I(T_1,T_2,V) =\\
    \frac{3\pi}{4} V g_M^2 \left[1 + \left( C_1(\theta,\phi) + 4\ln 2  -\frac{1}{2}\gamma_2 \right) g_M \right] \\
    + O(g_M^4).
\end{multline}
The leading term, which is the sum of the leading log terms of the series, yields
\beq
    I(T_1,T_2,V) = \frac{3\pi}{16 \ln^2\frac{M}{T_K} }V + \dots \qquad (M \gg T_K),
\eeq
and so
\beq
    G(T_1,T_2,V) = \frac{3\pi^2G_0}{16 \ln^2\frac{M}{T_K} } + \dots \qquad (M \gg T_K),\label{eq: LL conductance antiferro}
\eeq
where we have restored physical dimensions in the differential conductance $G\equiv \partial I/\partial V$ ($G_0 = 2e^2/h = 1/\pi$ is the unitarity limit of conductance).  This is a slight generalization of a well-known result, first found in Ref. \cite{KaminskiNazarovGlazman} (in the case of $T_1=T_2=0$ with $V$ as the variable, or $V=0$ with $T_1=T_2\equiv T$ as the variable); see also Ref. \cite{DoyonAndrei} for the case of equal temperatures and arbitrary voltage.

At the next approximation beyond leading log, the coefficient $\beta_3$ enters into the current as a term of the form $\beta_3 \frac{\ln \ln \frac{M}{T_K}}{\ln^3 \frac{M}{T_K}}$, and so our result cannot be fully correct (note that the coefficient of such a term cannot be adjusted by rescaling $T_K$).  It seems probable, based on a simpler calculation we have done (see Appendix \ref{sec: Cutoff artifact in the time-dependent magnetization}), that our unusual cutoff scheme has led to some extra ``cutoff artifact'' term in the current that changes the coefficient $\beta_3$.  For the moment, we can say that since the leading logs are correct in the small $g$ case, the leading logs of the large $|g|$ regime (see next section) should also be correct.

A calculation we \emph{can} do reliably, at the next order beyond, is the effect of temperature on the current; in particular, we consider the following quantity in the regime $V\gg T_K$:
\beq
    \Delta I (T_1,T_2,V) \equiv I(T_1,T_2,V) - I(T_1=0,T_2=0,V).\label{eq: deltaI def}
\eeq
The idea is that this subtraction eliminates the leading-order effect of $\beta_3$ (and of $\gamma_2$, which is sensitive to the same terms that affect $\beta_3$).  We note the following:
\begin{multline}
    g_M =g_V + \frac{1}{2 \ln \frac{V}{T_K}} \left[ \frac{\ln (\cos \theta)}{\ln \frac{V}{T_K}} + \frac{\beta_3}{2}\ln(\cos \theta) \frac{\ln \ln \frac{V}{T_K}}{\ln^2 \frac{V}{T_K}} \right] \\
    + O\left(\frac{1}{\ln^3 \frac{V}{T_K} }\right),
\end{multline}
hence,
\begin{multline}
    g_M^2 - g_V^2 = \ln(\cos \theta) \frac{1}{2\ln^3 \frac{V}{T_K}} + O\left(\frac{\ln \ln \frac{V}{T_K}}{\ln^4 \frac{V}{T_K} } \right),
\end{multline}
and so
\begin{multline}
    \Delta I(T_1,T_2,V) = \frac{3\pi}{32} \frac{V}{\ln^3 \frac{V}{T_K} } \Bigg[ C_1(\theta,\phi) - C_1(\theta =0, \phi) \\
    + 4 \ln(\cos \theta) \Bigg]
    + O\left(\frac{\ln \ln \frac{V}{T_K}}{\ln^4 \frac{V}{T_K} } \right),\label{eq: Delta I antiferro}
\end{multline}
where the $\phi$ coordinate in $C_1$ does not matter when $\theta=0$.  What we have calculated corresponds to the leading temperature-dependent term in the summation of the sub-leading logarithms [the second row of \eqref{eq: I(T1,T2,V)}]; the first contribution is temperature-independent and has been canceled, and higher contributions depend on the coefficient $\beta_3$.

Equation \eqref{eq: Delta I antiferro} is essentially a one-loop result.  It agrees with the calculations of Doyon and Andrei  in Ref. \cite{DoyonAndrei} (hereafter ``DA'').  Translating their calculation of the current into our notation and calculating the difference $\Delta I$, we find \footnote{The current in DA appears to be off by an overall factor of $2$, which we have corrected for; as written the DA result would seem to be off by a factor of $2$ from the standard result \eqref{eq: LL conductance antiferro}.}
\begin{multline}
    \Delta I_{\text{DA}}(T, V) = \frac{3\pi}{32} \frac{V}{\ln^3 \frac{V}{T_K} } \Biggr[4 \left(P(\cot \theta) - P(\infty) \right) \\
    + 4 \ln(\cos \theta) \Biggr] + O\left(\frac{\ln \ln \frac{V}{T_K}}{\ln^4 \frac{V}{T_K} } \right),
\end{multline}
where the function $P$ is given in an integral form in DA.  Specializing our result \eqref{eq: Delta I antiferro} to equal temperatures sets $\phi = \pi/4$, and we find numerically that our function $C_1(\theta,\phi=\pi/4) =  4 P(\cot \theta) + $ constant; thus, $C_1(\theta,\phi=\pi/4) - C_1(\theta=0, \phi=\pi/4) = 4 [P(\cot \theta) -   P(\infty)]$, so our result agrees with DA ($\Delta I = \Delta I_{\text{DA}}$).

We now turn to two special cases: the linear response conductance $G(T) = (\pd I /\pd V)|_{T_1=T_2=T,V=0}$ and the zero-temperature conductance $G(V) = (\pd I/\pd V)|_{T_1=T_2=0}$.  We find
\begin{widetext}
    \bseq
        \begin{alignat}{3}
            G(T)=  \frac{3\pi^2 G_0}{4} \Biggr\{ g^2 &+4 g^3 \ln\frac{D}{T}  &&+12 g^4 \ln^2 \frac{D}{T}  &&+32 g^5 \ln^3 \frac{D}{T} \notag\\
            &-4 \ln \frac{2\pi}{e^{1+\gamma}} g^3   &&-24 \ln \frac{2\pi}{e^{1+\gamma}} g^4\ln\frac{D}{T}  &&-32\left(\ln \frac{2\pi}{e^{1+\gamma}} +1 \right) g^5 \ln^2\frac{D}{T} \notag \\
            & &&-7.75 g^4  &&- 138.90 g^5 \ln\frac{D}{T}\notag\\
            & &&  &&+ 9.01 g^5 \qquad \qquad \qquad +O(g^6)  \Biggr\} ,\label{eq: G(T) small g series}\\
            G(V) = \frac{3\pi^2 G_0}{4} \Biggr\{ g^2 &+ 4 g^3 \ln\frac{D}{V}  &&+12 g^4 \ln^2 \frac{D}{V}  &&+32 g^5 \ln^3 \frac{D}{V} \notag\\
            &    &&  &&-32 g^5 \ln^2\frac{D}{V} \notag \\
            & &&-\frac{7}{4} \pi^2 g^4  &&-(64 + 17\pi^2)g^5 \ln\frac{D}{V}\notag\\
            & &&  &&+2 \left[\pi^2 -32 + 48 \ln 2 - 24 \zeta(3) \right]g^5 + O(g^6) \Biggr\} , \label{eq: G(V) small g series}
        \end{alignat}
    \eseq
    where $\zeta$ is the Riemann zeta function. Using the Callan-Symazik equation to take the scaling limit, we find the following results in the high energy regime ($T \gg T_K$ or  $V\gg T_K$):
    \bseq
        \begin{align}
            &G(T) =  \frac{3\pi^2 G_0}{16\ln^2\frac{T}{T_K}} \left[1+ 8 \frac{\ln \ln \frac{T}{T_K}}{\ln \frac{T}{T_K}} + \frac{\alpha_1^{(T)}}{\ln \frac{T}{T_K}} + \frac{48 \ln^2 \ln \frac{T}{T_K}}{\ln^2 \frac{T}{T_K} } + \frac{\alpha_2^{(T)} \ln \ln \frac{T}{T_K} }{\ln^2 \frac{T}{T_K} } + O \left( \frac{1}{\ln^2 \frac{T}{T_K} }\right)  \right], \label{eq: G(T) antiferro}\\
            &G(V) = \frac{3\pi^2 G_0}{16\ln^2\frac{V}{T_K}} \left[1+ 8 \frac{\ln \ln \frac{V}{T_K}}{\ln \frac{V}{T_K}} + \frac{\alpha_1^{(V)}}{\ln \frac{V}{T_K} } +\frac{48 \ln^2 \ln \frac{V}{T_K}}{\ln^2 \frac{V}{T_K} } + \frac{\alpha_2^{(V)} \ln \ln \frac{V}{T_K} }{\ln^2 \frac{V}{T_K} } + O \left( \frac{1}{\ln^2 \frac{V}{T_K} }\right)    \right], \label{eq: G(V) antiferro}
        \end{align}
    \eseq
\end{widetext}
where the $\alpha_j^{(T)},\alpha_j^{(V)}$ constants are
\bseq
    \begin{align}
        \alpha_1^{(T)} &= 8 \left( 1 + \ln 2 \right) - 2 \ln \frac{2\pi}{e^{1+\gamma}},\\
        \alpha_1^{(V)} &= 8 \left( 1 + \ln 2 \right),\\
        \alpha_2^{(T)} &= 4\left( 2+ 3\ln 2 \right) +3 \ln \frac{2\pi}{e^{1+\gamma}},\\
        \alpha_2^{(V)} &= 4\left( 2+ 3\ln 2 \right).
    \end{align}
\eseq
Note that the individual values of $\alpha_1^{(T)}$ and $\alpha_1^{(V)}$ can be changed by adjusting the normalization constant $\alpha$ in Eq. \eqref{eq: def of TK}.  In this high energy regime, one can define $T_K^{(T)}$ as the rescaling that sets $\alpha_1^{(T)}$ to zero, with a similar definition for $T_K^{(V)}$; then, the ratio $T_K^{(T)}/T_K^{(V)}= \exp\left[ \left(\alpha_1^{(T)} - \alpha_1^{(V)} \right)/2 \right]= \frac{e^{1+\gamma}}{2\pi}$ is independent of rescaling.

Let us compare these results with the literature.  The leading-order results for $G(T)$ and $G(V)$ are well known \cite{KaminskiNazarovGlazman}, and are special cases of Eq. \eqref{eq: LL conductance antiferro}.  For a higher-order check, we compare to the real-time renormalization group calculation of Pletyukhov and Schoeller (PS) \cite{PletyukhovSchoeller}.  While these authors calculated the full conductance curves numerically, we are concerned for the moment with comparing to the analytical expressions they find for the first two terms ($g_R^2$ and $g_R^3$) of  $G(T)$ and $G(V)$ as power series in the running coupling $g_R$.  Re-expressing their answers in terms of bare quantities, we note that the $D$-independent $g^3$ terms of our series [the distinctive number $\ln \frac{2\pi}{e^{1+\gamma}}$ for $G(T)$ and zero for $G(V)$] are in exact agreement with PS.  This in turn means we have exact agreement for the ratio $T_K^{(T)}/T_K^{(V)}$.   Our scaling differs from theirs at higher order, seeing as they find the conventional expression ($\beta_3=2$).  Conventional scaling would have been obtained in our calculation had an additional contribution $3\pi^2 G_0 \left( g^4 \ln \frac{D}{T} - g^5 \ln^2 \frac{D}{T} \right)$ been present in $G(T)$ [or the same term in $G(V)$ with $V$ replacing $T$], but extensive checks (see Appendix \ref{sec: Additional checks}) have not detected any such contribution.  However, we can show in the calculation of a different observable that such a term can arise as an ``artifact'' of the unusual cutoff scheme we have used; furthermore, in that other observable we can show that a modification of our scheme removes the artifact.  See Appendix \ref{sec: Cutoff artifact in the time-dependent magnetization} for details.

The first terms in the final answers Eq. \eqref{eq: G(T) antiferro} and Eq. \eqref{eq: G(V) antiferro} that $\beta_3$ affects are the double log terms $\frac{\ln \ln \frac{T}{T_K}}{\ln^3 \frac{T}{T_K} }$ and $\frac{\ln \ln \frac{V}{T_K}}{\ln^3 \frac{V}{T_K} }$ ; with the conventional $\beta_3=2$, the coefficient $8$ would instead be $1$.  [Note that the coefficients of the leading terms, $1/\ln^2 (T/T_K)$ and $1 / \ln^2(V/T_K)$, are unaffected.]

We therefore conclude that our approach yields the correct leading log answer in the high-energy regime, with the higher-order corrections being affected by an artifact of our cutoff scheme.  By subtracting the zero-temperature current, we can reliably calculate the effect of temperature at the first approximation beyond leading logs.  In the next section, we repeat the calculation in the strong coupling regime, focusing on the quantities that came out correctly in the antiferromagnetic case.

\subsection{Ferromagnetic regime: Universality}\label{sec: Ferromagnetic regime: Universality}
Our approach reveals another universal regime of the Kondo model: strong ferromagnetic coupling ($g<0$, $|g| \gg 1$).  We note that there are several proposed mesoscopic realizations  \cite{KuzmenkoEtAl, MitchellEtAl, BaruselliEtAl} of the weak ferromagnetic model; it may be possible to realize strong ferromagnetism by modifying these proposals to use the charge Kondo effect \cite{MitchellPrivateCommunication}.

We find that the strong ferromagnetic model generates a Kondo temperature given at leading order by $T_K^{(F)} = D e^{\frac{3\pi^2}{8} g}$.  A very similar discussion applies in this case as in the antiferromagnetic regime.  (Indeed, the quantity $-1/g$, which is small and positive, plays much the same role as a small antiferromagnetic coupling, though the parallel is not exact.)  The scaling limit in this regime consists of taking $D\to\infty$ and $g\to-\infty$ with $T_K$ fixed; the resulting universal functions are expected to agree with the low-energy results from a more realistic Hamiltonian.

We begin in the same way as in the antiferromagnetic case, by examining the scaling.  The same integrals appear again; the only change we need to make is to expand the spin sums $W_n^{(\sigma)}(J)$ about $J=-\infty$ instead of $J=0$.  (We can actually expand the spin sums about $|J|=\infty$ with the same result for either sign of $J$; we discuss the case of large positive $J$ in Sec. \ref{sec: RG discussion}.)  We find the following scaling form at large bandwidth:
\begin{widetext}
\begin{alignat}{3}
    I(T_1,T_2,V)=\notag\\
    \frac{1}{\pi}V \Biggr\{ 1 - \frac{4}{9\pi^2} \biggr[ \frac{7}{g^2} &-\frac{16}{\pi^2 g^3} \ln\frac{2D}{M}  &&+\frac{64}{\pi^4 g^4}\ln^2 \frac{2D}{M}  &&-\frac{2048}{9\pi^6 g^5} \ln^3 \frac{2D}{M} \notag\\
    &-C_1\frac{16}{\pi^2 g^3}   &&+C_1\frac{128}{\pi^4 g^4}  \ln\frac{2D}{M}  &&+\left(4-12C_1 \right)\frac{512}{\pi^6 g^5}\ln^2\frac{2D}{M} \notag \\
    & &&+
    \left(3 C_2+ 6\pi \widetilde{C}_1 - 22\pi^2
    \right)
    \frac{16}{9\pi^4 g^4} &&+
    \begin{pmatrix}
    32- 8 C_2
    +16 C_1 \\ - 12\pi \widetilde{C}_1 + 11\pi^2 
    \end{pmatrix}
    \frac{64}{9\pi^6 g^5} \ln \frac{2D}{M} \notag\\
    & &&  &&+C_4 \frac{1}{g^5}\qquad \qquad \qquad +O\left(\frac{1}{g^6}\right)  \biggr] \Biggr\}
\end{alignat}
\normalsize
where $C_1$, $C_2$, $\widetilde{C}_1$, and $C_4$ depend on the ratios $T_1/V$ and $T_2/V$; the first two have been defined already in Eqs. \eqref{eq: C1} and \eqref{eq: C2}, $\widetilde{C}_1$ is the imaginary part of the same quantity that appears in $C_1$:
\begin{align}
    \widetilde{C}_1\left(\theta,\phi \right) = 4\ \text{Im}\left\{ \gamma - \int_0^\infty du\ \ln u \frac{\partial}{\partial u} \left[ f\left(\theta,\phi;u\right) h\left(\theta,\phi;-u\right) \right] \right\},
\end{align}
\end{widetext}
and $C_4$ is given by a lengthy sum of integrals over $f$ and $h$, which we omit.  This expansion is valid for either sign of $g$, though we focus on the ferromagnetic case $g<0$ for now.

For $T_1=T_2$, we find that the Callan-Symanzik equation holds with a non-zero anomalous dimension $\gamma(g)$:
\bseq
    \begin{align}
        \beta(g) &= -\frac{8}{3\pi^2}\left[ 1 + \frac{32}{9 \pi^2g}  + \frac{\widetilde{\beta}_2}{\pi^4 g^2}+  O\left(\frac{1}{g^3}\right) \right],\\
        \gamma(g) &= \frac{256}{27\pi^4 g^3}\biggr\{ 1 + \frac{56}{9\pi^2 g}\notag\\
        &+ \frac{1}{\pi^4}\left[\frac{7}{4} \widetilde{\beta}_2 - \frac{115}{9\pi^2} +\frac{64}{3\pi^4} \right]\frac{1}{g^2}
         +O\left( \frac{1}{g^3}\right) \biggr\},
    \end{align}
\eseq
where the constant $\widetilde{\beta}_2$ would be determined by the next order ($1/g^6$).  The scaling invariant is the Kondo temperature for this regime \footnote{In the Bethe ansatz solution of the equilibrium problem, with the regularization scheme and notation used in Ref. \cite{AndreiFuruyaLowenstein}, the limits $J\to-\infty$ and $J\to 0^+$ both correspond to $c\to 0^+$.  We can thus read off (noting that the conventions are related by $J= 2 J_{\text{Bethe ansatz}}$) the \emph{same} strong ferromagnetic $T_K^{(F)}$ that appears in Eq. \eqref{eq: TK strong ferro}, aside from the $\ln |g|$ term in the exponent.  The absence of such log terms in the Bethe ansatz is well known in the antiferromagnetic case.}:
\beq
    T_K^{(F)} \equiv D e^{\frac{3\pi^2}{8}g- \frac{4}{3} \ln |g|}.\label{eq: TK strong ferro}
\eeq
Let us emphasize that the non-zero anomalous dimension $\gamma(g)$ for the current operator is necessary in this case to resum even the \emph{leading} logarithms.  Concretely, this means that one would \emph{not} obtain the correct beta function by compensating a change in coupling constant in the $1/g^2$ term by a change of bandwidth in the $(1/g^3)\ln\frac{2D}{M}$ term; the resulting beta function would not be consistent with the next term, $\sim(1/g^4)\ln^2\frac{2D}{M}$.  One is forced rescale the whole observable as well, which is equivalent to introducing $\gamma(g)$.

\begin{figure}[htp]
    \centering
    \includegraphics[width=\linewidth]{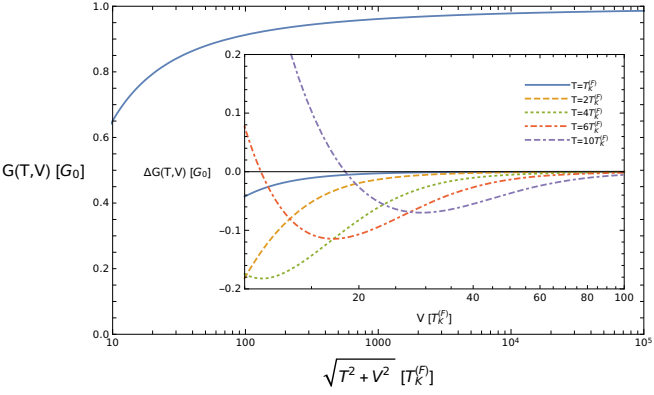}
    \caption{The universal conductance $G \equiv \pd I_{\text{steady-state}}/\pd V$ in the strong ferromagnetic regime at leading log approximation.  In contrast to the antiferromagnetic case in which $G$ is known to reach the unitarity limit $G_0 \equiv 2e^2/h$ at $T=V=0$ \cite{GlazmanRaikh}, here the unitarity limit is reached asymptotically at \emph{large} voltage or temperature.  As the external scale is lowered to $T_K^{(\text{F})}$ and below, the series in $1/g$ breaks down and another method is needed.  Inset: the first correction beyond leading log in the quantity $\Delta G \equiv G(T,V) - G(T = 0,V)$ for $V\gg T_K^{(\text{F})}$, with various values of $T$.}\label{fig: strong ferro combined}
\end{figure}

Notice that we can take the scaling limit $D\to\infty$, $g\to-\infty$ with $T_K^{(\text{F})}$ held fixed, indicating that the strong ferromagnetic regime is universal.  Resumming the leading logs, we find that the conductance approaches the unitarity limit asymptotically at \emph{high} voltage or temperature (Fig. \ref{fig: strong ferro combined}):
\beq
    G(T,V) = G_0 \left( 1 - \frac{3\pi^2}{16 \ln\frac{ \sqrt{T^2 +V^2}}{T_K^{(\text{F})}} } + \dots\right).\label{eq: G leading log strong ferro}
\eeq
In analogy to the antiferromagnetic case, we expect that the coefficient $-\frac{4}{3}$ of $\ln |g|$ in Eq. \eqref{eq: TK strong ferro} is affected by our cutoff scheme and may not be reliable; however, this only affects higher-order corrections to Eq. \eqref{eq: G leading log strong ferro}.  We expect that in the first correction beyond leading logs, the difference $\Delta G$ is reliable (see inset of Fig. \ref{fig: strong ferro combined}), again by analogy to the antiferromagnetic case.

Curiously, the scaling breaks down if the lead temperatures are different ($T_1 \ne T_2$).  The problem term, $\sim (1/g^5)\ln(2D/M)$ is in the sub-sub-leading log part of the series, and may possibly be affected by cutoff artifacts.

For the special cases $G(T)$ and $G(V)$, we obtain
\begin{widetext}
    \bseq
        \begin{alignat}{3}
            G(T) = G_0 \Biggr\{ 1- \frac{4}{9 \pi^2} \biggr[ \frac{7}{g^2} &-\frac{16}{\pi^2 g^3} \ln\frac{D}{T}  &&+\frac{64}{\pi^4 g^4} \ln^2 \frac{D}{T}  &&-\frac{2048}{9\pi^6 g^5} \ln^3 \frac{D}{T} \notag\\
            &+\frac{16}{\pi^2 g^3} \ln \frac{2\pi}{e^{1+\gamma}} &&-\frac{128}{\pi^4 g^4} \ln \frac{2\pi}{e^{1+\gamma}} \ln\frac{D}{T}  &&+\frac{2048}{9\pi^6 g^6}\left( 3 \ln \frac{2\pi}{e^{1+\gamma}} + 1 \right) \ln^2\frac{D}{T} \notag \\
            & &&-4.39 \frac{1}{g^4}  &&+ 1.61 \frac{1}{g^5} \ln\frac{D}{T}\notag\\
            & &&  &&-0.22 \frac{1}{g^5} \qquad \qquad + O\left(\frac{1}{g^6}\right)  \biggr]  \Biggr\},\label{eq: G(T) large g series}\\
            G(V) = G_0 \Biggr\{ 1- \frac{4}{9 \pi^2} \biggr[ \frac{7}{g^2} &-\frac{16}{\pi^2 g^3} \ln\frac{D}{V}  &&+\frac{64}{\pi^4 g^4} \ln^2 \frac{D}{V}  &&-\frac{2048}{9\pi^6 g^5} \ln^3 \frac{D}{V} \notag\\
            &  &&   &&+\frac{2048}{9\pi^6 g^5} \ln^2\frac{D}{V} \notag \\
            & &&-  \frac{436}{9 \pi^2 g^4}  &&+ \frac{64}{9\pi^2}\left( 64 + 25\pi^2 \right) \frac{1}{g^5} \ln\frac{D}{V}\notag\\
            & &&  &&+\frac{16}{27\pi^6 g^5}[192\left(4 - 6 \ln2 + 3 \zeta(3)  \right)\notag\\
            & && && \qquad \qquad -24\pi^2   ] \qquad  + O\left(\frac{1}{g^6}\right) 
            \biggr] \Biggr\} , \label{eq: G(V) large g series}
        \end{alignat}
    \eseq
    In the high energy regime ($T\gg T_K$ or $V\gg T_K$), the running coupling constant is large and negative, and we can use the Callan-Symanzik equation to find the following universal results:
    \bseq
        \begin{align}
            &G(T) =  G_0 \left\{ 1- \frac{3\pi^2}{16\ln^2\frac{T}{T_K}} \left[1+ \frac{8}{3} \frac{\ln \ln \frac{T}{T_K}}{\ln \frac{T}{T_K}} + \frac{\widetilde{\alpha}_1^{(T)}}{\ln \frac{T}{T_K}} + \frac{16}{3}\frac{ \ln^2 \ln \frac{T}{T_K}}{ \ln^2 \frac{T}{T_K} } + \frac{\widetilde{\alpha}_2^{(T)} \ln \ln \frac{T}{T_K} }{\ln^2 \frac{T}{T_K} } + O \left( \frac{1}{\ln^2 \frac{T}{T_K} }\right) \right] \right\}, \label{eq: G(T) strong ferro}\\
            &G(V) = G_0 \left\{ 1- \frac{3\pi^2}{16\ln^2\frac{V}{T_K}} \left[1+ \frac{8}{3} \frac{\ln \ln \frac{V}{T_K}}{\ln \frac{V}{T_K}} + \frac{\widetilde{\alpha}_1^{(V)}}{\ln \frac{V}{T_K}} +\frac{16}{3}\frac{\ln^2 \ln \frac{V}{T_K}}{\ln^2 \frac{V}{T_K} } + \frac{\widetilde{\alpha}_2^{(V)} \ln \ln \frac{V}{T_K} }{\ln^2 \frac{V}{T_K} } + O \left( \frac{1}{\ln^2 \frac{V}{T_K} }\right) \right] \right\}, \label{eq: G(V) strong ferro}
        \end{align}
    \eseq
\end{widetext}
where the $\widetilde{\alpha}_j^{(T)},\widetilde{\alpha}_j^{(V)}$ constants are
\bseq
    \begin{align}
        \widetilde{\alpha}_1^{(T)} &= \frac{8}{9} - \frac{8}{3}\ln\frac{3\pi^2}{8} -2 \ln \frac{2\pi}{e^{1+\gamma}},\\
        \widetilde{\alpha}_1^{(V)} &= \frac{8}{9} - \frac{8}{3}\ln\frac{3\pi^2}{8},
    \end{align}
\eseq
and
\bseq
    \begin{align}
        \widetilde{\alpha}_2^{(T)} &= -8\left(\frac{4}{9} \ln \frac{27\pi^6}{512} + \ln \frac{2\pi}{e^{1+\gamma}} \right),\\
        \widetilde{\alpha}_2^{(V)} &= -\frac{32}{9} \ln \frac{27\pi^6}{512}.
    \end{align}
\eseq
Notice that the unitarity limit is reached asymptotically at high energy (Fig. \ref{fig: strong ferro combined}).
This is the main prediction of our method so far.  Ultimately, the unitary conductance traces back to the fact that the bare $\mathcal{S}$ matrix of the model becomes a single-particle phase shift of $\pi/2$ in the limit $|J|\to \infty$ (see Sec. \ref{sec: Solution in an alternate basis}).

To see the predicted rise towards unitarity experimentally, one would need a hierarchy of scales $T_K^{(F)} \ll V \ll E_{\text{max}}$ or $T_K^{(F)} \ll T \ll E_{\text{max}}$, where $E_{\text{max}}$ is the lowest energy scale at which the Kondo model is no longer an accurate description of the system.

Defining $T_K^{(F,T)}$ and $T_K^{(F,V)}$ in the same way as in the antiferromagnetic case [see \eqref{eq: G(V) antiferro} and below], we find that the universal ratio is the same in this regime: $T_K^{(F,T)}/T_K^{(F,V)} =  \frac{e^{1+\gamma}}{2\pi}$.

\subsection{RG discussion}\label{sec: RG discussion}

The basic picture of scaling in the antiferromagnetic Kondo model is that the theory is effectively strongly coupled at low energies ($T,V\ll T_K$), even though the coupling constant that appears in the original Hamiltonian is small ($0 < g \ll 1 $).  Loosely speaking, one says that the coupling constant increases as one reduces the measurement scale, reaching infinity at zero energy.  It is tempting to suggest, then, that a calculation using the Kondo Hamiltonian with \emph{large} $g$ (expanding in powers of $1/g$) would reproduce the low-energy regime of the model with \emph{small} $g$.  In this section, we show that this is not so, both by general arguments and by examining our explicit answers in the large-$g$ regime.  Starting from weak coupling and flowing to strong coupling at low energy is not the same as starting the theory at strong coupling.

Our statement does not contradict the many successes of the effective field theory approach to the low-energy regime (of the model with small $g$), which refers to the leading irrelevant operators around the strong coupling fixed point.  Instead, the conclusion is that the effective field theory approach is more sophisticated than the simple idea of taking $g$ to be large in the original Hamiltonian.

To clarify the point, we must carefully set up the field theoretic version of the renormalization group.  For definiteness, we consider a dimensionless observable $\mathcal{O}(D,g,T)$ with temperature $T$ as the only external scale.  Our analysis is not confined to equilibrium, though, and $T$ can be replaced by any single energy scale (such as a bias voltage).  Suppose the observable is calculated as a power series in $g$, with the leading term being $g^2$; then a series expansion in $g$ must take the form
\beq
    \mathcal{O}(D,g,T) = g^2 + \sum_{n=3}^\infty  g^n F_n(D/T),
\eeq
where $F_n(D/T)$ are some functions.  As discussed in \cite{Delamotte}, these functions are constrained by the perturbative renormalizability of the model to take a logarithmic form in the $T \ll D$ regime:
\beq
    F_n(D/T) = \sum_{m=0}^{n-2} a_{nm} \ln^m \frac{D}{T} + \dots,
\eeq
where the $a_{nm}$ coefficients are pure numbers that depend on the observable being evaluated.  The logarithmic terms define the ``scaling form'' part of the observable:
\beq
    \mathcal{O}_{\text{scaling form}}(D,g,T) = g^2 + \sum_{n=3}^\infty \sum_{m=0}^{n-2} a_{nm} g^n\ln^m \frac{D}{T} .
\eeq
The scaling form satisfies the RG scaling (or Callan-Symanzik) equation
\beq
    \left[ D\frac{\partial}{\partial D} + \beta(g) \frac{\partial }{\partial g} + \gamma(g) \right]\mathcal{O}_{\text{scaling form}} = 0.
\eeq
Assuming (as we find for the current) that the leading order of the anomalous dimension term $\gamma(g)$ starts at the same order or higher as the leading order of the beta function, the solution of the Callan-Symanzik equation then implies that the scaling form can be written as a function of $T/T_K$ only [where $T_K$ is the scaling invariant defined by $( D\frac{\partial}{\partial D} + \beta(g) \frac{\partial }{\partial g})T_K =0$], up to corrections that vanish as $g\to 0^+$:
\beq
    \mathcal{O}_{\text{scaling form}}(D,g,T) = f_{\text{universal}}(T/T_K) \left[ 1 + O\left( g\right) \right].
\eeq
In the Kondo model, the leading order of the beta function has negative sign.  This implies that $T_K$ can be held fixed while taking the limit $D\to\infty$ and $g\to 0^+$, which means that the function $f_{\text{universal}}(T/T_K)$ is a universal result for the observable $\mathcal{O}$.  In contrast, the scaling invariant \emph{cannot} be held fixed in the limit $D\to\infty$ and $g\to 0^-$, so the function $f_{\text{universal}}(T/T_K)$ in the ferromagnetic case only represents what would happen if the simplified (wide-band) model itself were realized.

Let us focus on the antiferromagnetic $(g>0)$ case for now.  The procedure for calculating the asymptotic behavior of $f_{\text{universal}}(T/T_K)$ for $T\gg T_K$ using the first few series coefficients $a_{nm}$ is well known.  One finds that the solution of the Callan-Symanzik equation is characterized by a running coupling ($g_R = \frac{1}{2\ln T/T_K}$ at the leading approximation) which is found to grow as $T$ is reduced.  As $T$ approaches $T_K$ from above, one finds that infinitely many series coefficients are needed; however, non-perturbative techniques confirm that the running coupling keeps growing as $T$ is reduced.  If one ignores momentarily the distinction between the running coupling and the bare coupling, one can imagine that a series in $1/g$ would provide information about the low-temperature behavior of $f_{\text{universal}}(T/T_K)$, much in the same way that a series in $g$ yields the high-temperature behavior.

The basic problem with this approach is that if one repeats the same steps with the $1/g$ series, i.e., expand each order of the series for large bandwidth and declare the logarithmic part to be the ``scaling form,'' one arrives at a scaling form that may not be the same as the one found from the $g$ series.  Since the ultimate goal is to take $g\to0^+$ with $T_K$ fixed, the scaling form of the $g$ series is the correct one.  But the parts of this scaling form that describe the small $T/T_K$ behavior of the function $f_{\text{universal}}(T/T_K)$ may appear to be negligible in the $1/g$ series.

A simple example illustrates the point.  It is known that the universal conductance curve $G(T)$ reaches unitarity at $T=0$ with corrections of the form $T^2/T_K^2$.  Thus, the scaling form for the conductance must include a contribution of the form $\frac{1}{g} \frac{T^2}{D^2}$, seeing as this term becomes $T^2/D^2$ in the $g\to 0^+$ scaling limit (we assume the conventional expression $T_K = D e^{ -\frac{1}{2 g} + \frac{1}{2} \ln g }$ in this discussion).  Since this term vanishes for large bandwidth rather than diverging logarithmically, it is exactly the type of term that is dropped in determining the scaling form of the $1/g$ series.  The logarithmically diverging terms, on the other hand, can easily be negligible in the $g\to0^+$ scaling limit; consider, e.g., the expansion $\frac{1}{g +  \ln D/T} = \frac{1}{g} - \frac{1}{g^2} \ln \frac{D}{T} + \dots$ in powers of $1/g$.  Thus, no finite number of terms of the $1/g$ series will yield the low-temperature behavior, since there is no obvious way to identify which contributions are important in the $g\to 0^+$ scaling limit.

The scaling form of the $1/g$ series describes a different physical problem: one in which the bare coupling constant is large in magnitude.  The sign of the beta function then indicates that the strong \emph{ferromagnetic} regime is universal and the strong antiferromagnetic regime is non-universal.  The quantity $-\frac{1}{g}$ behaves much like $g$ does in the antiferromagnetic case; that is, the $g = -\infty$ point behaves like $g = 0^+$, and $g = 0^-$ behaves like $g = \infty$.  Let us state this more definitely.  A system with large negative bare coupling $g$ has a running coupling that is also large and negative at high energies, so an RG-improved power series in $\frac{1}{g}$ produces accurate results.  At low energies, a more powerful technique is needed; neither a series in $\frac{1}{g}$ nor a series in the \emph{inverse} parameter $g$ gives any information about the low-energy behavior (unless one has \emph{all} terms of the series), because in this case the correct scaling form is the one generated by the $1/g$ series (which can differ from the scaling form generated by the $g$ series).

\begin{figure}[htp]
    \centering
    \includegraphics[width=\linewidth]{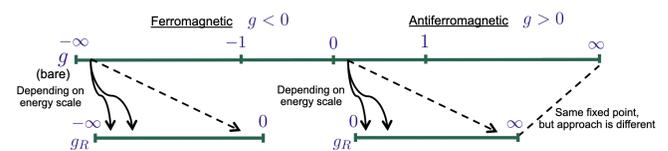}
    \caption[Kondo scaling picture]{Kondo scaling picture.  The two universal regimes are weak antiferromagnetic \emph{bare} coupling ($0 < g \ll 1$, $T_K = D e^{-1/(2g)}$) and strong ferromagnetic \emph{bare} coupling ($g<0$, $|g| \gg1$, $T_K^{(\text{F})} = D e^{- 3\pi^2 |g| /8}$).  The former has been much studied, and the latter is predicted by our calculations.  In either case, the running coupling $g_R$ is close to the bare coupling if the system is probed at a high energy scale (high relative to $T_K$ but always small compared to the bandwidth), but moves away from the bare coupling as the energy scale is reduced.}
    \label{fig: Kondo scaling}%
\end{figure}

Our calculation yields the beginning of the RG flow in the strong ferromagnetic regime (see Fig. \ref{fig: Kondo scaling}): starting at the unstable fixed point $g_R=-\infty$, the running coupling constant becomes \emph{smaller} in magnitude according to $g_R = - \frac{8}{3\pi^2} \ln \frac{T}{T_K^{(F)}}$ (at leading order).
As $T$ approaches $T_K^{(F)}$ from above, $|g_R|$ becomes too small for our calculation to be valid.  We expect that $g_R$ continues to flow to the stable fixed point $g_R=0^-$ without any other fixed points in between (much like the corresponding antiferromagnetic flow from $g_R = 0^+$ to $g_R=\infty$).  The ground state of the system would flow from a triplet at high energy, with entropy $\ln 3$, to a free spin at low energy, with entropy $\ln 2$.  We emphasize again that perturbation theory in small, bare, ferromagnetic $g$ provides no information at all about the low-energy behavior of a system with strong ferromagnetic $g$ except the extreme point.  In other words, the conductance in the universal strong ferromagnetic regime should be zero at $T=V=0$, but calculating the approach to zero requires another method (such as an analysis of leading irrelevant operators, or NRG).

\section{Conclusion and outlook}\label{sec: Conclusion and outlook}
We have provided an exact, explicit solution for the time-evolving wavefunction in a many-body problem, and found the corresponding NESS in the long-time limit.  In the thermodynamic limit, we have found a series expression for the current which can be expanded either for weak coupling or for strong coupling, and shown that either expansion converges to all orders in the steady-state limit.  Our series predicts a universal strong ferromagnetic regime in which the conductance approaches the unitarity limit asymptotically at large voltage or temperature.   We expect that the same basic picture of RG flow will be found if the calculation can be repeated using a conventional cutoff scheme.

There are a number of possible directions to take with this work in the future.  One is the evaluation of the $\mathcal{S}$ matrix, not the bare $\mathcal{S}$ matrix that we used in our calculations, but the physical $\mathcal{S}$ matrix for excitations above a filled Fermi sea.  The NESS we obtained in the Kondo model is a many-body scattering ``in'' state, and it is straightforward to obtain the corresponding ``out'' state by considering evolution to large negative times.  Since the initial quantum numbers are arbitrary, we are free to construct a state consisting of a Fermi sea with one electron above it with momentum $p$ and spin $a$; schematically, $|\text{FS}, p a\rangle_{\text{in}}$.  The $\mathcal{S}$ matrix for elastic single-particle scattering is then given by $_\text{out}\langle \text{FS}, p a'  |\text{FS}, p a\rangle_{\text{in}}$.  The calculation of the $\mathcal{S}$ matrix can proceed using some of the same technology developed here, such as the reduction of a general overlap to a sum of normal-ordered overlaps.  If necessary, the calculation could be done by considering finite time first and then taking the limit of large time.  More complicated scattering processes involving particle-hole pairs could be considered by making different choices of the initial and final quantum numbers.

Another direction would be to adapt either the self-consistent rate equation used in \cite{RoschEtAlPRL} or the Dyson equation used in \cite{ParcolletHooley} to the many-body wavefunction approach presented here, in order to repeat the calculation of the electric current in the presence of a non-zero magnetic field on the dot (particularly in the strong ferromagnetic regime).

It would be interesting to see if our general method for calculating local quenches can be useful in a wider class of problems.  As we have mentioned, the usual signatures of integrability in the Kondo model, such as the Yang-Baxter equation, do not appear in any obvious way in our calculations. 

To take full advantage of the fact that the wavefunction for a fixed number of electrons is exact, it is essential to find a different way of taking the thermodynamic limit of observables other than the approach we took here of expanding in powers of $J$ or $1/J$.  We hope that the technology for using these wavefunctions to calculate observable quantities in the thermodynamic limit can eventually reach the advanced state of development found in equilibrium calculations with the Bethe ansatz.

\begin{acknowledgments}
    We are grateful to Chung-Hou Chung, Piers Coleman, Garry Goldstein, Yashar Komijani, Yigal Meir, Andrew Mitchell, Achim Rosch, and Hubert Saleur for helpful discussions.  We have benefited from working on related problems with Huijie Guan, Paata Kakashvili, Christopher Munson, and Roshan Tourani.  A.B.C. acknowledges support from the Samuel Marateck Fellowship in Quantum Field Theory Physics and the Excellence Fellowship (both from Rutgers University).  This material is based upon work supported by the National Science Foundation under Grant No. 1410583.
\end{acknowledgments}

\appendix

\begin{widetext}

    \section{Notation for calculations}\label{sec: Notation for calculations}
    We present a compact notation for manipulating the many-body wavefunction and its matrix elements.  This notation allows us to do calculations that would be excessively lengthy if all indices were written out in full.  It will be used throughout the remaining appendices.
    
    We use boldface letters to stand for lists of indices:  $\mathbf{m} = (2,5,6)$, for example.  We use $m_j$ and $m(j)$ interchangeably to refer to individual list elements, such as $m_2 = m(2) =5$.  Boldface letters in subscripts indicate products in the manner of the following examples [in which $\mathbf{m}$ has length $n$, a small circle stands for composition, and $\sigma\in \text{Sym}(n)$]:
    \begin{align}
        c_{\alpha_{\mathbf{m}}} &= c_{\alpha_{m(1)}}\dots c_{\alpha_{m(n)} },\ c_{\alpha_{\mathbf{m}\circ \sigma} } = c_{\alpha_{m(\sigma_1)}}\dots c_{\alpha_{m(\sigma_n)}},\\
        c_{\alpha_{\mathbf{m}}}^\dagger &= c_{\alpha_{m(n)} }^\dagger \dots c_{\alpha_{m(1)}}^\dagger,\ c_{\alpha_{\mathbf{m}\circ \sigma} }^\dagger = c_{\alpha_{m(\sigma_n)}}^\dagger\dots c_{\alpha_{m(\sigma_1)}}^\dagger.
    \end{align}
    Given any list $\mathbf{m}$ of increasing indices ($m_1 < \dots < m_n$), we define $\mathcal{I}_j(\mathbf{m})$ to be the set of increasing lists of length $j$ chosen from $\mathbf{m}$:
    \beq
        \mathcal{I}_j(\mathbf{m}) = \{\bm{\ell} = (\ell_1,\dots, \ell_j)\subset \mathbf{m}\ |\ \ell_1 < \dots < \ell_j \}.
    \eeq
    It is often convenient to write a sum over a single index $\ell_1$ as a sum over lists $\bm{\ell}$ of length $1$ [i.e., $\bm{\ell} \in \mathcal{I}_1(\mathbf{m})$ ] in order to use the notation we define in the next paragraph.
    
    Given $\bm{\ell} \in \mathcal{I}_j(\mathbf{m})$, we define $\overleftarrow{perm}[\bm{\ell}]$ to be the permutation of $\mathbf{m}$ that brings all the entries of $\bm{\ell}$ to the left of all the remaining entries of $\mathbf{m}$; we define $\overrightarrow{perm}[\bm{\ell}]$ similarly.  For example, if $\mathbf{m} = (1,3,6,7)$ and $\bm{\ell}= (1,6)$, then $\overleftarrow{perm}[\bm{\ell}]$ maps $(1,3,6,7)\to (1,6,3,7)$ and $\overrightarrow{perm}[\bm{\ell}]$ maps $(1,3,6,7)\to (3,7,1,6)$.  Note that $\overleftarrow{perm}[\bm{\ell}]$ and $\overrightarrow{perm}[\bm{\ell}]$ depend implicitly on the list $\mathbf{m}$ from which the entries in $\bm{\ell}$ are chosen.  We write the sign factors for these permutations in the following way:
    \bseq
        \begin{align}
            \sgnleft\bm{\ell} &\equiv \sgn \overleftarrow{perm}[\bm{\ell}], \\
            \sgnright\bm{\ell} &\equiv \sgn \overrightarrow{perm}[\bm{\ell}]. 
        \end{align}
    \eseq
    The slash notation $\mathbf{m}/\bm{\ell}$ indicates the list $\mathbf{m}$ with the indices belonging to $\bm{\ell}$ all removed; in the example given above, $\mathbf{m} / \bm{\ell} = (3,7)$.  The same slash notation also applies for removing a single entry of list: for instance, $\mathbf{m} / 3 = (1,6,7)$.
    
    Using this notation, the many-body wavefunction \eqref{eq: wavefn construction} can be written more compactly as
    \beq
        |\Psi(t)\rangle = \sum_{n=0}^N \sum_{\mathbf{m}\in\mathcal{I}_n(\mathbf{N})} \left( \sgnleft\mathbf{m} \right) c_{\alpha_{\mathbf{N}/\mathbf{m} }}^\dagger(t) \sum_{\sigma\in \text{Sym}(n)} (\sgn \sigma) |\chi_{\alpha_{\mathbf{m}\circ \sigma},\beta}(t)\rangle.\label{eq: Psi(t) compact form}
    \eeq
    
    \section{Proof of general formalism}\label{sec: Proof of general formalism}
    We demonstrate that the ``inverse problem'' conditions \eqref{eq: chi condition 1} and \eqref{eq: chi condition 2} imply that the construction \eqref{eq: Psi(t) compact form} satisfies $|\Psi(t)\rangle = e^{-i H t} |\Psi\rangle$.  The second condition \eqref{eq: chi condition 2} immediately implies that $|\Psi(t=0)\rangle = |\Psi\rangle$.  The main task is to show that the first condition \eqref{eq: chi condition 1} implies that $(H- i \frac{d}{dt} )|\Psi(t)\rangle =0 $.  On any given term within $|\Psi(t)\rangle$, we bring $H-i \frac{d}{dt}$ to the right past all of the $c^\dagger(t)$ operators to hit the $|\chi(t)\rangle$ state, at the cost of generating an $A(t)$ operator for each $c^\dagger(t)$ operator that is passed.  Since $|\chi_{,\beta}(t)\rangle \equiv |\beta(t)\rangle$ is annihilated by $(H-i\frac{d}{dt})$, Eq. \eqref{eq: Psi(t) compact form} yields
    \begin{multline}
        \left( H - i \frac{d}{dt} \right) | \Psi(t)\rangle = \sum_{n=1}^N \sum_{\mathbf{m} \in \mathcal{I}_n(\mathbf{N} ) } \left( \sgnleft\mathbf{m} \right) c_{\alpha_{\mathbf{N} / \mathbf{m} } }^\dagger(t) \sum_{\sigma \in \text{Sym}(n) }(\sgn \sigma) \left( H - i\frac{d}{dt} \right) |\chi_{\alpha_{\mathbf{m} \circ \sigma} ,\beta}(t) \rangle\\
        +\sum_{n=0}^{N-1} \sum_{\mathbf{m} \in \mathcal{I}_n(\mathbf{N} ) } \left( \sgnleft\mathbf{m} \right) \sum_{\bm{\ell} \in \mathcal{I}_1(\mathbf{N} / \mathbf{m} ) } \left( \sgnleft\bm{\ell} \right)  c_{\alpha_{\mathbf{N} / \mathbf{m} / \bm{\ell}} }^\dagger(t) A_{\alpha_{\ell(1)} }(t)\sum_{\sigma\in \text{Sym}(n)} (\sgn \sigma) |\chi_{\alpha_{\mathbf{m} \circ \sigma} ,\beta}(t). \rangle \label{eq: H-id/dt on Psi}
    \end{multline} 
    Using the condition \eqref{eq: chi condition 1}, we find that the first term becomes
    \bseq
        \begin{align}
            1\text{st term of }&\eqref{eq: H-id/dt on Psi} = - \sum_{n=1}^N \sum_{\mathbf{m} \in \mathcal{I}_n(\mathbf{N} ) } \left( \sgnleft\mathbf{m} \right) c_{\alpha_{\mathbf{N} / \mathbf{m} } }^\dagger(t)    \sum_{\sigma \in \text{Sym}(n) }(\sgn \sigma)  A_{\alpha_{m(\sigma_n)}}(t) |\chi_{\alpha_{(\mathbf{m}  \circ \sigma) / m(\sigma_n )} ,\beta}(t) \rangle\\
            &= - \sum_{n=1}^N \sum_{\mathbf{m} \in \mathcal{I}_n(\mathbf{N} ) } \left( \sgnleft\mathbf{m} \right) c_{\alpha_{\mathbf{N} / \mathbf{m} } }^\dagger(t) \sum_{\bm{\ell} \in \mathcal{I}_1(\mathbf{m} )}\left( \sgnright\bm{\ell} \right) A_{\alpha_{\ell(1)}}(t)  \sum_{\sigma \in \text{Sym}(n-1) }(\sgn \sigma)  |\chi_{\alpha_{(\mathbf{m}/\bm{\ell}) \circ \sigma} ,\beta}(t) \rangle,
        \end{align}
    \eseq
    where the second line follows from relabelling $m_{\sigma_n} \to \ell_1$.
    
    For the second term of \eqref{eq: H-id/dt on Psi}, we note the following relabelling of summations, which is valid for any function $X$:
    \beq
        \sum_{\mathbf{m} \in \mathcal{I}_n(\mathbf{N} ) } \left( \sgnleft\mathbf{m} \right) \sum_{\bm{\ell} \in \mathcal{I}_1(\mathbf{N} / \mathbf{m} ) } \left( \sgnleft\bm{\ell} \right) X(\mathbf{m}, \bm{\ell}) = \sum_{\mathbf{m} \in \mathcal{I}_{n+1}(\mathbf{N} ) } \left( \sgnleft\mathbf{m} \right) \sum_{\bm{\ell} \in \mathcal{I}_1( \mathbf{m} ) } \left( \sgnright\bm{\ell} \right) X(\mathbf{m} / \bm{\ell}, \bm{\ell}).
    \eeq
    Thus,
    \beq
        2\text{nd term of }\eqref{eq: H-id/dt on Psi} =  \sum_{n=0}^{N-1} \sum_{\mathbf{m} \in \mathcal{I}_{n+1}(\mathbf{N} ) } \left( \sgnleft\mathbf{m} \right) c_{\alpha_{\mathbf{N} / \mathbf{m} } }^\dagger(t) \sum_{\bm{\ell} \in \mathcal{I}_1(\mathbf{m} )}\left( \sgnright\bm{\ell} \right) A_{\alpha_{\ell(1)}}(t)  \sum_{\sigma \in \text{Sym}(n) }(\sgn \sigma)  |\chi_{\alpha_{(\mathbf{m}/\bm{\ell}) \circ \sigma} ,\beta}(t) \rangle,
    \eeq
    which is precisely what is needed to cancel the first term of \eqref{eq: H-id/dt on Psi} (once we relabel the summation variable $n\to n-1$).  This completes the proof that \eqref{eq: Psi(t) compact form} satisfies the time-dependent Schrodinger equation.
    
    \section{Kondo crossing states in the general case}\label{sec: Kondo crossing states in the general case}
    We calculate the $n=1$ crossing state for $|t|<L/2$, finding that the negative time solution is related to the positive time solution by a simple transformation.  We then show that the formula \eqref{eq: Kondo crossing state} for the crossing states $|\chi_{e k_\mathbf{n} a_{\mathbf{n}} ,a_0}(t)\rangle$ solves the appropriate inverse problem for arbitrary $n$.  We also present the solution in a more general Hamiltonian with an anisotropic Kondo interaction and a potential scattering term.
    
    We generalize the ansatz \eqref{eq: n=1 crossing state} for the $n=1$ crossing state to
    \begin{multline}
        |\chi_{e k_1 a_1, a_0}(t)\rangle = \frac{1}{\sqrt{L} } \int_{-L/2}^{L/2} dx\ \left( F_{k_1 a_1,a_0}^{b_1,b_0}(t-x_1) \Theta(0<x_1<t) + G_{k_1 a_1,a_0}^{b_1,b_0}(t-x_1) \Theta(t<x_1<0) \right) \psi_{e b_1}^\dagger(x) e^{i b_0 B t}|b_0\rangle, \label{eq: n=1 crossing state general t}
    \end{multline}
    where $F$ is given by Eq. \eqref{eq: F} and $G$ is another smooth function.  For $|t| < L/2$, we obtain
    \begin{multline}
        \left( H - i \frac{d}{dt} \right) |\chi_{e k_1 a_1 ,a_0}(t)\rangle = 
        \frac{1}{\sqrt{L}} \biggr[ \left( - i I_{d_1 d_0}^{b_1 b_0} + \frac{1}{4} J \bm{\sigma}_{b_1 d_1}\cdot \bm{\sigma}_{b_0 d_0} \right) F_{k_1 a_1,a_0}^{d_1,d_0}(t)e^{i d_0 B t} \Theta(t) \\
        + \left(  i I_{d_1 d_0}^{b_1 b_0} + \frac{1}{4} J \bm{\sigma}_{b_1 d_1}\cdot \bm{\sigma}_{b_0 d_0} \right) G_{k_1 a_1,a_0}^{d_1,d_0}(t)e^{i d_0 B t} \Theta(-t) \biggr] \psi_{e b_1}^\dagger(0) |b_0\rangle.
    \end{multline}
    Inserting a factor of $1= \Theta(t) + \Theta(-t)$ into Eq. \eqref{eq: A(t) on a_0(t)} yields
    \beq
        A_{e k_1 a_1}(t)|a_0(t)\rangle = \frac{1}{\sqrt{L}} \frac{1}{2} J e^{-i k_1 t}e^{i a_0 B t} \left[ \Theta(t) + \Theta(-t)\right] \bm{\sigma}_{b_1 a_1} \cdot \bm{\sigma}_{b_0 a_0} \psi_{e b_1}^\dagger(0)|b_0 \rangle.
    \eeq
    The differential equation $\left( H- i \frac{d}{dt} \right) |\chi_{e k_1 a_1,a_0}(t) \rangle = - A_{e k_1 a_1}(t) |a_0\rangle$ then separates into a $\Theta(t)$ part and a $\Theta(-t)$ part.  The $\Theta(t)$ part has already been considered in the main text, leading to the condition \eqref{eq: F(t) condition} on the function $F$.  The $\Theta(-t)$ part leads to the following condition on the function $G$:
    \beq
        \left( i I_{d_1 d_0}^{b_1 b_0} + \frac{1}{4} J \bm{\sigma}_{b_1 d_1}\cdot \bm{\sigma}_{b_0 d_0} \right) G_{k_1 a_1,a_0}^{d_1,d_0}(t) e^{i d_0 B t}=
        -\frac{1}{2} J e^{-i k_1 t} e^{i a_0 B t}\bm{\sigma}_{b_1 a_1}\cdot \bm{\sigma}_{b_0 a_0},
    \eeq
    from which we conclude [comparing to Eq. \eqref{eq: F(t) condition}] that $G(-t)= F^*(t)$.
    
    Our next task is to show that $|\chi_{e k_\mathbf{n} a_{\mathbf{n}} ,a_0}(t) \rangle$ as given in \eqref{eq: Kondo crossing state} satisfies
    \begin{align}
        \left( H - i \frac{d}{dt} \right) |\chi_{e k_\mathbf{n} a_{\mathbf{n}} ,a_0}(t) \rangle &= - A_{e k_n a_n}(t) |\chi_{e k_{\mathbf{n} /n } a_{\mathbf{n}/n} ,a_0}(t) \rangle, \label{eq: chi condition 1 for Kondo general n}\\
        |\chi_{e k_\mathbf{n} a_\mathbf{n} ,a_0}(t=0)\rangle &= 0. \label{eq: chi condition 2 for Kondo general n}
    \end{align}
    The crossing state \eqref{eq: Kondo crossing state} vanishes at $t=0$ by construction.  To show that the differential equation \eqref{eq: chi condition 1 for Kondo general n} holds, we need the $n$-variable generalization of the delta-Heaviside regularization \eqref{eq: delta Theta simplest}, namely,
    \beq
        \delta(x_n)\Theta(0<x_n < \dots < x_1 < t) = \frac{1}{2} \delta(x_n)\Theta(0< x_{n-1} < \dots < x_1 < t ).
    \eeq
    By computations very similar to the $n=1$ case discussed in the main text, we obtain
    \begin{multline}
        \left( H - i \frac{d}{dt} \right) |\chi_{k_\mathbf{n} a_{\mathbf{n}} ,a_0 }(t) \rangle = L^{-n/2} \int_{-L/2}^{L/2} dx_{\mathbf{n}/n}\ \delta_{a_0}^{c_0} \left( \prod_{j=1}^{n-1} F_{k_j a_j, c_{j-1}}^{b_j , c_j}(t - x_j ) \right) \left( -i I_{d_n d_0}^{b_n b_0} + \frac{1}{4} J \bm{\sigma}_{b_n d_n}\cdot \bm{\sigma}_{b_0 d_0}   \right)\\
        \times F_{k_n a_n,c_{n-1}}^{d_n , d_0}(t)
        \Theta(0< x_{n-1}<\dots < x_1 < t) \psi_{e b_n}^\dagger(0) \psi_{e b_{\mathbf{n}/n} }^\dagger(x_{\mathbf{n}/n }) e^{i b_0 B t}|b_0\rangle,
    \end{multline}
    and
    \begin{multline}
        A_{e k_n a_n}(t) |\chi_{e k_{\mathbf{n}/n} a_{\mathbf{n}/n} ,a_0 }(t)\rangle = L^{-n/2} \int_{-L/2}^{L/2} dx_{\mathbf{n}/n}\ \delta_{a_0}^{c_0} \left( \prod_{j=1}^{n-1} F_{k_j a_j, c_{j-1}}^{b_j , c_j}(t- x_j) \right) \frac{1}{2}J e^{- i k_{n+1} t} \bm{\sigma}_{b_n a_n}\cdot \bm{\sigma}_{b_0 c_{n-1}}  \\ 
        \times \Theta(0< x_{n-1} < \dots < x_1< t)\psi_{e b_n}^\dagger(0) \psi_{e b_{\mathbf{n}/n}}^\dagger(x_{\mathbf{n}/n} )e^{i b_0 B t}|b_0\rangle.
    \end{multline}
    Comparing, we see that the differential equation \eqref{eq: chi condition 1 for Kondo general n} holds due to the same condition \eqref{eq: F(t) condition} that $F$ was required to satisfy in order to solve the $n=1$ problem.  This confirms that Eq. \eqref{eq: Kondo crossing state} is the correct $n$-electron crossing state for the Kondo model.  The case of negative $t$ can be done similarly.  
    
    A more general form of the Kondo Hamiltonian can be solved by essentially the same calculations, with the only change being a modification of the $\mathcal{T}$ matrix.  In particular, we can allow anisotropy and potential scattering:
    \beq
        H = -i \int_{-L/2}^{L/2 } dx\ \sum_{\gamma=1,2} \psi_{\gamma a}^\dagger(x) \frac{d}{dx}\psi_{\gamma a}(x) 
        + \sum_{\gamma,\gamma' =1,2}\frac{1}{2}  \psi_{\gamma a}^\dagger(0)\left[ \sum_{j=1}^3 J_j \sigma_{a a'}^j S^j+ J' \delta_{a a'} \right]\psi_{\gamma' a'}(0)  - B S^z. \label{eq: H for two-lead Kondo generalized}
    \eeq
    Following the same steps, we find that the condition \eqref{eq: F(t) condition} that the function $F$ is required to satisfy (in the $J=0$ basis) generalizes to
    \begin{multline}
        \left[ -i I_{d_1 d_0}^{b_1 b_0} + \frac{1}{2} \left(\frac{1}{2} \sum_{j=1}^3 J_j \sigma_{b_1 d_1}^j \sigma_{b_0 d_0}^j +J'\delta_{b_1 a_1}\delta_{b_0 a_0} \right) \right] F_{k_1 a_1,a_0}^{d_1, d_0}(t)e^{i d_0 B t}=\\
        - e^{-i k_1 t}e^{i a_0 B t}\left[\frac{1}{2} \sum_{j=1}^3 J_j \sigma_{b_1 a_1}^j \sigma_{b_0 a_0}^j + J' \delta_{b_1 a_1}\delta_{b_0 a_0} \right].
    \end{multline}
    Only the spin part has changed (not the time-dependent part).  The same solution \eqref{eq: F} works with a more general $\mathcal{T}$ matrix that is found by matrix inversion.  Here, we present the solution in the partially anisotropic case, in which we fix $m=1,2,$ or $3$ and declare that the remaining two Kondo couplings are equal to $J_{\perp}$.  (We allow $m$ to be general so that the special direction may or may not coincide with $z$-axis, which is the direction of the $B$ field.)  The $\mathcal{T}$ matrix is given by  
    \begin{multline}
        \mathcal{T} = i \Biggr[ -2 I+  \frac{1}{1 +i\frac{1}{2} ( \frac{1}{2} J_m + J' )}P_+ \left( I+ \sigma^m \otimes \sigma^m \right) +  \frac{1}{1+i\frac{1}{2} (J_{\perp} -\frac{1}{2} J_m +  J' )} P_+ \left( I - \sigma^m \otimes \sigma^m \right) \\
        +  \frac{1}{1-i \frac{1}{2}(2J_{\perp} -\frac{1}{2} J_m - J' )}   P_-\left( I + \sigma^m \otimes \sigma^m \right) +  \frac{1}{1 -i \frac{1}{2}(  J_{\perp} + \frac{1}{2} J_m -J' )}  P_-\left( I - \sigma^m \otimes \sigma^m \right) \Biggr],\label{eq: general T-matrix J=0 basis}
    \end{multline}
    where $P_{\pm} = \frac{1}{2}\left( I \pm P\right)$.
    
    In the fully isotropic case ($J_x=J_y=J_z \equiv J)$ with potential scattering included, we obtain
    \beq
        \mathcal{T}= 2 i\left(-I +  \frac{1}{1 +i\frac{1}{2} ( \frac{1}{2} J + J' )}P_+  +\frac{1}{1-i\frac{1}{2} (\frac{3}{2}J - J' )}   P_- \right), 
    \eeq
    which provides another check; a short calculation confirms that the corresponding bare $\mathcal{S}$ matrix $\mathcal{S}= I - i\mathcal{T}$ agrees exactly with that found in the Bethe ansatz solution of the one-lead model (see \cite{Andrei_lecture_notes}, for example, bearing in mind that the conventions are related by $J = 2 J_{\text{Bethe ansatz}}$).
    
    We can also solve the quench problem for the Hamiltonian \eqref{eq: H for two-lead Kondo generalized} in the $|J|=\infty$ basis.
    
    \section{Evaluation of bilinears}\label{sec: Evaluation of bilinears}
    We derive Eq. \eqref{eq: bilinear main text}, the formula for the expectation value of $\psi_{oa}^\dagger(x) \psi_{ea}(x)$.  For most of the proof, it is convenient to work in a more general setting; hence, we consider the expectation value of the product $\mathcal{O}_1^\dagger \mathcal{O}_2$ of two fermionic operators, and return to the notation of $c_\alpha^\dagger$ operators and impurity states  $|\beta\rangle$ (see Sec. \ref{sec: Time evolution: General formalism}).  The time-dependent operators $c_\alpha^\dagger(t)$ behave the same as $c_\alpha^\dagger$ operators under normal ordering and satisfy the same anticommutation relations ($\{ c_{\alpha'}(t), c_\alpha^\dagger(t) \} = \{ c_{\alpha'}, c_\alpha^\dagger \} = \delta_{\alpha \alpha'}$).

    We begin by proving a useful relation for rearranging sums:
    \begin{multline}
        \sum_{n,n' = 0}^N \sum_{\mathbf{m} \in \mathcal{I}_n(\mathbf{N}) } \left( \sgnleft\mathbf{m} \right) \sum_{\mathbf{m}' \in \mathcal{I}_{n'}(\mathbf{N}) } \left( \sgnleft\mathbf{m}' \right) \sum_{p=0}^{\textup{min} \{N-n, N-n'   \}   } \sum_{\bm{\ell} \in \mathcal{I}_p (\mathbf{N}/\mathbf{m} ) }
        \left( \sgnright\bm{\ell} \right) \sum_{\bm{\ell}' \in \mathcal{I}_p (\mathbf{N}/\mathbf{m}' ) } \left( \sgnright\bm{\ell}' \right) X\left(\mathbf{m}, \mathbf{m}', \bm{\ell},\bm{\ell}'\right) =\\
        \sum_{p=0}^N \sum_{\bm{\ell},\bm{\ell}' \in \mathcal{I}_p(\mathbf{N} ) } \left( \sgnleft\bm{\ell}\right)\left( \sgnleft\bm{\ell}'\right) \sum_{n,n'=0}^N 
        \sum_{\mathbf{m} \in \mathcal{I}_n( \bm{\ell}) } \left( \sgnleft\mathbf{m} \right)\sum_{\mathbf{m}' \in \mathcal{I}_{n'}(\bm{\ell}') } \left( \sgnleft\mathbf{m}' \right) X\left(\mathbf{m}, \mathbf{m}', \mathbf{N}/\bm{\ell},\mathbf{N}/ \bm{\ell}' \right), \label{eq: useful relation}
    \end{multline}
    where $X$ is any function.  \emph{Proof.} On the left-hand side, do the $p$ sum before the $n$,$n'$ sums and the $\bm{\ell}$, $\bm{\ell}'$ sums before the $\mathbf{m}$,$\mathbf{m}$' sums.  This yields
    \beq
        \sum_{p=0}^N  \sum_{\bm{\ell},\bm{\ell}' \in \mathcal{I}_p(\mathbf{N}) }  \left( \sgnright\bm{\ell} \right)  \left( \sgnright\bm{\ell}' \right) \sum_{n,n'=0}^{N-p} \sum_{\mathbf{m} \in \mathcal{I}_n(\mathbf{N}/\bm{\ell} )} \left( \sgnleft\mathbf{m} \right) \sum_{\mathbf{m}' \in \mathcal{I}_n(\mathbf{N}/\bm{\ell}' ) } \left( \sgnleft\mathbf{m}' \right) 
        X(\mathbf{m}, \mathbf{m}', \bm{\ell},\bm{\ell}').
    \eeq
    Then we need only relabel $p\to N-p$, $\bm{\ell}\to \mathbf{N} /\bm{\ell}$, and $\bm{\ell}'\to \mathbf{N} /\bm{\ell}'$, noting that this changes each $\sgnright$ to $\sgnleft$.
    
    The next preparatory step is to show that the normal-ordered overlap of states evolving from any initial quantum numbers is zero (except for the trivial case of time-evolving impurity states with no creation operators):
    \beq
        :\langle \Psi_{\alpha_{\mathbf{m}}' ,\beta'}(t) | \Psi_{\alpha_\mathbf{m} ,\beta}(t)\rangle : \ \ =
        \begin{cases}
            \delta_{\beta \beta'} & \mathbf{m} \text{ is the empty list,}\\
            0 & \text{otherwise.}
        \end{cases}\label{eq: NO overlap is zero}
    \eeq
    We can show this by direct calculation in the Kondo model, but the following proof is simpler and more general.  We use Wick's theorem:
    \begin{multline}
        c_{\alpha_{\mathbf{m}'}'}(t)  c_{\alpha_\mathbf{m}}^\dagger(t) = \sum_{p=0}^{\text{min} \{ |\mathbf{m}|,|\mathbf{m}'| \}  }\sum_{\bm{\ell} \in \mathcal{I}_p(\mathbf{m}) }\left( \sgnright\bm{\ell} \right) \sum_{\bm{\ell}' \in \mathcal{I}_p(\mathbf{m}') } \left( \sgnright\bm{\ell}'  \right)\\
        \times \sum_{\sigma \in \text{Sym}(p) } \left(\sgn \sigma\right) \left( \prod_{j=1}^p \{ c_{\alpha_{\ell'(\sigma_j)}'}(t), c_{\alpha_{\ell(j)}}^\dagger(t)  \}  \right) 
        :  c_{\alpha_{\mathbf{m}' / \bm{\ell}'} '}(t) c_{\alpha_{\mathbf{m} / \bm{\ell} }}^\dagger(t)  :, \label{eq: Wick's theorem no insertion}
    \end{multline}
    and the relation \eqref{eq: useful relation} to obtain the following expression for the overlap of two states as a sum of normal-ordered overlaps:
    \begin{multline}
        \langle \Psi_{\alpha_{\mathbf{N}}' ,\beta' } (t)  | \Psi_{\alpha_{\mathbf{N}},\beta}(t)\rangle = \sum_{n=0}^N \sum_{\mathbf{m},\mathbf{m}' \in \mathcal{I}_n(\mathbf{N} ) } \left( \sgnleft\mathbf{m} \right) \left( \sgnleft\mathbf{m}' \right) \sum_{\sigma\in \textup{Sym}(N-n)} \left(\sgn \sigma  \right)\\
        \times \left( \prod_{j=1}^{N-n }\{ c_{\alpha_{(\mathbf{N} / \mathbf{m}' )(\sigma(j))   } '} (t), c_{\alpha_{(\mathbf{N} / \mathbf{m})(j) } }^\dagger(t) \} \right)\  :  \langle \Psi_{\alpha_{\mathbf{m}'}',\beta'}(t)| \Psi_{\alpha_{\mathbf{m}} ,\beta} (t)\rangle :,\label{eq: overlap as sum of NO overlaps}
    \end{multline}
    where the $n=0$ term on the right-hand side is $\left( \prod_{j=1}^N \{ c_{\alpha_{\sigma(j)}'}(t), c_{ \alpha_j }^\dagger(t) \} \right)  \langle \beta' (t)| \beta (t) \rangle$.  The left-hand side is exactly equal to this $n=0$ term; to see this, consider the left-hand side at $t=0$ (it is independent of time) and recall that the $c_\alpha^\dagger(t)$  operators have the same anticommutation relations as the $c_\alpha^\dagger$ operators.  Thus, the sum from $n=1$ to $N$ on the right-hand side yields zero.  Taking $N=1$, we obtain
    \beq
        0 =\ : \langle \Psi_{\alpha_{N(1)}' , \beta'}(t) | \Psi_{\alpha_{N(1)} ,\beta}(t) \rangle :, 
    \eeq
    which is the first non-trivial case of the identity \eqref{eq: NO overlap is zero}.  Since the $\alpha$,$\alpha'$ labels are arbitrary, we see that the $n=1$ contribution on the right-hand side of Eq. \eqref{eq: overlap as sum of NO overlaps} vanishes for any $N$.  Taking $N=2$ yields
    \beq
        0 =\ : \langle \Psi_{\alpha_{N(1) }' \alpha_{N(2)}' , \beta'}(t) | \Psi_{\alpha_{N(1)} \alpha_{N(2)} ,\beta}(t) \rangle :,
    \eeq
    and so on up to arbitrary $N\ge1$ by induction.  This completes the proof of Eq. \eqref{eq: NO overlap is zero}.
    
    We can now consider the bilinear $\mathcal{O}_1^\dagger \mathcal{O}_2$.  Wick's theorem with the bilinear states
    \begin{multline}
        c_{\alpha_{\mathbf{m}'}' }(t) \mathcal{O}_1^\dagger \mathcal{O}_2  c_{\alpha_{\mathbf{m}}}^\dagger(t)  = \sum_{p=0}^{\text{min} \{ |\mathbf{m} |,|\mathbf{m}'| \}  }\sum_{\bm{\ell} \in \mathcal{I}_p(\mathbf{m}) }\left( \sgnright\bm{\ell} \right) \sum_{\bm{\ell}' \in \mathcal{I}_p(\mathbf{m}') } \left( \sgnright\bm{\ell}'  \right)\\
        \times \sum_{\sigma \in \text{Sym}(p) } \left(\sgn \sigma\right) \left( \prod_{j=1}^p \{ c_{\alpha_{\ell'(\sigma_j)}'}(t), c_{\alpha_{\ell(j)}}^\dagger(t) \}    \right) \Biggr[ :  c_{\alpha_{\mathbf{m}' / \bm{\ell}'}'}(t)  \mathcal{O}_1^\dagger \mathcal{O}_2  c_{\alpha_{\mathbf{m} / \bm{\ell}}}^\dagger(t)  :\\
        + \sum_{\mathbf{s} \in \mathcal{I}_1(\mathbf{m} / \bm{\ell} )} \left( \sgnright\mathbf{s} \right)  \{ \mathcal{O}_2, c_{\alpha_{s(1)}}^\dagger(t) \} \ :  c_{\alpha_{\mathbf{m}' / \bm{\ell}'}'}(t)  \mathcal{O}_1^\dagger  c_{\alpha_{\mathbf{m} / \bm{\ell} / \mathbf{l}}}^\dagger(t)  : \\
        + \sum_{\mathbf{s}' \in \mathcal{I}_1( \mathbf{m}'/ \bm{\ell}' )} \left( \sgnright\mathbf{s}' \right)  \{ c_{\alpha_{s'(1)}'}(t), \mathcal{O}_1^\dagger \} \ : c_{\alpha_{\mathbf{m}' / \bm{\ell}' / \mathbf{s}'}'}(t)  \mathcal{O}_2 c_{\alpha_{\mathbf{m} / \bm{\ell}}}^\dagger(t)  : \\
        + \sum_{\mathbf{s} \in \mathcal{I}_1( \mathbf{m} / \bm{\ell} )} \left( \sgnright\mathbf{s} \right)  \{ \mathcal{O}_2 , c_{\alpha_{s(1)}}^\dagger(t) \}  \sum_{\mathbf{s}' \in \mathcal{I}_1(\mathbf{m}'/ \bm{\ell}' )} \left( \sgnright\mathbf{s}' \right)  \{ c_{\alpha_{s'(1)}'}(t), \mathcal{O}_1^\dagger \} 
        : c_{\alpha_{\mathbf{m}' / \bm{\ell}' / \mathbf{s}'}'}(t) c_{\alpha_{\mathbf{m} / \bm{\ell} / \mathbf{s}}}^\dagger(t) :  \Biggr].\label{Wick's theorem bilinear insertion}
    \end{multline}
    Using this and the relation \eqref{eq: useful relation}, we obtain
    \begin{multline}
        \langle \Psi_{\alpha_{\mathbf{N}}' ,\beta' } (t)|\ \mathcal{O}_1^\dagger \mathcal{O}_2 |\Psi_{\alpha_{\mathbf{N}} ,\beta}(t) \rangle =  \sum_{n=1}^N \sum_{\mathbf{m},\mathbf{m}' \in \mathcal{I}_n(\mathbf{N}) } \left( \sgnleft\mathbf{m} \right) \left( \sgnleft\mathbf{m}' \right) \sum_{\sigma\in \textup{Sym}(N-n) } \left( \sgn \sigma \right)\\
        \times \left( \prod_{j=1}^{N-n }\{ c_{\alpha_{(\mathbf{N} / \mathbf{m}' )(\sigma_j)   } '} (t), c_{\alpha_{(\mathbf{N} / \mathbf{m})(j) } }^\dagger(t) \} \right) 
        \Biggr[ : \langle \Psi_{\alpha_{\mathbf{m}' }' ,\beta'} (t)|  \mathcal{O}_1^\dagger \mathcal{O}_2 | \Psi_{\alpha_{\mathbf{m}},\beta } (t) \rangle :\ \\
        + \sum_{\bm{\ell} \in \mathcal{I}_1(\mathbf{m}) } \left( \sgnright\bm{\ell} \right) \{ \mathcal{O}_2, c_{\alpha_{\ell(1)}}^\dagger(t) \}\   : \langle \Psi_{\alpha_{\mathbf{m}'}',\beta' }(t) | \mathcal{O}_1^\dagger |\Psi_{\alpha_{\mathbf{m}/ \bm{\ell}} ,\beta} (t)\rangle : \\
        + \sum_{\bm{\ell}' \in\mathcal{I}_1( \mathbf{m}') }\left( \sgnright\bm{\ell}' \right) \{ c_{\alpha_{\ell'(1)}'}(t), \mathcal{O}_1^\dagger \}\ : \langle \Psi_{\alpha_{\mathbf{m}' / \bm{\ell} '} ,\beta'}(t) | \mathcal{O}_2  | \Psi_{\alpha_{\mathbf{m}},\beta} (t)\rangle :\\
        + \sum_{\bm{\ell} \in\mathcal{I}_1( \mathbf{m}) }\left( \sgnright\bm{\ell} \right)\{ \mathcal{O}_2, c_{\alpha_{\ell(1)}}^\dagger(t) \}\sum_{\bm{\ell}' \in\mathcal{I}_1( \mathbf{m}') }\left( \sgnright\bm{\ell}' \right) \{ c_{\alpha_{\ell'(1)}'}(t), \mathcal{O}_1^\dagger \} 
        : \langle \Psi_{\alpha_{\mathbf{m}' / \bm{\ell} '}' ,\beta'}(t)  | \Psi_{\alpha_{\mathbf{m} / \bm{\ell}},\beta} (t)\rangle : \Biggr] .
    \end{multline}
    Due to the identity \eqref{eq: NO overlap is zero}, the last term in the brackets is zero unless $n=1$.  A further simplification occurs when we set $\alpha_{\mathbf{N}}= \alpha_{\mathbf{N}}'$ and $\beta=\beta'$: the product of anticommutators is then equal to unity if $\mathbf{m}'=\mathbf{m}$ and $\sigma$ is the identity permutation, and zero otherwise.  We also take advantage of the fact that the fermionic antisymmetry of the bra and ket vectors under exchange of quantum numbers remains valid in a normal-ordered inner product (even with $\mathcal{O}_1^\dagger$ and/or $\mathcal{O}_2$ inserted); this allows us to replace the sums over increasing lists of indices by unrestricted sums, at the cost of combinatorial factors.  After some relabelings of indices, we obtain
    \begin{multline}
        \langle \Psi_{\alpha_{\mathbf{N}} ,\beta } (t)|\ \mathcal{O}_1^\dagger \mathcal{O}_2 |\Psi_{\alpha_{\mathbf{N}} ,\beta}(t) \rangle = \sum_{n=1}^N  \sum_{m_1,\dots, m_n=1}^N  
        \Biggr[\frac{1}{n!}  : \langle \Psi_{\alpha_{\mathbf{m} } ,\beta} (t)|  \mathcal{O}_1^\dagger \mathcal{O}_2 | \Psi_{\alpha_{\mathbf{m}},\beta } (t) \rangle : \\
        + \frac{1}{(n-1)!}\{ \mathcal{O}_2, c_{\alpha_{m(n)}}^\dagger(t) \}\   : \langle \Psi_{\alpha_{\mathbf{m}},\beta }(t) | \mathcal{O}_1^\dagger  |\Psi_{\alpha_{\mathbf{m}/ m(n)} ,\beta} (t)\rangle : 
        + \frac{1}{(n-1)!} \{ c_{\alpha_{m(n)}}(t), \mathcal{O}_1^\dagger \}\ : \langle \Psi_{\alpha_{\mathbf{m} / m(n)} ,\beta}(t) | \mathcal{O}_2  | \Psi_{\alpha_{\mathbf{m}},\beta} (t)\rangle :\Biggr]\\
        + \sum_{j=1}^N \{ c_{\alpha_j}(t), \mathcal{O}_1^\dagger \}\{ \mathcal{O}_2, c_{\alpha_j}^\dagger(t) \}.
    \end{multline}
    Let us specialize to the two-lead Kondo model and take the inserted operators to be $\mathcal{O}_1^\dagger = \psi_{o a}^\dagger(x)$, $\mathcal{O}_2= \psi_{e a}(x)$.  Then, since the crossing states are built from even operators only, the $\psi_{o a }^\dagger(x)$ operator must be in an anticommutator (since otherwise the normal ordering symbol makes it annihilate a crossing state); this eliminates two terms. Writing the electron quantum numbers as $\alpha \equiv \gamma k a$, we obtain
    \begin{multline}
        \langle \Psi_{\alpha_{\mathbf{N}} ,a_0 } (t)|\ \psi_{oa}^\dagger(x) \psi_{ea}(x) |\Psi_{\alpha_{\mathbf{N}} ,a_0}(t) \rangle = \sum_{n=1}^N  \sum_{m_1,\dots,m_n=1}^N  
        \frac{1}{(n-1)!} \{ c_{\alpha_{m(n)}}(t), \psi_{oa}^\dagger(x) \}\\
        \times : \langle \Psi_{\alpha_{\mathbf{m} / m(n)} ,\beta}(t) | \psi_{ea}(x) | \Psi_{\alpha_{\mathbf{m}},\beta} (t)\rangle : 
        + \sum_{j=1}^N \{ c_{\alpha_j}(t), \psi_{oa}^\dagger(x) \}\{ \psi_{ea}(x), c_{\alpha_j}^\dagger(t) \}.
    \end{multline}
    This is Eq. \eqref{eq: bilinear main text} in the main text,  once the compact notation is written out in full.
    
    \section{Evaluation of the normal-ordered overlap}\label{sec: Evaluation of the normal-ordered overlap}
    We derive the result \eqref{eq: Omega in terms of Omega off-diag} for the normal-ordered overlap in the even sector that appears in the calculation of the electric current.  We need the following identity for rearranging the types of sums that arise in normal-ordered overlaps:
    \begin{multline}
        \sum_{\mathbf{m} \in \mathcal{I}_j(\mathbf{n} )} \left( \sgnright\mathbf{m} \right) \sum_{\sigma\in \text{Sym}(j)}\left( \sgn \sigma\right)\int_0^t dx_\mathbf{m}\ X_{k_{\mathbf{m}\circ\sigma}a_{\mathbf{m}\circ\sigma}}^{b_\mathbf{m}}\left(t, x_\mathbf{m} \right)\Theta( x_{m_j} < \dots < x_{m_1} ) \psi_{ b_\mathbf{m}}^\dagger(x_\mathbf{m} )\\
        \times \sum_{w\in \text{Sym}(n-j)}\left( \sgn w\right) \int_0^t dx_{\mathbf{n} /\mathbf{m}}\ Y_{k_{(\mathbf{n} / \mathbf{m})\circ w}a_{(\mathbf{n}/ \mathbf{m} )\circ w}}^{b_{\mathbf{n} / \mathbf{m} }}\left(t, x_{\mathbf{n} / \mathbf{m}}  \right)\Theta(x_{(\mathbf{n}/\mathbf{m} )_{n-j}} < \dots < x_{(\mathbf{n} /\mathbf{m} )_1}  )\psi_{ b_{\mathbf{n}/\mathbf{m}} }^\dagger(x_{\mathbf{n}/\mathbf{m} })=\\
        \sum_{\sigma\in \text{Sym}(n)} \left( \sgn \sigma\right) \sum_{\mathbf{m} \in \mathcal{I}_j(\mathbf{n} )} \int_0^t dx_\mathbf{n}\ X_{k_{\sigma \circ \mathbf{m} } a_{\sigma \circ \mathbf{m} }}^{b_\mathbf{m} } (t,x_\mathbf{m} )Y_{k_{\sigma \circ (\mathbf{n} / \mathbf{m} )} a_{\sigma \circ (\mathbf{n} / \mathbf{m}) }}^{b_{\mathbf{n} / \mathbf{m}} } (t,x_{\mathbf{n} / \mathbf{m}} )\Theta(x_n<\dots < x_1) \psi_{b_\mathbf{n} }^\dagger(x_\mathbf{n} ),\label{eq: overlap resummation identity}
    \end{multline}
    where $1\le j \le n$, and $X$ and $Y$ are any functions.  To prove this identity, we note that the product of two Heavside functions can always be written as a sum over Heaviside functions, with the summation including all orderings consistent with the two original Heaviside functions.  For instance, $\Theta(x_1 < x_2)\Theta(x_3 < x_4) = \Theta(x_1 < x_2 < x_3 < x_4) + \Theta(x_3 < x_1 < x_4 < x_2 )+ $ (four more terms), that is, all the orderings of the four variables such that $x_1 < x_2$ and $x_3 < x_4$.  We assume that no two of the $x$ variables are ever equal (so that orderings are always unambiguous); this amounts to ignoring sets of measure zero, which make no difference as the $x$ variables are always integrated.  The generalization of this example is
    \beq
        \Theta(x_{m_j} < \dots < x_{m_1})\Theta( x_{(\mathbf{n}/\mathbf{m} )_{n-j}} < \dots < x_{(\mathbf{n} /\mathbf{m} )_1} ) = \sum_{\bm{\ell} \in \mathcal{I}_j(\mathbf{n})} \Theta( x_{\iota[\mathbf{m}, \bm{\ell}](n)} < \dots < x_{\iota[\mathbf{m}, \bm{\ell}](1)}),
    \eeq
    where the permutation $\iota[\mathbf{m},\bm{\ell}] \in \text{Sym}(n)$ is defined via
    \beq
        \iota[\mathbf{m},\bm{\ell}] \circ \overrightarrow{perm}[\mathbf{m}]= \overrightarrow{perm}[\bm{\ell}]. \label{eq: iota}
    \eeq
    The meaning of this permutation becomes more clear if we note that $\iota[\mathbf{m}, \bm{\ell}] \circ \bm{\ell} = \mathbf{m}$ and $\iota[\mathbf{m}, \bm{\ell}] \circ (\mathbf{n} / \bm{\ell}) = \mathbf{n} / \mathbf{m}$; in other words, $\iota[\mathbf{m},\bm{\ell}]$ puts $\mathbf{m}$ at spots $\bm{\ell}$ and leaves $\mathbf{n}/\mathbf{m}$ in the original order.  Making the change of variables $x_p \to x_{\iota[\mathbf{m},\bm{\ell}]^{-1}(p)}$ and $b_p \to b_{\iota[\mathbf{m},\bm{\ell}]^{-1}(p)}$,  we find that the left-hand side of Eq. \eqref{eq: overlap resummation identity} is equal to
    \begin{multline}
        \sum_{\bm{\ell} ,\mathbf{m} \in \mathcal{I}_j(\mathbf{n} )} \left( \sgnright\mathbf{m} \right) \sum_{\sigma\in \text{Sym}(j),w\in \text{Sym}(n-j) }\left( \sgn \sigma\right) \left( \sgn w\right) \int_0^t dx_\mathbf{n}\ X_{k_{\bm{\ell} \circ \sigma} a_{\bm{\ell} \circ \sigma} }^{b_{\bm{\ell}}} (t, x_{\bm{\ell}} ) \\
        \times Y_{k_{(\mathbf{n} /\bm{\ell}) \circ w} a_{(\mathbf{n} /\bm{\ell}) \circ w} }^{b_{\mathbf{n} / \bm{\ell} }} (t, x_{\mathbf{n} / \bm{\ell}} )
        \Theta(x_n<\dots<x_1)\psi_{b_{\bm{\ell}} }^\dagger(x_{\bm{\ell}} ) \psi_{b_{\mathbf{n} /\bm{\ell} }}^\dagger(x_{\mathbf{n} / \bm{\ell} } ).
    \end{multline}
    We rearrange the creation operators---$\psi_{b_{\bm{\ell}} }^\dagger(x_{\bm{\ell}} )\psi_{b_{\mathbf{n} /\bm{\ell} }}^\dagger(x_{\mathbf{n} / \bm{\ell}} ) = \left( \sgnright\bm{\ell} \right) \psi_{b_\mathbf{n}}^\dagger(x_\mathbf{n})$---and note that $\left( \sgnright\mathbf{m} \right)\left( \sgnright\bm{\ell} \right) = \sgn \iota[\mathbf{m},\bm{\ell}]$.  To complete the proof, we relabel several of the summations as a single sum over permutations $\sigma'$:
    \beq
        \sum_{\mathbf{m} \in \mathcal{I}_j(\mathbf{n} )} \sum_{\sigma\in \text{Sym}(j),w\in \text{Sym}(n-j) }\left(\sgn \iota[\mathbf{m},\bm{\ell} ] \right) \left( \sgn \sigma\right) \left( \sgn w\right) \longleftrightarrow \sum_{\sigma' \in \text{Sym}(n)} \left( \sgn \sigma'\right),
    \eeq
    where the permutation $\sigma' \in \text{Sym}(n)$ is defined via $\sigma' \circ \bm{\ell} = \mathbf{m} \circ \sigma$ and $\sigma' \circ (\mathbf{n} / \bm{\ell}) = (\mathbf{n}/ \mathbf{m} )\circ w$.  The right-hand side of Eq. \eqref{eq: overlap resummation identity} is then obtained once we relabel $\sigma'$ as $\sigma$ and $\bm{\ell}$ as $\mathbf{m}$.
    
    Our task is to evaluate the normal-ordered inner product of
    \beq
        | \Psi_{ek_{\mathbf{n}}a_{\mathbf{n} },a_0}(t) \rangle = \sum_{j=0}^n \sum_{\mathbf{m} \in \mathcal{I}_j (\mathbf{n} )} \left( \sgnright\mathbf{m} \right)
        c_{e k_{\mathbf{m} }a_{\mathbf{m} }}^\dagger(t) \sum_{\sigma\in \text{Sym}(\ell)} \left(\sgn \sigma\right) |\chi_{e k_{(\mathbf{n} / \mathbf{m}) \circ \sigma} a_{(\mathbf{n} / \mathbf{m}) \circ \sigma},a_0}(t)\rangle 
    \eeq
    and
    \beq
        \langle \Psi_{k_{\mathbf{n}/n}'a_{\mathbf{n}/n }',a_0'}(t) | c_{e k_n' a_n'}(t) = \sum_{j'=1}^n \sum_{\substack{ \mathbf{m}' \in \mathcal{I}_{j'} (\mathbf{n} ) \\ n \in \mathbf{m}'} }\left( \sgnright\mathbf{m}' \right)
        \sum_{\sigma\in \text{Sym}(j')} \left(\sgn \sigma\right) \langle \chi_{e k_{(\mathbf{n} / \mathbf{m}') \circ \sigma}' a_{(\mathbf{n} / \mathbf{m}') \circ \sigma}',a_0'}(t) | c_{e k_{\mathbf{m}' }'a_{\mathbf{m}' }'}(t). 
    \eeq
    Note that we have changed the labeling (via $\mathbf{m}\to \mathbf{n}/\mathbf{m}$, $\mathbf{m}'\to \mathbf{n}/\mathbf{m}'$) so that we are summing over which subsets of the original quantum numbers are put into momentum operators (rather than into crossing states).  The key point is that normal ordering forces each $c^\dagger(t)$ to contract with a $\psi(x)$ operator inside a $\langle \chi(t)|$ state, and each $c(t)$ operator to contract with a $\psi^\dagger$ operator inside a $|\chi\rangle$ state; $c^\dagger(t)$ and $c(t)$ operators never contract with each other.  We can therefore drop the part of $c^\dagger(t)$ that is outside the forward light cone (in position space).  Our strategy is to bring each half of the inner product to a more suitable form using the identity \eqref{eq: overlap resummation identity}, then impose normal ordering on the overlap by requiring that the $c^\dagger(t)$ and $c(t)$ operators do not contract.
    
    Performing some relabelings of indices and using the identity \eqref{eq: overlap resummation identity}, we obtain
    \begin{multline}
        | \Psi_{ek_{\mathbf{n}}a_{\mathbf{n} },a_0}(t) \rangle = L^{-n/2} \sum_{j=0}^n \sum_{\mathbf{m} \in \mathcal{I}_j(\mathbf{n} ) } \left(\sgnright\mathbf{m} \right)\int_0^t dx_\mathbf{n} \sum_{\sigma\in \text{Sym}(j) } (\sgn \sigma)  e^{-i k_{\mathbf{m}\circ w} (t- x_\mathbf{m} ) } I_{a_{\mathbf{m}\circ \sigma } }^{b_\mathbf{m}} 
        \Theta(x_{m(j)} < \dots < x_{m(1)})\\
        \times \psi_{e b_\mathbf{m} }^\dagger(x_\mathbf{m}) \sum_{\sigma \in \text{Sym}(n-j) }(\sgn \sigma)\  e^{-i k_{(\mathbf{n} / \mathbf{m})\circ \sigma} (t- x_{\mathbf{n} / \mathbf{m}} ) } \mathcal{M}_{a_{(\mathbf{n} / \mathbf{m})\circ \sigma } ,a_0}^{b_{\mathbf{n} / \mathbf{m}} ,b_0}\Theta(x_{(\mathbf{n}/\mathbf{m})(n-j)} < \dots < x_{(\mathbf{n} / \mathbf{m})(1)})
        \psi_{b_{\mathbf{n} / \mathbf{m}} }^\dagger(x_{\mathbf{n} /\mathbf{m} })|b_0\rangle + \dots\\
        = L^{-n/2} \sum_{\sigma \in \text{Sym}(n)} (\sgn \sigma) \sum_{j=0}^n \sum_{\mathbf{m} \in \mathcal{I}_j (\mathbf{n} ) } \int_0^t dx_\mathbf{n}\ e^{-i k_{\sigma\circ \mathbf{n} } (t-x_\mathbf{n})} I_{ a_{\sigma \circ \mathbf{m} } }^{b_\mathbf{m} } \mathcal{M}_{a_{\sigma \circ (\mathbf{n} / \mathbf{m} )} ,a_0}^{b_{\mathbf{n} / \mathbf{m} },b_0} \Theta(x_n< \dots < x_1) \psi_{b_\mathbf{n}}^\dagger(x_\mathbf{n} ) |b_0\rangle + \dots,\label{eq: right side of overlap}
    \end{multline}
    where $\mathbf{m}$ are the indices that were assigned to $c^\dagger(t)$ operators (which have been truncated to include only the part that survives inside a normal-ordered product), and where we have used the notation
    \beq
        \mathcal{M}_{a_\mathbf{n} ,a_0}^{b_\mathbf{n}, b_0} = \delta_{a_0}^{c_0}\delta_{c_n}^{b_0}\prod_{j=1}^n \left(-i \mathcal{T} \right)_{a_j c_{j-1}}^{b_j c_j}.
    \eeq
    A similar calculation for the other half of the inner product [requiring a slight generalization of the identity \eqref{eq: overlap resummation identity} to accommodate the condition $n\in \mathbf{m}'$] yields
    \begin{multline}
        c_{e k_n' a_n'}^\dagger(t) | \Psi_{ek_{\mathbf{n}/n}'a_{\mathbf{n}/n }',a_0'}(t) \rangle = L^{-n/2} \sum_{\sigma' \in \text{Sym}(n)} (\sgn \sigma') \sum_{j'=1}^n \sum_{\substack{\mathbf{m}' \in \mathcal{I}_{j'}(\mathbf{n} )  \\ n \in \sigma' \circ\mathbf{m}'}}  \int_0^t dx_{\mathbf{n}}\ e^{-i k_{\sigma'\circ \mathbf{n}' } (t-x_\mathbf{n} )} I_{ a_{\sigma' \circ \mathbf{m}' }' }^{b_{\mathbf{m}' }}\\
        \times \mathcal{M}_{a_{\sigma' \circ (\mathbf{n} / \mathbf{m}' )}' ,a_0'}^{b_{\mathbf{n} / \mathbf{m}' },b_0}
        \Theta(x_n< \dots < x_1) \psi_{b_\mathbf{n}}^\dagger(x_\mathbf{n} ) |b_0\rangle + \dots,\label{eq: left side of overlap}
    \end{multline}
    where $\mathbf{m}'$ are the indices assigned to $c^\dagger(t)$ operators.  The overlap of \eqref{eq: right side of overlap} and \eqref{eq: left side of overlap} can then be put into normal order by requiring that the lists $\mathbf{m}$ and $\mathbf{m}'$ have no entries in common.  The Heaviside functions force the $\psi$ and $\psi^\dagger$ operators to contract in the simplest way, and so we obtain
    \begin{multline}
        :\langle \Psi_{k_{\mathbf{n}/n}'a_{\mathbf{n}/n }',a_0'}(t) | c_{e k_n' a_n'}(t) | \Psi_{ek_{\mathbf{n}}a_{\mathbf{n} },a_0}(t) \rangle:\  = L^{-n} \sum_{\sigma,\sigma'\in \text{Sym}(n)}(\sgn \sigma) (\sgn \sigma')\sum_{j = 0 }^n \sum_{j' = 1 }^n \sum_{\substack{\mathbf{m} \in \mathcal{I}_j(\mathbf{n}), \mathbf{m}' \in \mathcal{I}_{j'}(\mathbf{n}) \\ |\mathbf{m} \cap \mathbf{m}' | =0 ,\ n \in \sigma' \circ \mathbf{m}'  }  } I_{ a_{ \mathbf{m}' } '}^{b_{\mathbf{m}' }} \\
        \times \mathcal{M}_{\ a_{ \mathbf{n} / \mathbf{m}' }' ,a_0'}^{*b_{\mathbf{n} / \mathbf{m}' } ,b_0} \mathcal{M}_{a_{ \mathbf{n} / \mathbf{m} } ,a_0}^{b_{\mathbf{n} / \mathbf{m} } ,b_0} I_{ a_{ \mathbf{m} } }^{b_\mathbf{m} } \int_0^t dx_\mathbf{n}\ e^{-i (k_{\sigma\circ \mathbf{n}} -k_{\sigma' \circ \mathbf{n} }')(t-x_\mathbf{n} )} \Theta(x_n<\dots<x_1).
    \end{multline}
    Using the unitarity of the bare $\mathcal{S}$ matrix ($\mathcal{S}_{\ c_1 c_0}^{* b_1 b_0} \mathcal{S}_{a_1 a_0}^{c_1 c_0} = I_{a_1 a_0}^{b_1 b_0})$, we further simplify this expression to
    \begin{multline}
        :\langle \Psi_{k_{\mathbf{n}/n}'a_{\mathbf{n}/n }',a_0'}(t) | c_{e k_n' a_n'}(t) | \Psi_{ek_{\mathbf{n}}a_{\mathbf{n} },a_0}(t) \rangle:\  = L^{-n} \sum_{\substack{\sigma,\sigma'\in \text{Sym}(n) \\ \sigma'(n)=n }}(\sgn \sigma) (\sgn \sigma') \Xi_{n-1}[a_{\sigma' \circ (\mathbf{n} /n) }'; a_{\sigma \circ(\mathbf{n}/n )} ]_{a_0' a_0}^{ b_0 c_{n-1} }\\
        \times \mathcal{M}_{a_{\sigma(n) } c_{n-1}}^{a_n b_0  } \int_0^t dx_\mathbf{n}\ e^{-i (k_{\sigma\circ \mathbf{n}} -k_{\sigma' \circ \mathbf{n} }')(t-x_\mathbf{n} )} \Theta(x_n<\dots<x_1).
    \end{multline}
    Equation \eqref{eq: Omega in terms of Omega off-diag} in the main text is then obtained by setting each $k_j'=k_j$ and $a_j'=a_j$, and writing out the indices.
    
    A very similar calculation confirms Eq. \eqref{eq: NO overlap is zero}, which was shown earlier by general arguments; one finds that the requirement $n\in \mathbf{m}'$ is absent, and that the inner product vanishes due to the unitarity of the bare $\mathcal{S}$ matrix.
    
    \section{Properties of spin sums}\label{sec: Properties of spin sums}
    In this appendix, we prove that for $n\ge 2$, the spin sum $W_n^{(\sigma)}(J)$ has at least $n+1$ powers of $\coeffP$ (which demonstrates that the current series in the main text can be read as a series in $J$ or in $1/J$).  We then prove the spin sum identity \eqref{eq: spin sum identity for convergence Kondo} from the main text, which confirms that all orders of either series ($J$ or $1/J$) converge in the long-time limit.
    
    From the definition \eqref{eq: Xi}, we have the following rule for generating $\Xi_{n+1}$ from $\Xi_n$:
\begin{multline}
    \Xi_{n+1}[a_\mathbf{n}', a_{n+1}'; a_\mathbf{n}, a_{n+1}]_{a_0' a_0}^{c' c} = - |\coeffP|^2 \Xi_n[ a_\mathbf{n}'; a_\mathbf{n}]_{a_0' a_0}^{c' c}\delta_{a_{n+1}}^{a_{n+1}'} \\
    + \coeffI \coeffP^* \left(\Xi_n[ a_\mathbf{n}'; a_\mathbf{n}]_{a_0' a_0}^{a_{n+1} c} \delta_{a_{n+1}'}^{c'} - \Xi_n[ a_\mathbf{n}'; a_\mathbf{n}]_{a_0' a_0}^{c' a_{n+1} } \delta_{a_{n+1}}^c \right).\label{eq: Xi update rule Appendix}
\end{multline}
The base case, $n=1$, can be found by a short calculation:
\beq
    \sum_{a_0} \Xi_1[a_1';a_1]_{a_0 a_0}^{c' c} = |\coeffP|^2 \left( 2 I_{a_1 c}^{a_1' c'} - I_{a_1 c}^{a_1' c'}. \right).\label{eq: Xi base case Appendix}
\eeq
Consider $n\ge2$.  From Eq. \eqref{eq: W}, $\mathcal{S} =  \coeffI I + \coeffP P$, and the fact that the tensor $\Xi_n$ vanishes when its upper two indices are contracted, we obtain
\beq
    W_n^{(\sigma)}(J) = -\frac{1}{2^{n+1}}(\sgn \sigma)\sum_{a_0,a_1,\dots,a_n}   \coeffP \Xi_n[ a_{\mathbf{n} /n} ; a_{(\mathbf{n} / n)\circ \sigma} ]_{a_0 a_0}^{a_{\sigma_n} a_n }.
\eeq
The base case and the update rule \eqref{eq: Xi update rule Appendix} then confirm that $W_n^{(\sigma)}(J)$ has at least $n+1$ powers of $\coeffP$.
    
    We proceed to prove the identity \eqref{eq: spin sum identity for convergence Kondo} from the main text, repeated here for reference:
\beq
    \Xi_n[a_\mathbf{n}; a_{\mathbf{n} \circ \sigma} ]_{a_0 a_0}^{c' c} = 0 \qquad [n\ge 1,\ \sigma \in \text{Sym}(n)],\label{eq: spin sum identity for convergence Appendix}
\eeq
with implied summation over \emph{any} repeated spin indices.  Rather than use the explicit forms of the coefficients $\coeffI$ and $\coeffP$, we only use the fact that they are constrained by the unitarity of the bare $\mathcal{S}$ matrix:
\bseq
    \begin{align}
        |\coeffI|^2 + |\coeffP|^2 &= 1,\\
        \coeffI \coeffP^* + \coeffI^* \coeffP &= 0.
    \end{align}
\eseq
The proof uses the update rule \eqref{eq: Xi update rule Appendix} and the base case \eqref{eq: Xi base case Appendix}.  To give a sense of the pattern for $\Xi_n$, we present the $n=2$ case, as well:
\beq
    \Xi_2[a_1',a_2';a_1 ,a_2]_{a_0 a_0}^{c' c} = |\coeffP|^2 \left[ |\coeffP|^2 \left( I_{a_1 a_2 c}^{a_1' a_2' c'} - 2 I_{a_2 a_1' a_1}^{a_2' c' c} \right)
    + 2 \coeffI \coeffP^* \left( I_{a_2 a_2' a_1}^{a_1' c' c} - I_{a_1 a_1' a_2}^{a_2' c' c} \right) \right].
\eeq
The pattern is the following: a sum of identity tensors multiplied by some function of $\coeffI$ and $\coeffP$.  In each identity tensor, we can either have (1) $c'$ contracts with $c$ and each $a_j'$ contracts with $a_j$, or (2) $c'$ contracts with some $a_{j'}'$, $c$ contracts with some $a_j$, and the remaining $a_m$ and $a_m'$ indices contract in some way (always pairing a primed with an unprimed index).  To be precise, we will show by induction the following general form:
\beq
    \Xi[a_\mathbf{n}'; a_\mathbf{n}]_{a_0 a_0}^{c' c} = X_n I_{a_\mathbf{n} c}^{a_\mathbf{n}' c'} + \sum_{\sigma' \in \text{Sym}(n-1)} \sum_{j,j'=1}^n Y_{n,j,j'}^{(\sigma')} I_{a_{(\mathbf{n} / j)\circ \sigma'}}^{a_{\mathbf{n} / j'}' }I_{a_{j'}' a_j}^{c' c},\label{eq: Xi general form appendix}
\eeq
where the coefficients $X_n$ and $Y_{n,j,j'}^{(\sigma')}$ depend on $\coeffI$ and $\coeffP$.  The base case is of this form, with $-X_1 = \frac{1}{2}Y_{1,1,1}^{(1)} = |\coeffP|^2$.  For the induction step, we assume this general form for some $n\ge 1$ and use the update rule to obtain
\begin{multline}
    \Xi_{n+1}[a_\mathbf{n}', a_{n+1}'; a_\mathbf{n}, a_{n+1}]_{a_0 a_0}^{c' c} = -|\coeffP|^2 X_n I_{a_\mathbf{n} a_{n+1} c}^{a_\mathbf{n}' a_{n+1}' c'} -  \sum_{\sigma' \in \text{Sym}(n-1)} \sum_{j,j'=1}^n Y_{n,j,j'}^{(\sigma')} I_{a_{(\mathbf{n} / j)\circ \sigma'}}^{a_{\mathbf{n} / j'} } \\
    \times\left[- |\coeffP|^2 \delta_{a_{n+1}}^{a_{n+1}'}\delta_{a_{j'}' }^{ c' } \delta_{ a_j}^c + \coeffI \coeffP^* \left( \delta_{a_{n+1}}^{a_{j'}'} \delta_{a_{n+1}' }^{c' }\delta_{a_j}^c -  \delta_{a_j}^{a_{n+1}'} \delta_{a_{j'}' }^{ c' }\delta_{ a_{n+1}}^c \right) \right]. 
\end{multline}
As claimed, this expression is of the general form \eqref{eq: Xi general form appendix}.  We can read off $X_{n+1} = -|\coeffP|^2 X_n$.  While extracting $Y_{n+1,j,j'}^{(\sigma')}$ would be messy, we can see that the remaining terms all include $I_{a_{j'}' a_j}^{c' c}$ with $j,j' \in \{1,\dots, n+1\}$, with the remaining $a_m$ indices contracted with the remaining $a_{m'}'$ indices in some order.  [Note that we have written some contractions as Kronecker deltas for typographical clarity.  Also, we can put the unprimed indices in the canonical order $a_{\mathbf{n} / j'}$ that appears in \eqref{eq: Xi general form appendix} simply by rearranging the corresponding unprimed indices below, which is just some choice of the permutation $\sigma' \in \text{Sym}(n)$.] 

We proceed to prove the main result by induction.  For the base case, we note that setting $c'=c$ yields zero in Eq. \eqref{eq: Xi base case Appendix}.  Next, we assume that \eqref{eq: spin sum identity for convergence Appendix} holds for some $n\ge1$ [and for any $\sigma \in \text{Sym}(n)$], and we let $w\in \text{Sym}(n+1)$.  Then, the update rule \eqref{eq: Xi update rule Appendix} yields
\begin{multline}
    \Xi_{n+1}[a_\mathbf{n}, a_{n+1}; a_{w\circ \mathbf{n}}, a_{w_{n+1}}]_{a_0 a_0}^{c' c} = - |\coeffP|^2 \Xi_n[a_\mathbf{n}; a_{w\circ \mathbf{n}}]_{a_0 a_0}^{c' c}\delta_{a_{w_{n+1}}}^{a_{n+1}} \\
    +\coeffI \coeffP^* \left(\Xi_n[a_\mathbf{n}; a_{w\circ \mathbf{n}}]_{a_0 a_0}^{a_{w_{n+1}} c}\delta_{a_{n+1}}^{c'} - \Xi_n[a_\mathbf{n}; a_{w\circ \mathbf{n}}]_{a_0 a_0}^{c' a_{n+1}}\delta_{a_{w_{n+1}}}^c  \right).
\end{multline}
The first term on the right-hand side vanishes due to the induction assumption.  This is particularly clear if $w_{n+1}=n$; but even if $w_{n+1}\le n$, we are free to relabel the summation indices to obtain the same form \eqref{eq: spin sum identity for convergence Appendix} that vanishes by assumption.  To deal with the second term on the right-hand side, we use the general form \eqref{eq: Xi general form appendix} to find
\begin{multline}
    \text{coeff. of }\coeffI \coeffP^* = X_n I_{a_{w\circ \mathbf{n}}}^{a_\mathbf{n}} \left(  \delta_{a_{n+1} }^{c' }\delta_{a_{w_{n+1}}}^c- \delta_{ a_{w_{n+1}} }^{c' }\delta_{  a_{n+1}}^c\right)\\
    + \sum_{\sigma' \in \text{Sym}(n-1)} \sum_{j,j'=1}^n Y_{n,j,j'}^{(\sigma')}  I_{a_{w\circ (\mathbf{n} / j)\circ \sigma'}}^{a_{\mathbf{n} / j'} } \left( \delta_{a_j' }^{a_{w_{n+1}}}\delta_{a_{n+1}}^{ c' }\delta_{ a_{w_j}}^{ c}- \delta_{a_{w_j} }^{a_{n+1} }\delta_{ a_{j'} }^{ c'}\delta_{ a_{w_{n+1}}}^c \right).
\end{multline}
In the $X_n$ term, we get zero immediately if $w_{n+1}=n+1$; if instead $w_{n+1} \le n$, then $I_{a_{w\circ\mathbf{n}}}^{a_\mathbf{n}} = (\text{const}) \delta_{a_{n+1}}^{a_{w_{n+1}}}$ (where the constant is some number obtained from summing all the other spin indices), yielding zero once we sum over $a_{n+1}$ and $a_{w_{n+1}}$.  Similarly, in each $Y_{n,j,j'}^{(\sigma')}$ term, we will have to contract either (1) $a_{w_j}$ with $a_{j'}$ and $a_{w_{n+1}}$ with $a_{n+1}$, or (2) $a_{w_j}$ with $a_{n+1}$ and $a_{w_{n+1}}$ with $a_{j'}$, and either way, the two terms in parentheses cancel once the spin indices are summed.  For instance,  if $w_j = j'$ and $w_{n+1} = n+1$, then we are in case (1) immediately; if instead $w_j = j'$ but $w_{n+1} \le n$, then the identity tensor in front yields $(\text{const})\delta_{a_{n+1}}^{a_{w_{n+1}}}$, and we are again in case (1); and so on \footnote{If $w_j \ne j'$ and $w_{n+1} \le n$, then the identity tensor in front becomes either $(\text{const}) \delta_{a_{j'}}^{a_{w_j}} \delta_{a_{n+1} }^{ a_{w_{n+1}} }$ [yielding case (1)] or $(\text{const})\delta_{a_{j'}}^{a_{w_{n+1}}} \delta_{a_{n+1} }^{ a_{w_j} }$ [yielding case (2)], depending on how the remaining spin indices are contracted.}. Thus, we have shown that Eq. \eqref{eq: spin sum identity for convergence Appendix} holds for $n+1$, completing the induction proof.
    
    \section{Asymptotic evaluation of integrals}\label{sec: Asymptotic evaluation of integrals}
    We study the asymptotic behavior as $\lambda\to\infty$ of the general form \eqref{eq: general form}, namely,
    \beq
        R^{(\sigma)}[\{ f,h \}, \lambda] \equiv \int_0^\infty du_1 \dots du_{n-1}\ \left[ \prod_{j=1}^{n-1} \frac{e^{i \lambda v_j^{(\sigma)} } -f\left( v_j^{(\sigma)} \right) }{ v_j^{(\sigma)} } \right] h\left( v_n^{(\sigma)} \right),\label{eq: R}
    \eeq
    where $\sigma \in \text{Sym}(n)$ and the $v_j^{(\sigma)}$ variables are the following linear combinations of the integration variables
    \beq
        v_j^{(\sigma)} = \sum_{m = j}^{n-1} u_m - \sum_{m = \sigma^{-1}(j) }^{n-1} u_m \qquad (1\le j \le n).
    \eeq
    These linear combinations are listed in Table \ref{tab: v_j variables} for all of the $11$ permutations $\sigma$ that we need in order to evaluate the current up to and including the $J^5$ or $1/J^5$ term.
    
    \begin{table}[htp]
        \caption{\label{tab: v_j variables}
        Linear combinations $v_1^{(\sigma)}, \dots, v_n^{(\sigma)}$.
        }
        \begin{ruledtabular}
        \begin{tabular}{ l c  c c c}
        $\sigma \equiv (\sigma_1,\dots,\sigma_n)$&
        $v_1^{(\sigma)}$ & $v_2^{(\sigma)}$ & $v_3^{(\sigma)}$ & $v_4^{(\sigma)}$\\
        \colrule
        $(1)$ & 0 &  &  & \\
        $(2,1)$ & $u_1$ & $-u_1$  &  & \\
        $(3,1,2)$ & $u_1$ & $u_2$ & $-u_1 - u_2$ & \\
        $(2,3,1)$ & $u_1 + u_2$ & $-u_1$ & $-u_2$ &  \\
        $(3,2,1)$ & $u_1 + u_2$ & $0$ & $-u_1 - u_2$ & \\
        $(2,3,4,1)$ & $u_1 +u_2 +u_3$ & $-u_1$ & $-u_2$ & $-u_3$\\
        $(2,4,1,3)$ & $u_1+u_2$ & $-u_1$ & $u_3$ & $-u_2 -u_3$\\
        $(3,1,4,2)$ & $u_1$ & $u_2+u_3$ & $-u_1 -u_2$ &  $-u_3$\\
        $(3,4,1,2)$ & $u_1 + u_2$ & $u_2 + u_3$ & $-u_1 - u_2$ & $-u_2 - u_3$\\
        $(4,1,2,3)$ & $u_1$ & $u_2$ & $u_3$ & $-u_1 - u_2 -u_3$\\
        $(4,3,2,1)$ & $u_1 + u_2 + u_3$ & $u_2$  & $-u_2$ & $-u_1 -u_2- u_3$
    \end{tabular}
    \end{ruledtabular}
    \end{table}
    
    We use brackets to indicate that $R^{(\sigma)}[\{ f,h\}, \lambda]$ is a \emph{functional} of $f$ and $h$ and a function of the real parameter $\lambda$.  As discussed in the main text, $\lambda$ is essentially the bandwidth divided by a dimensional scale, and the functions $f$ and $h$ take various forms depending on which case is being considered.
    
    We have found the asymptotic form as $\lambda\to\infty$ of $R^{(\sigma)}[\{ f,h \}, \lambda]$ for all $11$ of the necessary permutations.  By leaving $f$ and $h$ unspecified, we can cover all cases discussed in the main text at once. 
    
    We will not attempt to characterize exactly what properties of $f$ and $h$ are necessary for our calculations below to be valid.  At the very least, we assume that $f$ and $h$ are both analytic with poles only along the imaginary axis (but no pole at the origin), that $f(0)=1$ (otherwise $R^{(\sigma)}[\{ f,h \}, \lambda]$ would be ill-defined due to the denominators), and that $h(v)$ decays like $1/v$ or faster as $v\to\infty$; we also assume that $f'(0)=0$ and that $h(0)$ is real, although these conditions could easily be relaxed.  All of these assumptions hold for the particular $f$ and $h$ functions defined in the main text.
    
    Before presenting the full results, we show one more example.  We have already given the simplest non-trivial example in \eqref{eq: R21 asymptotic} in the main text, which is the asymptotic expansion of $R^{(2,1)}[\{ f, h\}, \lambda]$.  An example result from the next order ($n=3$) is
    \bseq
    \begin{align}
        R^{(2,3,1)}[\{ f, h \}, \lambda] \equiv &\int_0^\infty du_1 du_2\ \frac{e^{i\lambda(u_1 +u_2)} - f(u_1+u_2) }{u_1 +u_2} \frac{e^{-i \lambda u_1} - f(-u_1 )}{-u_1}h(-u_2)\\
        &\overset{\lambda\to\infty}{\longrightarrow} -\frac{1}{2}h(0) \ln^2 \lambda +\biggr[ -h(0)\left( \gamma + i\frac{\pi}{2} \right)
        + \int_0^\infty du\ \ln u \frac{d}{du} \left( f(u) h(-u) \right) \biggr] \ln \lambda - \left(\frac{7\pi^2}{24} + \frac{1}{2}\gamma^2 + i \frac{1}{2}\pi \gamma \right)h(0) \notag\\
        &+\left( \gamma + i \frac{\pi}{2}\right) \int_0^\infty du\ \ln u \frac{d}{du} \left[  f(u) h(-u) \right]+ \frac{1}{2}\int_0^\infty du\ \ln^2 u\frac{d}{du}\left[ f(u)h(-u) \right]\notag\\
        &-\int_0^\infty du_1 du_2\ \frac{1}{u_2} \ln \frac{u_1 +  u_2}{u_1} \frac{\partial}{\partial u_1} \left[ f(u_1 + u_2) f(-u_1) h(-u_2) \right],
    \end{align}
    \eseq
    where $\gamma$ is the Euler constant.  Notice that here and in the simpler example \eqref{eq: R21 asymptotic}, the asymptotic expansion consists of powers of $\ln \lambda$ with coefficients that are functionals of $f$ and $h$; higher powers of $\ln \lambda$ are multiplied by simpler functionals, and the highest power is $\ln^{n-1} \lambda$.

    We have shown analytically that for all of the $11$ necessary permutations, the asymptotic form of $R^{(\sigma)}[\{f,h\}, \lambda]$ is a sum of logarithmic terms (including a constant term, i.e., $\ln^0 \lambda$) and a linear term.  That is, we have shown
    \beq
        R^{(\sigma)}[\{ f,h \}, \lambda] \overset{\lambda\to \infty}{\longrightarrow} z_\text{linear}^{(\sigma)}[\{f,h \}]  \lambda+ \sum_{j=0}^{n-1} z_j^{(\sigma)}[\{f,h \}] \ln^j \lambda, \label{eq: asymptotic form of R}
    \eeq
    where $z_\text{linear}^{(\sigma)}[\{f,h \}]$ and $z_j^{(\sigma)}[\{f,h \}]$ are complex numbers (functionals of $f$ and $h$).  Let us first discuss the coefficient $z_{\text{linear}}^{(\sigma)}[\{f,h\}]$ of the linear term.  This coefficient vanishes for all of the $11$ permutations except for $(3,2,1)$ and $(4,3,2,1)$; for these two permutations, we find
    \beq
        z_{\text{linear}}^{(3,2,1)}[\{ f, h\} ] = - \frac{i}{\pi} z_{\text{linear}}^{(4,3,2,1)}[\{ f, h \}]= -i \int_0^\infty du\ f(u)h(u).
    \eeq
    In the current, these linear terms cancel at the order we are working to ($J^5$ or $1/J^5$), so we can ignore them.
    
    We proceed to the logarithmic terms.  It turns out that for all $11$ permutations, the coefficients $z_j^{(\sigma)}[\{f, h\}]$ can be expressed entirely in terms of the following three functionals:
    \bseq
        \begin{align}
            \rho_1[\{ f, h \} ] &=  \left(- \gamma +i \frac{\pi}{2} \right)h(0)+  \int_0^\infty du\ \ln u \frac{d}{du} \left[  f(u) h(-u) \right],\\
            \rho_2[\{ f, h \} ] &= - \left( \frac{7\pi^2}{24} + \frac{1}{2}\gamma^2 + i \frac{1}{2}\pi \gamma \right) h(0) + \left( \gamma + i \frac{\pi}{2}\right) \int_0^\infty du\ \ln u \frac{d}{du} \left[  f(u) h(-u) \right] \notag\\
            &\qquad + \frac{1}{2} \int_0^\infty du\ \ln^2 u \frac{d}{du} \left[  f(u) h(-u) \right] -\int_0^\infty du_1 du_2\ \frac{1}{u_2} \ln \frac{u_1 +  u_2}{u_1} \frac{\partial}{\partial u_1} \left[ f(u_1 + u_2) f(-u_1) h(-u_2) \right],\\
            \rho_3[\{ f, h \} ] &= \left( \gamma - i \frac{1}{2}\pi \right)^2 h(0) -2 \left( \gamma - i\frac{1}{2}\pi \right) \int_0^\infty du\ \ln u \frac{d}{du} \left[  f(u) h(-u) \right] \notag\\
            &\qquad+ \int_0^\infty du_1 du_2\ \ln u_1 \ln u_2 \frac{\partial }{\partial u_1}\frac{\partial }{\partial u_2} \left[ f(u_1 ) f(u_2) h(-u_1 -u_2 ) \right].
        \end{align}
    \eseq
    Table \ref{tab: asymptotic coefficients} contains our results for the coefficients $z_j^{(\sigma)}[\{ f, h\} ]$ of the asymptotic expansion.
    \begin{table}[htp]
        \caption{\label{tab: asymptotic coefficients}
        Leading log, sub-leading log, and sub-sub-leading log terms in $R^{(\sigma)}[\{f, h \}, \lambda]$ [see Eq. \eqref{eq: asymptotic form of R}].
        }
        \begin{ruledtabular}
        \begin{tabular}{ L C C C}
        \sigma \equiv (\sigma_1,\dots,\sigma_n)&
        z_{n-1}^{(\sigma)}[\{ f, h\} ] & z_{n-2}^{(\sigma)}[\{ f, h\} ] & z_{n-3}^{(\sigma)}[\{ f, h\} ]\\
        \colrule
        (1) & h(0) &  & \\
        (2,1) & -h(0) & \rho_1  & \\
        (2,3,1) & -\frac{1}{2} h(0) & - i \pi h(0) + \rho_1 &  \rho_2   \\
        (3,1,2) & h(0) & -2\rho_1 & \rho_3 \\
        (3,2,1) & 0 & 0 & -h(0)\\
        (2,3,4,1) & -\frac{1}{3} h(0) & -i \pi h(0) +\rho_1 & \frac{2}{3}\pi^2 h(0) + 2 \rho_2 \\
        (2,4,1,3) & \frac{1}{2}h(0) & i \pi h(0) -\frac{3}{2}\rho_1 & -i \pi \rho_1-\rho_2+\rho_3\\
        (3,1,4,2) & \frac{1}{6}h(0) & -\frac{1}{2}\rho_1 & \frac{2}{3}\pi^2 h(0) +i \pi \rho_1- \rho_2\\
        (3,4,1,2) & 0 & h(0) & ( 2 + i \pi )h(0) -2 \rho_1 \\
        (4,1,2,3) & -h(0) & 3 \rho_1 & -3 \rho_3\\
        (4,3,2,1) & 0 & -h(0) & -(2+ i \pi )h(0) + 2 \rho_1
        \end{tabular}
        \end{ruledtabular}
    \end{table}
    These results completely specify the integrals we need for $n=1$, $2$, and $3$, while for $n=4$, they provide the complete expansion except for the coefficient $z_0^{(\sigma)}[\{f, h\}]$ of the smallest term (the $\lambda$-independent constant); these remaining coefficients can also be written as lengthy functionals of $f$ and $h$ (including triple integrals), and we list their approximate numerical values in Table \ref{tab: z0 coefficients} for the two special cases corresponding to the zero bias conductance $G(T)$ and $I(T_1=0,T_2=0,V)$.
    
    \begin{table}[htp]
        \caption{\label{tab: z0 coefficients}
        Constant terms $z_0^{(\sigma)}[\{f,h\}, \lambda]$ for $n=4$ integrals in two special cases.
        }
        \begin{ruledtabular}
        \begin{tabular}{ L C C}
        \sigma \equiv (\sigma_1,\dots,\sigma_n)&
        z_0^{(\sigma)}[\{f,h\} ] \text{ for }f(v)=h(v) = v /\sinh v & z_0^{(\sigma)}[\{f, h \}] \text{ for } f(v) = \sinc v,\ h(v) = \cos v \\
        \colrule
        (2,3,4,1) & 2.24 + 1.06 i & 1.14 + 3.51 i \\
        (2,4,1,3) & 4.50 -3.12 i & 1.35 - 1.76 i  \\
        (3,1,4,2) & 1.48 - 7.24 i & 0.97 -6.02 i \\
        (3,4,1,2) & 3.51 -3.14 i & 0.37 - 3.14 i \\
        (4,1,2,3) & 6.76 - 3.20 i & 1.90 - 2.62 i   \\
        (4,3,2,1) & -3.95 & -1.49 
        \end{tabular}
        \end{ruledtabular}
    \end{table}
    Our asymptotic results are in good agreement with Monte Carlo evaluation \footnote{The authors acknowledge the Office of Advanced Research Computing (OARC) at Rutgers, The State University of New Jersey, for providing access to the Amarel cluster. URL: \url{http://oarc.rutgers.edu}}.  An example of this agreement is shown in Fig. \ref{fig: Numerics_example}.
    \begin{figure}[htp]
        \includegraphics[width=.6\linewidth]{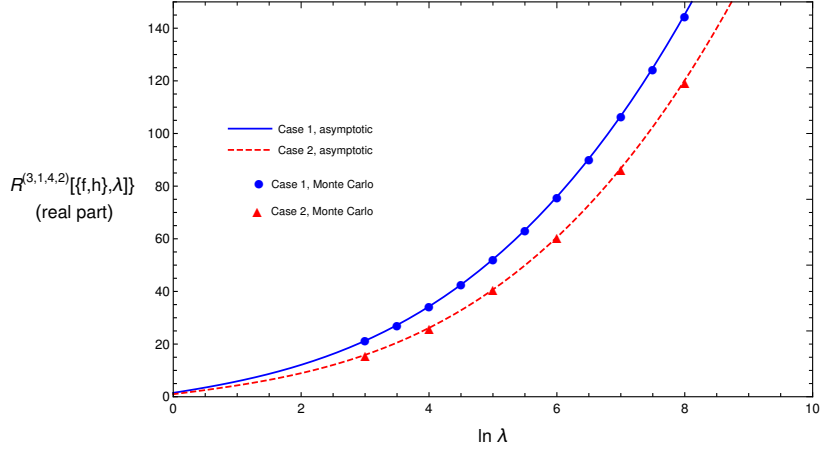}
        \caption{\label{fig: Numerics_example}Sample numerical checks of our asymptotic result for $R^{(3,1,4,2)}[\{f,h\},\lambda]$.  Case 1 is $f(v)=h(v)= v/ \sinh v$, which is used in the calculation of $G(T)$; case 2 is $f(v)= \cos v$ and $h(v) = \sinc v$, which is used in the calculation of $I(T_1=0,T_2=0, V)$ [and hence, $G(V)$].  Only the real part of $R^{(3,1,4,2)}[\{f,h\},\lambda]$ appears in the answer to the order we consider ($J^5$ or $1/J^5$), but the agreement for the imaginary part is similar.}
    \end{figure}
    
    The calculations that produce Table \ref{tab: asymptotic coefficients} are lengthy; to illustrate the method used, we derive the asymptotic expansion \eqref{eq: R21 asymptotic} in the main text.  The integral to be studied is
    \beq
        R^{(2,1)}[\{f, h \}, \lambda] = \int_0^\infty du_1\  \frac{e^{i \lambda u_1 } -f(u_1) }{u_1} h(-u_1). \label{eq: R12}
    \eeq
    We would like to separate the $\lambda$-dependent term of \eqref{eq: R12}, but cannot do so because $e^{i\lambda u_1}/u_1$ by itself diverges too strongly at $u_1 =0$.  We therefore integrate by parts, finding (note that $h$ falls off sufficiently rapidly at infinity so that the boundary contribution is zero)
    \beq
        R^{(2,1)}[\{f, h \}, \lambda] = R_1^{(2,1)}[ \{f, h \}, \lambda] + R_2^{(2,1)}[ \{f, h \} ],
    \eeq
    where
    \bseq
        \begin{align}
            R_1^{(2,1)}[ \{f, h \} ,\lambda] &= - \int_0^\infty du_1\ \ln u_1 \frac{d}{du_1} \left[ e^{i \lambda u_1 } h(-u_1)  \right],\\
            R_2^{(2,1)}[ \{f, h \} ] &=  \int_0^\infty du_1\ \ln u_1 \frac{d}{du_1} \left[ f(u_1) h(-u_1)  \right]. \label{eq: R2perm21}
        \end{align}
    \eseq
    We evaluate $R^{(2,1)}[ \{f, h \} ,\lambda]$ for large $\lambda$ using a contour argument based on example 1 in section 6.6 of Ref \cite{BenderOrszag}.  The essential idea is to turn the rapidly oscillating phase into a decaying exponential.
    
    Recall that any poles of $h$ are on the imaginary axis.  Write $C$ for the contour that starts at $0$ and extends to $i \infty$ going slightly to the right ($\text{Re}\ u_1 >0$) around each of the poles.  This contour $C$ taken in reverse, the original integration contour from $0$ to $\infty$, and a semicircular arc from $\infty$ to $i \infty$ form a closed contour that contains no poles.  Furthermore, it can be verified that the semicircular arc makes no contribution.  Therefore, the original contour can be replaced by $C$:
    \beq
        R_1^{(2,1)}[ \{f, h \} ,\lambda] = - \int_C du_1\  \ln u_1 \frac{d}{du_1} \left[ e^{i \lambda u_1 } h(-u_1)  \right].
    \eeq
    For large $\lambda$, the function $h$ can be replaced by its value at zero; the reason for this is that the difference $h(-u_1) - h(0)$  starts at linear order, which permits integration by parts:
    \begin{align}
        -\int_C du_1\ \ln u_1 \frac{d}{d u_1} \left[ e^{i \lambda u_1} (h(-u_1) - h(0) ) \right] &= \int_C du_1\ \frac{1}{u_1}(h(-u_1) - h(0) )  e^{i \lambda u_1} \\
        &= \int_C du_1\ \frac{d}{du_1} \left[ \frac{1}{u_1}(h(-u_1) - h(0) )  \frac{1}{i\lambda} e^{i \lambda u_1}  \right]\notag\\
        &\qquad - \int_C du_1\ \frac{d}{du_1} \left[ \frac{1}{u_1}(h(-u_1) - h(0) )\right]  \frac{1}{i\lambda} e^{i \lambda u_1}\\
        &= O\left( \frac{1}{\lambda} \right).
    \end{align}
    We have therefore shown
    \beq
        R_1^{(2,1)}[ \{f, h \} ,\lambda] = -\int_C du_1\ \ln u_1 \frac{d}{du_1} \left[ e^{i \lambda u_1} h(0) \right] + O\left(\frac{1}{\lambda}\right).
    \eeq
    Since there are no longer any poles, we can shift the contour $C$ to be exactly the positive imaginary axis; then the remaining integrals are elementary after the change of variables $s_1 = \lambda u_1$:
    \bseq
        \begin{align}
            R_1^{(2,1)}[\{ f, h\}, \lambda] &= -\int_0^\infty du_1\ \ln (i u_1) \frac{d}{du_1} \left[ e^{- \lambda u_1} h(0) \right] + O\left(\frac{1}{\lambda}\right)\\
            &= h(0) \left( - \ln \lambda - \gamma + i \frac{1}{2}\pi \right) + O\left( \frac{1}{\lambda} \right).
        \end{align}
    \eseq
    Adding this to Eq. \eqref{eq: R2perm21}, we obtain the second row of Table \ref{tab: asymptotic coefficients}.
    
    For the higher-order integrals, the basic strategy is the same: use integration by parts to rewrite the integral in a form that can be separated into a sum of simpler terms, shift integration contours to turn oscillating phases into decaying exponentials, and replace functions by their values at zero via integration by parts.  In the case of $\sigma=(4,3,2,1)$, this last step has to be done more carefully due to the linear divergence.

    \section{Additional checks}\label{sec: Additional checks}
    We begin this appendix by summarizing two alternate calculations we have done that yield the same series answer for the current that is obtained in the main text.  We then discuss some alternate ways of carrying out the integrals, again confirming our earlier answers.  Finally, we verify that we obtain the usual leading-order scaling of the anisotropic Kondo model.
    
    Rather than use the original definition \eqref{eq: I(t) basic def} of the time-evolving current $I(t)$ (as the time derivative of the number of electrons in one reservoir), we can instead calculate the expectation value of a local operator:
    \beq
        \widehat{I} = \text{Re}\left[ iJ \psi_{1a}^\dagger(0) \bm{\sigma}_{aa'} \psi_{2a'}(0) \cdot \mathbf{S}   \right].
    \eeq
    It can be shown by general arguments that $I(t) = \langle \Psi(t) | \widehat{I}  |\Psi(t)\rangle$.  We now present two equivalent ways of evaluating the right-hand side.
    
    The first check is to evaluate the expectation value $\langle \Psi(t) | \widehat{I} |\Psi(t) \rangle$ using the approach of Appendix \ref{sec: Evaluation of bilinears} (taking care to include the action of the impurity operator $\mathbf{S}$ on impurity states).  The result, for $N$ electrons, agrees with $I(t)$ as calculated in the main text.  The second check, which also confirms that $|\Psi(t)\rangle$ satisfies the Schrodinger equation, is to write $\langle \Psi(t) | \widehat{I} |\Psi(t) \rangle$ in an alternate form, as the derivative of an overlap between two states.  This is accomplished by means of the following simple result, which we present in a general setting.  Suppose the Hamiltonian $H$ consists of a ``reference'' Hamiltonian $H_{\text{ref}}$ plus terms that depend on a varying real parameter $\phi$:
    \beq
        H_{\phi} = H_{\text{ref}} + \sum_{j=1}^n f_j(\phi) \mathcal{O}_j,
    \eeq
    where the functions $f_j(\phi)$ and operators $\mathcal{O}_j$ are arbitrary.  We wish to calculate the expectation value of an operator (see below) in the time-dependent state $e^{-i H_{\phi_0} t}|\Psi\rangle$, where $|\Psi\rangle$ is an arbitrary initial state and $\phi = \phi_0$ corresponds to the physical Hamiltonian of interest.  Let $|\Psi_\phi \rangle$ be a family of states such that $|\Psi_{\phi_0}\rangle = |\Psi\rangle$.  We then have
    \beq
        \langle \Psi| e^{i H_{\phi_0} t} \left( \sum_{j=1}^n  f_j'(\phi_0) \mathcal{O}_j \right) e^{-i H_{\phi_0} t} |\Psi \rangle
        = i \frac{\partial}{\partial t}\frac{\partial}{\partial \phi}\biggr \rvert_{\phi=\phi_0}  \langle \Psi | e^{i H_{\phi_0} t} e^{-i H_\phi t} |\Psi_\phi \rangle,\label{eq: derivative formula}  
    \eeq
    as can be seen by doing the time derivative first on the right-hand side.  Thus, the time-dependent expectation value of a certain form of observable reduces to the calculation of an overlap between two states---one evolving with the physical value $\phi=\phi_0$, and the other with a varying value $\phi$.

    In the two-lead Kondo model, we calculate the current by introducing a varying parameter $\phi$ that is a relative phase between the tunneling terms $\psi_1^\dagger\psi_2$ and $\psi_2^\dagger \psi_1$.  To be precise, we set $f_1(\phi) = (e^{i\phi}-1), f_2(\phi)=(e^{-i \phi} -1), \mathcal{O}_1 = \psi_{1a}^\dagger(0) \bm{\sigma}_{a a'}  \psi_{2a'}(0)\cdot \mathbf{S}$, and $\mathcal{O}_2= \psi_{2a}^\dagger(0) \bm{\sigma}_{aa'}\psi_{1a'}(0)\cdot \mathbf{S} $ in Eq. \eqref{eq: derivative formula}.  The time-evolving wavefunction for arbitrary phase $\phi$ is found exactly using our formalism (essentially the only change is that the matrix that relates the lead $1$/lead $2$ basis to the odd/even basis depends on the varying phase), and the current is found as the derivative of the overlap.  The result for the current for $N$ electrons again agrees with the main text.  Note that this also provides confirmation that we have solved the time-dependent Schrodinger equation correctly, seeing as that is what is used in deriving the general formula \eqref{eq: derivative formula}.
    
    We proceed to some checks of our evaluations of integrals.  We have found the large-bandwidth asymptotic form of the basic steady-state integral $\varphi_n^{(\sigma)}(T_1=0,T_2=0,V)$ in an alternate way that agrees with the results of Appendix \ref{sec: Asymptotic evaluation of integrals} and also provides the analytical formula for the bandwidth-independent $g^5$ and $1/g^5$ terms in $G(V)$ in the main text.  We have also repeated the calculation of $G(T)$ in an alternate cutoff scheme in which the Fermi function smoothly drops to zero at large negative energies, rather than being sharply cut off.
    
    The basic integral that appears in our current series is given by Eq. \eqref{eq: varphi sharp cutoff Kondo} in the main text.  Using the notation of Appendix \ref{sec: Asymptotic evaluation of integrals}, the result obtained in the main text, in the special case of zero temperature, can be written as
    \beq
        \frac{1}{V} \lim_{t\to\infty} \varphi_n^{(\sigma)}(T_1=0,\mu_1=0; T_2=0,\mu_2=-V; t) \equiv \frac{1}{V} \varphi_n^{(\sigma)}(T_1= 0, T_2= 0, V) =  R^{(\sigma)}\left[\{ f, h\}, 2\frac{D}{V} -1 \right],
    \eeq
    where $f(v) = \sinc v$ and $h(v) = \cos v$.  The asymptotic expansion of $R^{(\sigma)}\left[\{ f, h\}, 2\frac{D}{V} -1 \right]$ for $D/V\gg1$ can be read off from Table \ref{tab: asymptotic coefficients} and the third column of Table \ref{tab: z0 coefficients}; our task is to calculate $\frac{1}{V} \varphi_n^{(\sigma)}(T_1= 0, T_2= 0, V)$ in an alternate way as a check.
    
    An alternate approach in this special case is to do the position integrals in Eq. \eqref{eq: varphi sharp cutoff Kondo} before the momentum integrals, arriving at the long-time limit by means of the Laplace transform.  Recall that the long-time limit of a function $F(t)$ is determined by the behavior of its Laplace transform near the origin:
    \beq
        \lim_{t\to\infty} F(t) = \lim_{s\to0^+} s \widetilde{F}(s),\qquad \text{where } \widetilde{F}(s) = \int_0^\infty dt\ e^{- s t} F(t).
    \eeq
    Taking the Laplace transform and doing the position integrals, we find
    \begin{multline}
        s \widetilde{\varphi}^{(\sigma)}(T_1,\mu_1; T_2,\mu_2; s) = \left(\frac{i}{2}\right)^{n-1}\int_{-D}^D dk_1\dots dk_n\  \left\{ \prod_{j=1}^{n-1} \left[ \fermifn_1(k_j) + \fermifn_2(k_j) \right]  \right\}
        \left[ \fermifn_1(k_n) - \fermifn_2(k_n) \right] \\
        \times \prod_{\ell=1}^{n-1} \frac{i}{ k_{\sigma_1} + \dots + k_{\sigma_\ell} - k_1 - \dots - k_\ell  + i s}.
    \end{multline}
    The point of these manipulations is that if we set $T_1=T_2=0$, we obtain a form that is tractable analytically.  After some relabelings of coordinates, we obtain
    \begin{multline}
        s \widetilde{\varphi}^{(\sigma)}(T_1=0,\mu_1=0;T_2= 0,\mu_2= -V; s) = i^{n-1} \sum_{m=0}^{n-1}\left(\frac{1}{2}\right)^m \binom{n-1}{m} \int_{-D}^{-V} dk_1 \dots dk_{n-m+1}\\
        \times \int_{-V}^0 dk_{n-m}\dots dk_n\ 
        S_{k_1 \dots k_{n-1}} \prod_{\ell=1}^{n-1} \frac{i}{ k_{\sigma_1} + \dots + k_{\sigma_\ell} - k_1 - \dots - k_\ell  + i s},
    \end{multline}
    where the symmetrizer $S_{k_1\dots k_{n-1}}$ acts on the first $n-1$ momenta of any function $X$ via
    \beq
        S_{k_1 \dots k_{n-1}} X(k_1,\dots,k_n)  = \frac{1}{(n-1)!}\sum_{\sigma' \in \text{Sym}(n-1)} X(k_{\sigma_1'}, \dots, k_{\sigma_{n-1}'},k_n ).
    \eeq
    By lengthy computer evaluation, these integrals were done analytically for all of the $11$ permutations; then the limit $s\to0^+$ was taken and an expansion was done for large $D/V$.  The final results are conveniently written in the following form:
    \begin{align}
        \frac{1}{V}\varphi_n^{(\sigma)}(T_1=0,T_2=0,V) &= \frac{1}{V} \lim_{s\to0^+}s \widetilde{\varphi}^{(\sigma)}(T_1=0,\mu_1=0;T_2= 0,\mu_2= -V;s)\\
        &\overset{D\gg V}{\longrightarrow} b_{\text{linear}}^{(\sigma)}\frac{D}{V} + \sum_{n=0}^3 b_n^{(\sigma)} \sum_{m=0}^n \frac{1}{m!}\ln^m \frac{D}{V}
    \end{align}
    where $b_{\text{linear}}^{(\sigma)}$ is zero for all $11$ permutations except for $b_{\text{linear}}^{(3,2,1)} = -\frac{i}{\pi}b_{\text{linear}}^{(4,3,2,1)} = -i \pi/2$, and where the remaining coefficients are listed in Table \ref{tab: Asymptotics for zero temperature}.
    \begin{table}[htp]
    \caption{\label{tab: Asymptotics for zero temperature}
    Asymptotic expansion of $R^{(\sigma)}[\{ f, h\}, \lambda]$ for $f(v)=\sinc v$, $h(v) = \cos v$. }
        \begin{ruledtabular}
        \begin{tabular}{L C C C C}
        \sigma \equiv (\sigma_1,\dots,\sigma_n) & b_3^{(\sigma)} & b_2^{(\sigma)} & b_1^{(\sigma)} & b_0^{(\sigma)} \\
        \hline
        (1) &  &  & & 1 \\
        (2,1) &  &  & -1 & i\pi/2  \\
        (2,3,1) &  & -1 & -i \pi/2 & -\pi^2/12  \\
        (3,1,2) &  & 2 & - i\pi & -\pi^2/3   \\
        (3,2,1) &  & 0 & 0 & -1 \\
        (2,3,4,1) & -2 & -i \pi & \pi^2/2 & \left[ \zeta(3) +i \pi^3/3 \right]/2 \\
        (2,4,1,3) & 3 & i \pi/2 & \pi^2/4 & -\left[ 3 \zeta(3)+ i \pi^3/3 \right]/4 \\
        (3,1,4,2) & 1 & -i \pi/2 & \pi^2/4 & - \left[ \zeta(3) + i 2\pi^3/3\right]/4 \\
        (3,4,1,2) & 0 & 2 & 2 &  2 -\pi^2/8 - 3 \ln2 - i \pi  \\
        (4,1,2,3) & -6 & 3i \pi & \pi^2 & \left[3 \zeta(3) -i\pi^3 /2  \right]/2  \\
        (4,3,2,1) & 0 & -2 & -2 & 3\ln2 -2 \\
        \end{tabular}
        \end{ruledtabular}
    \end{table}
    These results are in good agreement with Table \ref{tab: asymptotic coefficients} and the third column of Table \ref{tab: z0 coefficients}.

    Another check is provided by varying the cutoff scheme that regulates the UV divergences of the model.  The cutoff scheme we have used amounts to multiplying the Fermi function by a Heaviside function $\Theta(k + D)$ (the cutoff of large positive energies turns out to be unimportant due to the exponential suppression of the Fermi function there).   For an alternate cutoff scheme, we replace the Heaviside function by a smoothly decaying function (which is chosen for convenience to have the form of a Fermi function); the resulting Fourier transform for the cutoff Fermi function is
    \beq
        \int_{-\infty}^\infty dk\ \fermifn(T,\mu_, k) \frac{1}{e^{-\frac{1}{T}(D'+k)} +1}e^{-i k y} = \frac{\pi}{i}T \frac{ e^{i D' y} - e^{-i \mu y}  }{\sinh(\pi T y)},
    \eeq
    again with exponentially small corrections $O(e^{-(D'+\mu)/T})$.  We write the cutoff as $D'$ as a reminder that, while it plays the same role, it is not identical to the sharp cutoff $D$ except in the case $T=0$.  In this alternate cutoff scheme, we repeated the calculation of the integrals $R^{(\sigma)}[\{f, h\}, \lambda]$ by Monte Carlo integration at several logarithmically spaced values of $\lambda$.  The results indicate that $D'$ and $D$ yield equivalent answers in the large-bandwidth regime; we have shown this analytically for some of the integrals using the contour method described in Appendix \ref{sec: Asymptotic evaluation of integrals}. 
    
    Still another check is obtained by repeating the calculation allowing anisotropy in the Kondo interaction.  As shown in Appendix \ref{sec: Kondo crossing states in the general case}, the anisotropy changes the $\mathcal{T}$ matrix that appears in the wavefunction.  The same series answer for the current is obtained, with the only change being a modification of the spin sums $W_n^{(\sigma)}(J)$.  The leading log results are
    \begin{multline}
        G(V)  = \frac{3\pi^2}{4 }G_0 \Biggr[ \frac{2}{3} g_{\perp}^2 + \frac{1}{3} g_z^2 + 4 g_{\perp}^2 g_z \ln \frac{D}{V} + 12  \left( \frac{1}{3}g_{\perp}^4 + \frac{2}{3} g_{\perp}^2 g_z^2  \right)\ln^2 \frac{D}{V}
        + 32 \left( \frac{2}{3} g_{\perp}^4 g_z + \frac{1}{3} g_{\perp}^2 g_z^3  \right) \ln^3 \frac{D}{V} + O(g^6) \Biggr],
    \end{multline}
    and the same for $G(T)$ with with $V$ replaced by $T$.  The Callan-Symanzik equation is satisfied with the following beta functions at leading order:
    \bseq
        \begin{align}
            \beta_{g_{\perp}}(g_{\perp}, g_z) &= - 2  g_{\perp} g_z +O (g^3),  \\ 
            \beta_{g_z} (g_{\perp}, g_z ) &= - 2 g_{\perp}^2 + O(g^3),
        \end{align}
    \eseq
    which are standard \cite{Hewson}.

    \section{Cutoff artifact in the time-dependent magnetization}\label{sec: Cutoff artifact in the time-dependent magnetization}
    
    In this appendix, we show that the unconventional cutoff scheme we have used in this paper leads to an extra log divergence in a toy calculation (relative to the conventional scheme).  We also provide a way to modify our scheme to correct this, recovering the conventional answer.  We suspect that this same phenomenon occurs in the calculation of the current, leading to a cutoff ``artifact'' of the form $g^4 \ln D$ or $g^5 \ln^2 D$ that changes the third-order coefficient of the beta function [$\beta_3$ in Eq. \eqref{eq: beta small J}].  Details of the calculations of this section can be found in Ref. \cite{Culver_thesis}.
    
    We consider the time-evolving expectation value of the impurity magnetization $S^z$ evaluated with three different cutoffs: a cutoff $D_H$ on the Hamiltonian, a cutoff $D_{\rho}$ on the initial density matrix $\rho$, and a cutoff $D_{\text{proj}}$ on the time-evolving density matrix.  This last cutoff is implemented by replacing $e^{-i H t} \rho e^{i H t} \to P_{D_{\text{proj}}} e^{-i H t} \rho e^{i H t} P_{D_{\text{proj}}}$, where $P_{D_{\text{proj}}}$ is the projection operator onto the modes within $[-D_\text{proj}, D_\text{proj}]$. Conventional calculations have $D_H= D_\rho = D_{\text{proj}}$; indeed, once $D_H$ is finite, the other two cutoffs make no difference as long as neither is less than $D_H$.  Our calculation in the main text used $D_H = D_{\text{proj}}=\infty$ with $D_\rho$ finite, but this leads to a cutoff artifact in the impurity magnetization (as we show below).  Setting $D_H=\infty$ and $D_\rho = D_{\text{proj}}$ removes the artifact, recovering the conventional answer.  Presumably, repeating the calculation of the current in the main text with $D_\rho = D_{\text{proj}}$ (instead of $D_\rho = \infty$) should yield the correct coefficient $\beta_3$ in the beta function; however, this calculation appears to be considerably more difficult than the $D_\rho = \infty$ case.
    
    The time-dependent magnetization up to second order, starting from an initial state with impurity spin $a_0$ and working at zero temperature, is found to be
    \beq
        \langle S^z \rangle_t = S_{a_0 a_0}^z \left[ 1 - (2 \rho J_{\perp})^2 X(D t) + \dots \right],
    \eeq
where $X(Dt)$ is given in the different schemes by (for large bandwidth)
\beq
    X(D t) = 
    \begin{cases}
        \ln \left(D t\right) + 1+ \gamma - \ln 2  & \text{for }D_H = D_\rho = D_{\text{proj}}\equiv D\ (\text{conventional scheme}),\\
        2 \left[ \ln \left(D t\right) + 1+ \gamma \right] & \text{for }D_H  =D_{\text{proj}} =\infty,\ D_\rho \equiv D\ (\text{scheme used in main text}), \\
        \ln \left(D t\right) + 1+ \gamma - \ln 2 & \text{for }D_H = \infty,\ D_{\text{proj}} = D_\rho \equiv D\ \text{(projection scheme).}
    \end{cases}
\eeq
The first two cases can be read off with a slight generalization of a calculation done in Ref. \cite{AndersSchiller_PRB}; we have done the second two cases with our wavefunction method \cite{Culver_thesis} (thus the second case is done two different ways, and they agree).  Thus, we see that the cutoff scheme we use in the main text can lead to an extra $\ln D$ term in an observable.  Recall from the main text that this is exactly the kind of term that is missing from our calculation, that could change the coefficient $\beta_3$ of the beta function.  We also note here that having $D_H=\infty$ seems essential in our method at present, in order to get derivatives and delta functions in position space that make our ``inverse problems'' solvable.

\end{widetext}

\bibliography{references}

\end{document}